\documentclass[reprint,prd,amsmath,aps,floats,amssymb,floatfix, superscriptaddress,nofootinbib,twocolumn,preprintnumbers]{revtex4-1}  

\setcitestyle{authoryear,round}
\usepackage[T1]{fontenc}
\usepackage{aecompl}

\usepackage{newtxtext,newtxmath}
\usepackage{mdframed,bm}
\usepackage[T1]{fontenc}
\usepackage{ae,aecompl}
\usepackage{graphicx}	
\usepackage{amsmath}	
\usepackage{amssymb}	
\usepackage{booktabs}
\usepackage{color}
\usepackage{enumitem}

\usepackage{multirow}
\usepackage[pdftex,dvipsnames]{xcolor}

\usepackage[colorlinks=true]{hyperref}

\hypersetup{
    urlcolor=black,
    citecolor=blue,
    linkcolor=blue,
  }%

\usepackage{physics}
\usepackage{cleveref}
\usepackage{comment}
\usepackage{array}
\usepackage{slashed}
\usepackage[dvipsnames]{xcolor}

\usepackage{xspace}
\usepackage{xcolor}
\usepackage{float}
\restylefloat{table}

\usepackage{physics}

\newcommand{\aap}{A\&A}
\newcommand{\apjs}{ApJS}
\newcommand{\aj}{A.J.}
\newcommand{\mnras}{MNRAS}

\newcommand{\jcap}{JCAP}

\definecolor{bostonuniversityred}{rgb}{0.8, 0.0, 0.0}

\newcommand{\metacal}{\textsc{Metacalibration}}
\newcommand{\redmagic}{\textsc{redMaGiC}}
\newcommand{\buzzard}{\textsc{Buzzard}}
\newcommand{\cosmolike}{\textsc{CosmoLike}}

\allowdisplaybreaks
\begin{document}

\title{Dark Energy Survey Year 3 Results: \\ High-precision measurement and modeling of galaxy-galaxy lensing}

\author{J.~Prat} \email{Corresponding author: jprat@uchicago.edu}
\affiliation{Department of Astronomy and Astrophysics, University of Chicago, Chicago, IL 60637, USA}
\affiliation{Kavli Institute for Cosmological Physics, University of Chicago, Chicago, IL 60637, USA}
\author{J.~Blazek}
\affiliation{Department of Physics, Northeastern University, Boston, MA 02115, USA}
\affiliation{Laboratory of Astrophysics, \'Ecole Polytechnique F\'ed\'erale de Lausanne (EPFL), Observatoire de Sauverny, 1290 Versoix, Switzerland}
\author{C.~S{\'a}nchez}
\affiliation{Department of Physics and Astronomy, University of Pennsylvania, Philadelphia, PA 19104, USA}
\author{I.~Tutusaus}
\affiliation{Institut d'Estudis Espacials de Catalunya (IEEC), 08034 Barcelona, Spain}
\affiliation{Institute of Space Sciences (ICE, CSIC),  Campus UAB, Carrer de Can Magrans, s/n,  08193 Barcelona, Spain}
\author{S.~Pandey}
\affiliation{Department of Physics and Astronomy, University of Pennsylvania, Philadelphia, PA 19104, USA}
\author{J.~Elvin-Poole}
\affiliation{Center for Cosmology and Astro-Particle Physics, The Ohio State University, Columbus, OH 43210, USA}
\affiliation{Department of Physics, The Ohio State University, Columbus, OH 43210, USA}
\author{E.~Krause}
\affiliation{Department of Astronomy/Steward Observatory, University of Arizona, 933 North Cherry Avenue, Tucson, AZ 85721-0065, USA}
\author{M.~A.~Troxel}
\affiliation{Department of Physics, Duke University Durham, NC 27708, USA}
\author{L.~F.~Secco}
\affiliation{Department of Physics and Astronomy, University of Pennsylvania, Philadelphia, PA 19104, USA}
\affiliation{Kavli Institute for Cosmological Physics, University of Chicago, Chicago, IL 60637, USA}
\author{A.~Amon}
\affiliation{Kavli Institute for Particle Astrophysics \& Cosmology, P. O. Box 2450, Stanford University, Stanford, CA 94305, USA}
% ...

\collaboration{DES Collaboration}


\date{\today}

\label{firstpage}
\begin{abstract}
We present and characterize the galaxy-galaxy lensing signal measured using the first three years of data from the Dark Energy Survey (DES Y3) covering 4132 deg$^2$. These galaxy-galaxy measurements are used in the DES Y3 3$\times$2pt cosmological analysis, which combines weak lensing and galaxy clustering information. We use two lens samples: a magnitude-limited sample and the \textsc{redMaGic} sample, which span the redshift range $\sim 0.2-1$ with 10.7 M and 2.6 M galaxies respectively. For the source catalog, we use the \metacal \ shape sample, consisting of $\simeq$100 M galaxies separated into 4 tomographic bins. Our galaxy-galaxy lensing estimator is the mean tangential shear, for which we obtain a total S/N of $\sim$148 for \textsc{MagLim} ($\sim$120 for \textsc{redMaGic}), and $\sim$67 ($\sim$55) after applying the scale cuts of 6 Mpc/$h$. Thus we  reach  percent-level statistical precision, which requires that our modeling and systematic-error control be of comparable accuracy. The tangential shear model used in the 3$\times$2pt cosmological analysis includes lens magnification, a five-parameter intrinsic alignment model (TATT), marginalization over a point-mass to remove information from small scales and a linear galaxy bias model validated with higher-order terms. We explore the impact of these choices on the tangential shear observable and study the significance of effects not included in our model, such as reduced shear, source magnification and source clustering. We also test the robustness of our measurements to various observational and systematics effects, such as the impact of observing conditions, lens-source clustering, random-point subtraction, scale-dependent \metacal \ responses, PSF residuals, and B-modes.  \\
\end{abstract}

\preprint{DES-2020-0551}
\preprint{FERMILAB-PUB-21-248-AE}
\maketitle



\section{Introduction}
\label{sec:intro}

Gravitational lensing is caused by light traveling in a curved space time, according to some gravitational potential. When the light of background (source) galaxies passes close to foreground (lens or tracer) galaxies it gets perturbed, distorting the image of the source galaxies we observe. This distortion happens both for the shape and size of the source images, due to the effect of the shear and magnification, respectively. The amount of distortion is correlated with the properties of the lens sample and the underlying dark matter large scale structure it traces. In this work we measure the correlation between galaxy shapes and the lens galaxy positions, usually called galaxy-galaxy lensing or galaxy-shear correlations. A few estimators of this correlation have been explored in the literature, including the most basic stacked tangential shear estimator which was used in the first detection of galaxy-galaxy lensing by \citet{Brainerd_1996}, the surface mass density excess \citep{Sheldon_2004} which is independent of the source redshift distribution in the absence of photometric errors, the annular differential surface density estimator proposed by \citet{Baldauf_2010} which removes small-scale information that propagates to larger scales, the estimator proposed by \citet{park2020localizing} that involves a linear transformation of the tangential shear quantity, and 2D tangential shear estimators reviewed in \citet{Dvornik_2019} that use positions and ellipticities of individual source galaxies, rather than using the ensemble properties. The mean tangential shear is the estimator on which all the rest are based and the one we choose in this work due to its simplicity in the measurement and modeling, for instance dealing with source redshift uncertainties.

Galaxy-galaxy lensing and in particular the tangential shear can be used to extract cosmological information using their well-understood large scales in combination with other probes such as galaxy clustering and/or CMB lensing such as in \citet*{Kwan_2016}, \citet{Baxter_2016}, \citet{van_Uitert_2018}, \citet{Joudaki_2017}, \citet*{Y1GGL}, or in \citet{Baldauf_2010}, \citet{Mandelbaum_2013}, \citet{Singh_2019} using the annular differential surface density estimator. Galaxy-galaxy lensing can also be used to characterize the largely uncertain galaxy-matter connection at small scales  (e.g. \citealt{Choi2012}, \citealt{Yoo_2012}, \citealt{Clampitt_2016} or \citealt*{Park_2016}), and also to construct ratios of tangential shear measurements sharing the same lens sample to extract mostly geometrical information from small scales without having to model the galaxy-matter connection (e.g. \citealt{Jain2003}, \citealt{Mandelbaum_2005}, \citealt*{Y1GGL}, \citealt{Hildebrandt_2020}, \citealt{Giblin_2020}). Recently there have also been studies using small and large scales to obtain cosmological parameters in combination with other probes using emulators to model the small scales e.g. \citet{Wibking_2019}. 

In this work we present and characterize the galaxy-galaxy measurements obtained using the first three years of observations from the Dark Energy Survey (DES Y3). At large scales (>6~Mpc/$h$), these measurements are used in combination with galaxy clustering and cosmic shear measurements to constrain cosmological parameters \citep{y3-3x2ptkp}. At small scales  (<6~Mpc/$h$) they are used to construct ratios of tangential shear measurements sharing the same lens sample for the DES Y3 shear-ratio probe described in \citet*{y3-shearratio}. The DES Y3 shear-ratio probe is used as an additional independent likelihood to the three two-point correlation functions described above and is able to increase the self-calibration of systematics or nuisance parameters in our model, such as those corresponding to intrinsic alignments, source redshifts and shear calibration.

The combination of galaxy-galaxy lensing, cosmic shear and galaxy clustering, usually referred to as 3$\times$2pt, is a powerful combination which is very robust to systematics and is able to constrain cosmological parameters at the late-time Universe, such as the amount of matter in the Universe, $\Omega_m$, the parameter describing the amplitude of the clustering, $\sigma_8$ and the parameter describing the equation of state of dark energy, $w$. Galaxy-galaxy lensing is a key ingredient of this analysis, which (i) breaks the degeneracy between the galaxy bias --- the relation between the observable galaxies and the underlying dark matter density field --- and $\sigma_8$ together with galaxy clustering, (ii) provides cosmological information, both through the geometrical and power spectrum dependence, and (iii) improves the self-calibration of almost all the nuisance parameters in the analysis, being particularly crucial to constrain the Intrinsic Alignment parameters, for which we do not currently have a reliable way to put an external informative prior on. Within the DES Y3 3$\times$2pt release, this work is responsible for properly characterizing the galaxy-galaxy lensing measurements that will be used in this combination by performing a series of robustness and null tests, validating both the measurement and modeling pipelines (including comparing their outputs to independent codes), and testing the significance of higher-order effects not included in our fiducial model. Besides testing the large scales that will be used in the 3$\times$2pt combination, we also validate and characterize the tangential shear measurements in the whole range of scales between 2.5 and 250 arcmin, both to serve as testing for the DES Y3 shear-ratio analysis using small scales \citep*{y3-shearratio}, and also to facilitate potential subsequent analysis using this same data, e.g. \citet*{y3-HOD}, where a halo occupation distribution model is used to characterize the galaxy-matter connection.

The galaxy-galaxy lensing measurements presented here are the highest signal-to-noise measurements to date with a total S/N of $\sim$120 ($\sim$55 with scale cuts of >6~Mpc/$h$) for the \redmagic \ sample, which is a significant increase with respect to the total S/N of 73 obtained in the same range of scales for the DES Y1 galaxy-galaxy lensing analysis from \citet*{Y1GGL}. It is even larger using a denser flux limited lens sample \citep{y3-2x2maglimforecast}, the \textsc{MagLim} sample, with a S/N of $\sim$148 ($\sim$67 with scale cuts). Other recent galaxy-galaxy lensing measurements used in cosmological analyses include the galaxy-galaxy lensing power spectra results using BOSS and 2dFLenS lenses with KiDS-1000 sources \citep{Heymans_2020} or in \citet{van_Uitert_2018} using GAMA (Galaxy and Mass Assembly) lenses and KiDS-450 as sources. Given the improvement in S/N of the current measurements with respect to previous analyses, several advancements in the modeling have been required. Major differences with respect to the fiducial DES Y1 3$\times$2pt analysis consist of including lens magnification and a five-parameter Intrinsic Alignment model (the Tidal Alignment Tidal Torquing model known as TATT \citealt{Blazek_2019}, and used in \citealt{Samuroff_2019} using DES Y1 data) in the fiducial tangential shear modelling. Also, due to the non-locality of the tangential shear estimator, we have adopted the scheme proposed in \citet{MacCrann_2019}, which allows us to analytically marginalize over a point-mass by applying a transformation in the tangential shear covariance, effectively removing the small scales information that propagates to larger scales in the tangential shear measurement. 
In our measurements, we now include the boost factor correction in the fiducial estimator, which effectively corrects for the impact of lens-source clustering on the redshift distributions. Additionally, we measure the tangential shear around two different lens samples: the \redmagic \ sample constituted of photometrically selected luminous red galaxies (LRGs) \citep{y3-galaxyclustering}, and a four times denser flux limited sample described in \citet{y3-2x2ptaltlensresults}. The photometric redshift distributions of the lens samples are calibrated using cross-correlations with the BOSS sample and, in the case of the magnitude-limited sample, also using a SOMPZ scheme \citep{y3-lenswz, y3-2x2ptaltlenssompz}. Both the shear and source redshift calibrations have been largely improved keeping up with the decrease of statistical uncertainties, using image simulations \citep{y3-imagesims} to calibrate the \metacal\ shape measurements from \citet*{y3-shapecatalog} and a state-of-the art methodology to obtain and calibrate the source redshift distributions described in \citet*{y3-sompz} and \citet*{y3-sourcewz}.

This paper is organized as follows. Section \ref{sec:data} describes the different lens and source galaxy DES data catalogues that are used throughout this work. Section \ref{sec:measurement} describes the details of the galaxy-galaxy lensing measurements using those data, and discusses the impact of different choices and configurations in the measurement scheme. Next, in Section \ref{sec:theory} we present all the details regarding the fiducial model utilized to describe the measurements above, and we examine the relative contribution of different terms in the modeling. Section \ref{sec:model_validation} describes several modeling effects that are not included in the fiducial model, and determines their importance at different angular scales. In Section \ref{sec:sys_tests} we perform a series of tests at the data level, to ensure the robustness of the measurements to different potential sources of systematic errors. In Section~\ref{sec:discussion_future} we summarize the impact of each of the measurement and model components and their uncertainty. Finally we conclude in Section~\ref{sec:conclusions}.

\section{Data}
\label{sec:data}

The Dark Energy Survey is a photometric survey that  covers about one quarter  of the southern sky to a depth of $r>24$, imaging about 300 million galaxies in 5 broadband filters ($grizY$) up to redshift $z \sim 1.4$ \citep{Flaugher2015,DES2016}. In this work we use data from 4132 deg.$^2$ of the first three years of observations (DES Y3). Next we describe the lens and source galaxy samples used in this work, which are the same samples used in the DES Y3 3$\times$2pt analysis \citep{y3-3x2ptkp}, and their corresponding redshift distributions which are shown in Figure \ref{fig:nzs}. 

\subsection{Lens galaxy catalogs}

We use two different lens galaxy catalogs: the \textsc{redMaGiC} sample, described in detail and characterized in \citet{y3-galaxyclustering}, and a magnitude-limited sample, which is optimized in simulations in \citet{y3-2x2maglimforecast} and characterized and described on data in \citet{y3-2x2ptaltlensresults}. In Table~\ref{tab:samples} we include a summary description for each of the lens samples, with the number of galaxies in each redshift bin, number density, linear galaxy bias values and magnification parameters from \citet*{y3-2x2ptmagnification}. 

\subsubsection{redMaGiC sample}
One of the lens galaxy samples used in this work is a subset of the DES Y3 Gold Catalog \citep{y3-gold} selected by \textsc{redMaGiC} \citep{Rozo2015}, which is an algorithm designed to define a sample of luminous red galaxies (LRGs) with high quality photometric redshift estimates. It selects galaxies above some luminosity threshold based on how well they fit a red sequence template, calibrated using redMaPPer \citep{Rykoff2014,Rykoff2016} and a subset of galaxies with spectroscopically verified redshifts. The cutoff in the goodness of fit to the red sequence is imposed as a function of redshift and adjusted such that a constant comoving number density of galaxies is maintained. 

In the DES Y3 3$\times$2pt analysis, \textsc{redMaGiC} galaxies are used as a lens sample for the clustering and galaxy-galaxy lensing measurements. Weights are assigned to \textsc{redMaGiC} galaxies such that spurious correlations with observational systematics are removed. The methodology used to assign weights is described in \citet{y3-galaxyclustering}.
\textsc{redMaGiC} galaxies are split in five different tomographic bins, which are chosen prioritizing minimal redshift overlap between non-consecutive bins, and also taking into account that at $z=0.65$ the catalog changes from the so-called \textit{high density} sample to the so-called \textit{high luminosity} sample.  The high-density sample corresponds to a luminosity threshold of $L_{\min}=0.5L_\star$, where $L_\star$ is the characteristic luminosity of the luminosity function, and comoving number density of $\bar{n} = 10^{-3} \; (h/{\rm Mpc})^3$. The high luminosity sample is characterized by $L_{\min} = L_\star$ and $\bar{n} = 4 \times 10^{-4} \; (h/{\rm Mpc})^3$. Then, the first three redshift bins of the \textsc{redMaGiC} sample are obtained from the high density sample and the two higher redshift bins from the high luminosity sample. In comparison, in the DES Y1 3$\times$2pt analysis, the first three redshift bins of the \textsc{redMaGiC} sample were also obtained from the high density sample, the fourth $z$-bin also from the high luminosity sample but the fifth $z$-bin was obtained from an even \textit{higher-luminosity} sample, as was described in \citet{Elvin_Poole_2018}. Other differences with respect to the \textsc{redMaGiC} Y1 catalog include different limits in the redshift binning and the different photometry used to select the galaxies. In Y1 \texttt{mag\_auto} photometry  was used while in Y3 we employ \texttt{SOF} (single-object fitting), which could lead to different selection properties \citep{y3-galaxyclustering}. Both photometries are described in \citet{y3-gold}. Besides this, the photometric calibration process was also different: in Y1 we used the stellar locus regression code \citep{y1gold} while in the Y3 catalog we used the Forward Global Calibration Method \citep{Burke2018} and the dereddening maps described in \citet{Schlegel1998}. Finally, the new \redmagic \ code\footnote{\texttt{\scriptsize{https://github.com/erykoff/redmapper/releases/tag/v0.5.1}}} assumes that the correlation between intrinsic red sequence galaxy colors is very large. That is, if a galaxy is intrinsically redder than the mean red-sequence model in the $m_r-m_i$ color then it will also be intrinsically redder than the mean in $m_i-m_z$.

\subsubsection{\textsc{MagLim} sample}

In addition to the \textsc{redMaGiC} sample, we also use a magnitude-limited sample, which is chosen as fiducial in the 3$\times$2pt cosmological analysis. In this sample, galaxies are selected with a magnitude cut that evolves linearly with the photometric redshift estimate: $i < a z_{\rm phot} + b$. The optimization of this selection, using the DNF (Directional Neighbourhood Fitting) photometric redshift estimates \citep{DeVicente2016}, yields $a=4.0$ and $b=18$. This optimization was performed taking into account the trade-off between number density and photometric redshift accuracy, propagating this to its impact in terms of cosmological constraints obtained from galaxy clustering and galaxy-galaxy lensing in \citet{y3-2x2maglimforecast}. Effectively this selects brighter galaxies at low redshift while including fainter galaxies as redshift increases.  Additionally,  we apply a lower cut to remove the most luminous objects, $i > 17.5$. This sample has a galaxy number density of more than four times that of the \textsc{redMaGiC} sample but the redshift distributions are $\sim30\%$ wider on average. This sample is split into 6 redshift bins, as defined in Table.~\ref{tab:samples}, but the two highest redshift bins have been excluded from the 3$\times$2pt cosmological analysis as detailed in \citet{y3-3x2ptkp}. The redshift binning was chosen to minimize the overlap in the redshift distributions, and in \citet{y3-2x2maglimforecast} there is a test showing that changing the binning does not impact the cosmological constraints. See \citet{y3-2x2ptaltlensresults} for more details on this sample.

\subsection{Source galaxy catalog}

For the background sources we use the shape catalog described in \citet*{y3-shapecatalog} and \citet{y3-piff}, which is based on the \textsc{Metacalibration} technique \citet{Huff2017,Sheldon2017}, which is able to accurately measure weak lensing shear using the available imaging data. Remaining biases using this methodology are calibrated in \citet{y3-imagesims} using image simulations. 

\begin{figure}
\begin{center}
\includegraphics[width=0.45\textwidth]{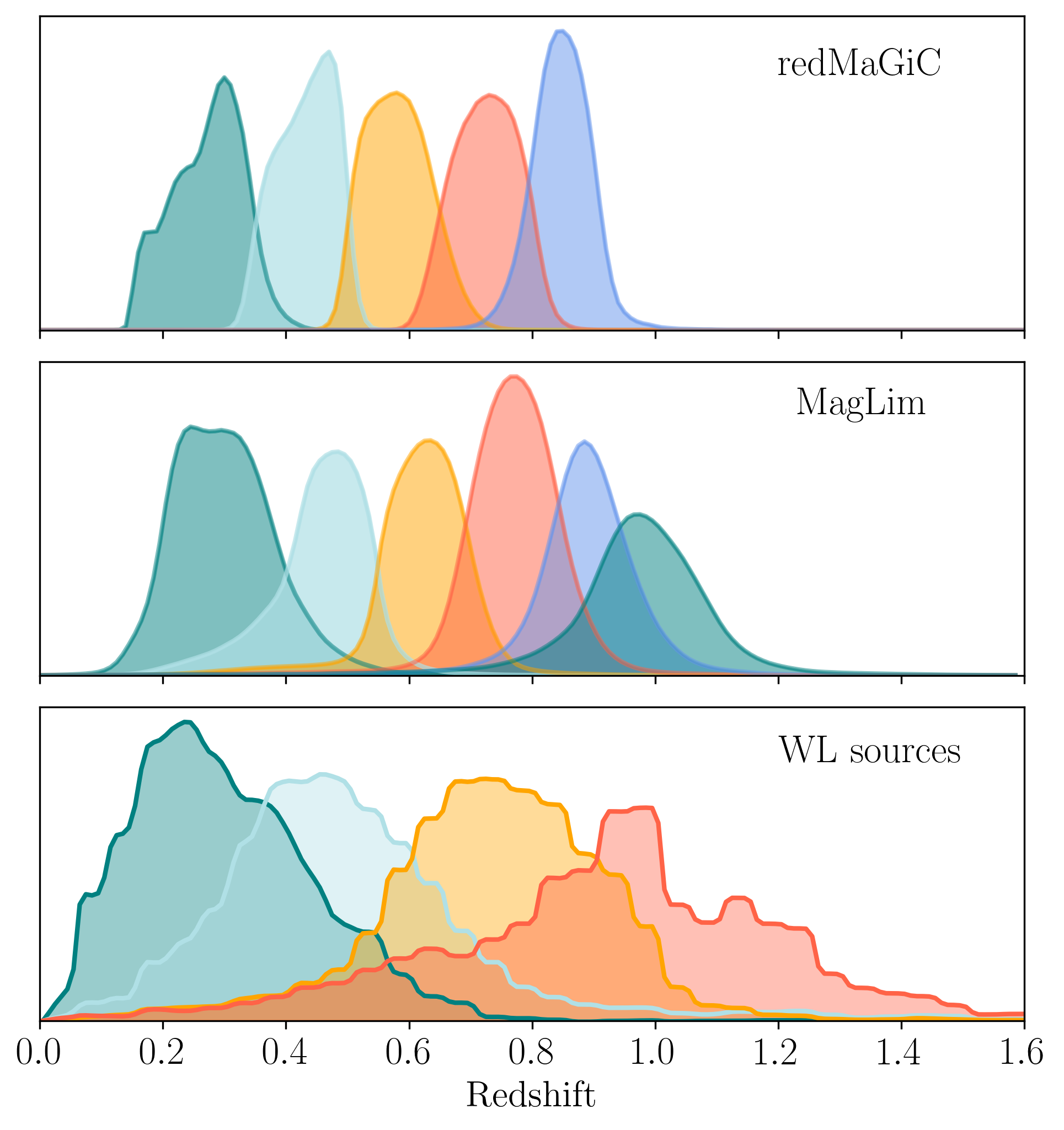}
\caption{Normalized redshift distributions of the lens and source samples. The top and middle panel show the redshift distributions of \redmagic \ and \textsc{MagLim} samples respectively, while the bottom panel shows that of source galaxies.}
\label{fig:nzs}
\end{center}
\end{figure}

The source redshift uncertainty has been calibrated in \citet*{y3-sompz} using the Self Organizing Maps Photometric Redshifts (SOMPZ) and the cross-correlation (WZ) method, further described in \citet*{y3-sourcewz}. SOMPZ is a scheme that provides a set of source redshift distributions and a characterization of their uncertainty, coming from sample variance, flux measurements and redshift errors using the deep fields \citep*{y3-deepfields} and the \textsc{Balrog} image simulations \citep{y3-balrog}. The WZ method is applied to this initial set of redshift distributions, removing the less \textit{likely} ones according to WZ data, which are the cross-correlations of the positions of the source sample with the positions of the \textsc{redMaGiC} sample. The outcome of these two methods combined is a set of realizations of the source redshift distributions, which is equivalent to using a mean $n(z)$ (as shown in Fig.~\ref{fig:nzs}) with a mean redshift prior of the order of $\sim 0.15$, as demonstrated in \citet{y3-hyperrank} using the \texttt{Hyperrank} method.

\begin{table}
\centering
\pmb{\textsc{redMaGiC} lens sample} \\
\vspace{2mm}
\setlength{\tabcolsep}{5pt}
\begin{tabular}{c|c|c|c|c}
\textbf{Redshift bin}  &   \textbf{$N^i_\text{gal}$} &  \textbf{$n^i_\text{gal}$ } & \textbf{$b^i$} & \textbf{$\alpha^i$} \\
\rule{0pt}{3ex} 
$0.15 < z < 0.35$ & 330243 &  0.022141 & 1.74  $\pm$ 0.12 & 1.31  \\
$0.35 < z < 0.50$ & 571551 &  0.038319 & 1.82  $\pm$ 0.11 & -0.52 \\
$0.50 < z < 0.65$ & 872611 &  0.058504 & 1.92  $\pm$ 0.11 & 0.34 \\
$0.65 < z < 0.80$ & 442302 &  0.029654 & 2.15  $\pm$ 0.12 & 2.25 \\
$0.80 < z < 0.90$ & 377329 &  0.025298 & 2.32  $\pm$ 0.14 & 1.97 \\
\end{tabular}
\\
\vspace{5mm}
\pmb{\textsc{MagLim} lens sample} \\
\vspace{2mm}
\begin{tabular}{c|c|c|c|c}
\textbf{Redshift bin} &   \textbf{$N^i_\text{gal}$} &  \textbf{$n^i_\text{gal}$ } & \textbf{$b^i$} & \textbf{$\alpha^i$} \\
\rule{0pt}{3ex} 
$0.20 < z < 0.40$ & 2236473 &  0.1499 & 1.49  $\pm$ 0.10 & 1.21 \\
$0.40 < z < 0.55$ & 1599500 &  0.1072 & 1.69  $\pm$ 0.11 & 1.15 \\
$0.55 < z < 0.70$ & 1627413 &  0.1091 & 1.90  $\pm$ 0.12 & 1.88 \\
$0.70 < z < 0.85$ & 2175184 &  0.1458 & 1.79  $\pm$ 0.13 & 1.97 \\
$0.85 < z < 0.95$ & 1583686 &  0.1062 & --  & 1.78 \\
$0.95 < z < 1.05$ & 1494250 &  0.1002 & --  & 2.48 \\

\end{tabular}
\\
\vspace{5mm}
\pmb{\textsc{Metacalibration} source sample} \\
\vspace{2mm}
\begin{tabular}{c|c|c|c|c}
\textbf{Redshift bin} &   \textbf{$N^j_\text{gal}$} &  \textbf{$n^j_\text{gal}$ } & \textbf{$\sigma_\epsilon^j$} & \textbf{$\alpha^j$} \\
\rule{0pt}{3ex} 
1 & 24940465 &  1.476 & 0.243 & 0.335 \\
2 & 25280405 &  1.479 & 0.262 & 0.685 \\
3 & 24891859 &  1.484 & 0.259 & 0.993 \\
4 & 25091297 &  1.461 & 0.301 & 1.458 \\
\end{tabular}

\caption{Summary description for each of the samples used in this work. $N_\text{gal}$ is the number of galaxies in each redshift bin, $n_\text{gal}$ is the effective number density in units of gal/arcmin$^{2}$ calculated with an area of 4143 deg$^2$, $b^i$ is the mean linear galaxy bias from the 3$\times$2pt combination, the \textbf{$\alpha$}'s are the magnification parameters as measured in \citet*{y3-2x2ptmagnification} and $\sigma_\epsilon^j$ is the weighted standard deviation of the ellipticity for a single component as computed in \citet*{y3-shapecatalog}.}
\label{tab:samples}
\end{table}

\section{Measurement: Tangential shear estimator}
\label{sec:measurement}

\begin{figure*}
\begin{center}

\includegraphics[width=0.99\textwidth]{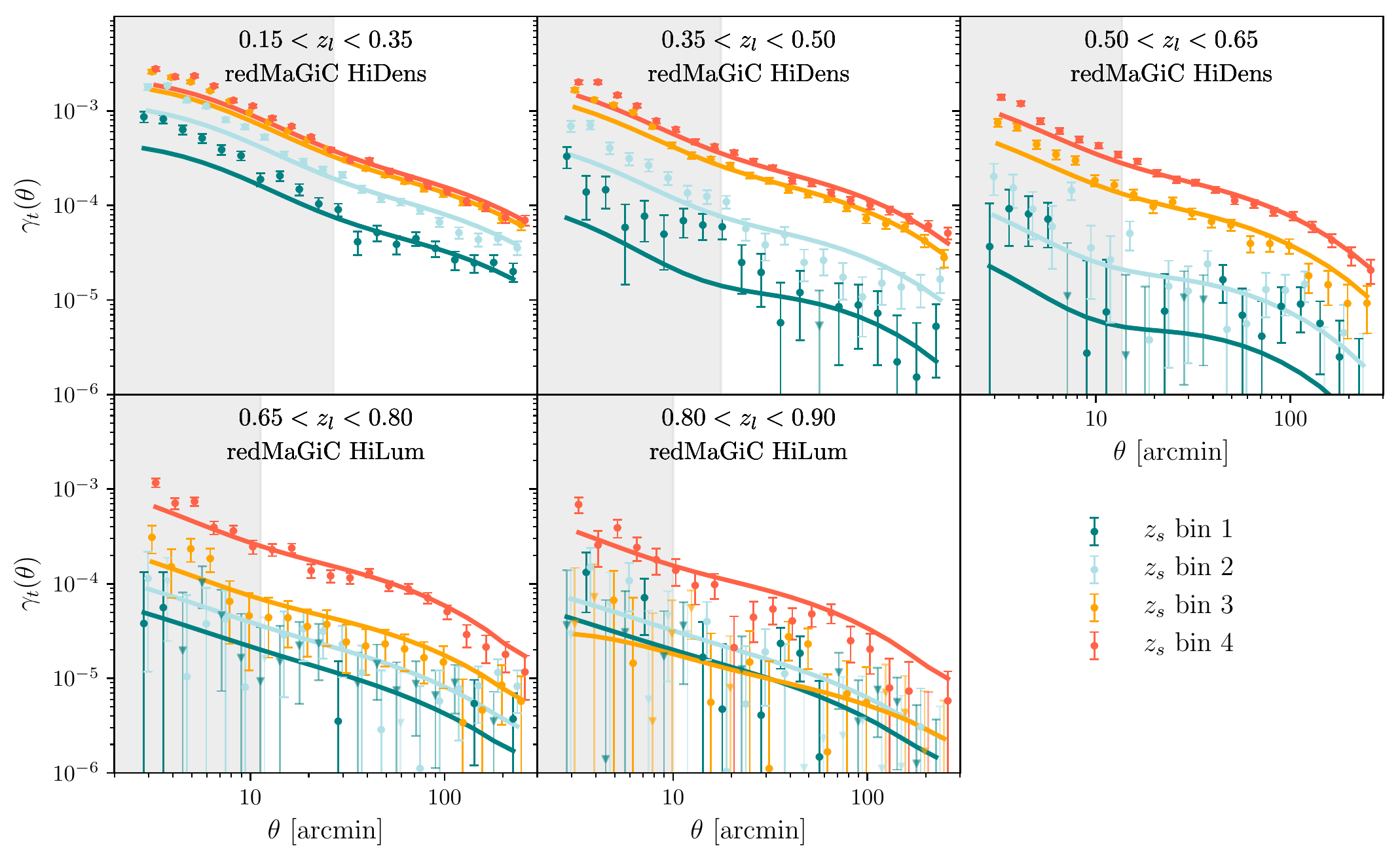}
\includegraphics[width=0.99\textwidth]{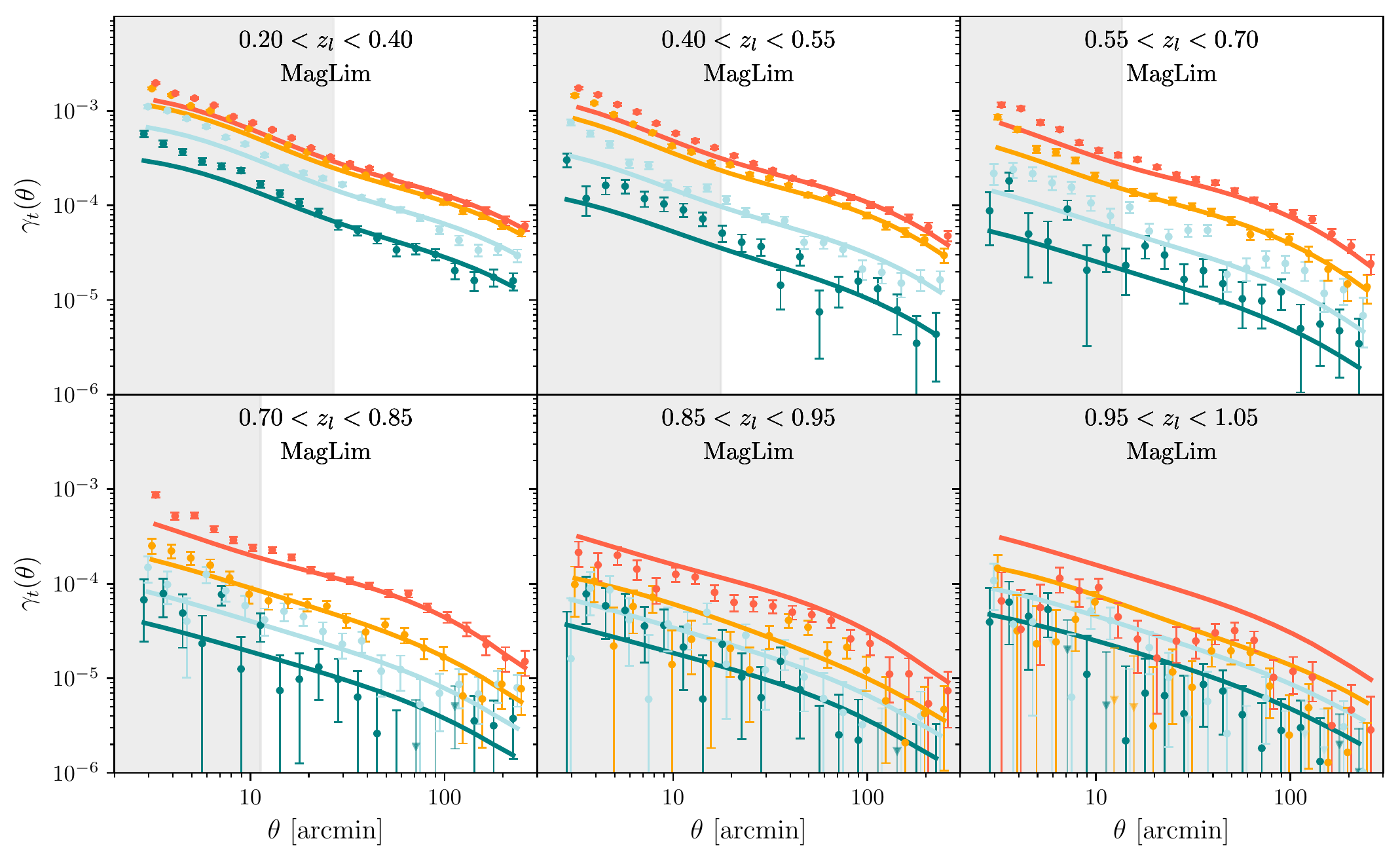}

\caption{Tangential shear measurements of the \redmagic \ (top) and \textsc{MagLim} (bottom) sample together with the best-fit theory line from the DES Y3 3$\times$2pt results. The shaded regions are excluded from the analysis; note that this includes the complete 5th and 6th \textsc{MagLim} lens redshift bins.}
\label{fig:gammat_measurements}
\end{center}
\end{figure*}

Galaxy-galaxy lensing is the cross-correlation of the shapes of background (source) galaxies with foreground galaxy positions, which trace the underlying matter field producing the lensing. The mean tangential shear around lens galaxies probes the azimuthally averaged projected mass distribution around them. In this section we describe the details of the \textit{mean tangential shear} measurement, or in short just \textit{tangential shear} from now on, which is the galaxy-galaxy lensing estimator we use in the DES Y3 3$\times$2pt cosmological analysis. In Fig.~\ref{fig:gammat_measurements} we show the final measurements together with the best-fit model from the 3$\times$2pt cosmological analysis. In this section we start by presenting the basic tangential shear estimator to then discuss several different measurement choices and refinements and their impact and significance on the measurement. 

\subsection{Basic tangential shear estimator}\label{sec:basic_gt_estimator}

Starting from the ellipticity measurements of the source galaxies in Equatorial coordinates $e_1, e_2$ we are able to extract the \textit{cosmic shear} $\gamma$, which we can link to cosmological parameters. Assuming spherical symmetry, the shear at any point will be oriented tangentially to the direction toward the center of the mass distribution causing the lensing. Thus, the tangential component of the shear captures all the cosmological signal and can be obtained by averaging the tangential component of the ellipticity over many lens-source galaxy pairs, canceling the intrinsic shape of the source galaxies, except in the presence of intrinsic alignments (IA). For a given lens-source galaxy pair $LS$ the tangential component of the ellipticity of the source
galaxy is:
\begin{equation}\label{eq:gammat_projection}
    e_{t, LS} = -e_1\cos{(2\phi)} -e_2\sin{(2\phi)}, 
\end{equation}
where $\phi$ is the position angle of the source galaxy with respect to the horizontal axis of the Cartesian coordinate system, centered at the lens galaxy. For a particular combination of lens and source tomographic bins, we perform a weighted average of the tangential component of the ellipticity of the source galaxies over all lens-source pairs $LS$ in our sample separated by some angular distance $\theta$ on the sky, grouping the pairs into 20 log-spaced angular bins between 2.5 and 250 arcmin:
\begin{equation}\label{eq:simplest_gt}
    \gamma_t \, (\theta) = \frac{\sum_{LS} w_{LS} \, e_{t, LS}(\theta)}{\sum_{LS} w_{LS}(\theta)} ,
\end{equation}
where $w_{LS} = w_l \, w_s$ is the weight factor for a given lens-source pair as a function of angular scale, where $w_l$ is the weight associated to the lens galaxy and $w_s$ the one associated to the source galaxy. Lens galaxy weights aim to remove correlations between density and observing conditions and have been computed in \citet{y3-galaxyclustering} and source galaxies weights are computed as the inverse variance of the ellipticity weighted by the shear response as detailed in \citet*{y3-shapecatalog}. We subtract the weighted mean ellipticity for each component $e_i$ before computing Eq.~(\ref{eq:simplest_gt}), as recommended by \citet*{y3-shapecatalog}. The values we subtract are shown in Table~\ref{tab:mean_shear}.

This is the simplest tangential shear estimator we can construct. However, due to several effects, such as lens-source clustering, mask effects and shape measurement biases our final estimator will include more components, that is, boost factors, random point subtraction and shear responses to address each of them respectively. We will add each component sequentially in the subsections below to reach our final tangential shear estimator given in Eq.~(\ref{eq:gt_fullestimator}).

\begin{table}
\centering
\begin{tabular}{ccc}
\hline 
Source redshift bin & $\left< e_1\right>$ & $ \left< e_2 \right>$  \\
\hline 

1 & $3.22 \times 10^{-4}$ & $1.60 \times 10^{-4}$ \\
2 & $3.36 \times 10^{-4}$ & $3.74 \times 10^{-5}$ \\
3 & $3.77 \times 10^{-4}$ & $8.75 \times 10^{-6}$ \\
4 & $4.06 \times 10^{-4}$ & $-2.68 \times 10^{-5}$ \\
\bottomrule

\end{tabular}
\caption{Mean weighted ellipticity per each component. We subtract these values from the ellipticity components of each galaxy before computing the tangential shear.}\label{tab:mean_shear}
\end{table}

\subsection{Lens-source clustering: Boost factors}\label{sec:boost_factors}

The model prediction for $\gamma_t$ assumes the mean $n(z)$ of the relevant lens and source bins, but does not account for the fact that source galaxies are preferentially located near lens positions due to the clustering between them whenever they overlap in redshift. There are several implications of this ``lens-source clustering'' which we explore here and in Sec.~(\ref{sec:model_validation}). Most notably, it leads to an excess number of lens-source pairs compared to what would be expected from the mean number densities. Because these pairs are physically nearby, the sources are unlensed, and the estimator in Eq.~(\ref{eq:simplest_gt}) is biased in a scale-dependent way compared to the theoretical prediction for $\gamma_t$. The impact of these excess lens-source pairs is an additional factor related to the projected lens-source correlation function, $\omega_{LS}(\theta)$. There are two possible approaches to remove this effect: 
\begin{itemize}
    \item Model the lens-source correlation function with sufficient accuracy for the scales under consideration, including the potential impact of nonlinear bias and magnification.
    \item Apply a ``boost'' factor to correct for the decrease of the measured lensing signal in the presence of lens-source clustering by measuring the excess of sources around tracers compared to random points as a function of scale, for every tracer-source bin combination. This was suggested for the first time in \citet{Sheldon_2004} and has since then been used in several analyses, such as in \citet{Mandelbaum_2005}, \citet{Mandelbaum_2006},  \citet{Miyatake_2015}, \citet{Singh_2017}, \citet{Luo_2018}, \citet{Amon_2018}, \citet{Singh_2020}, \citet{blake2020testing}, becoming part of the standard estimator for galaxy-galaxy lensing analyses. As demonstrated below, this approach is equivalent to using an unbiased estimator normalized using random positions rather than lens positions.
\end{itemize}

\begin{figure*}
\begin{center}
\includegraphics[width=0.99\textwidth]{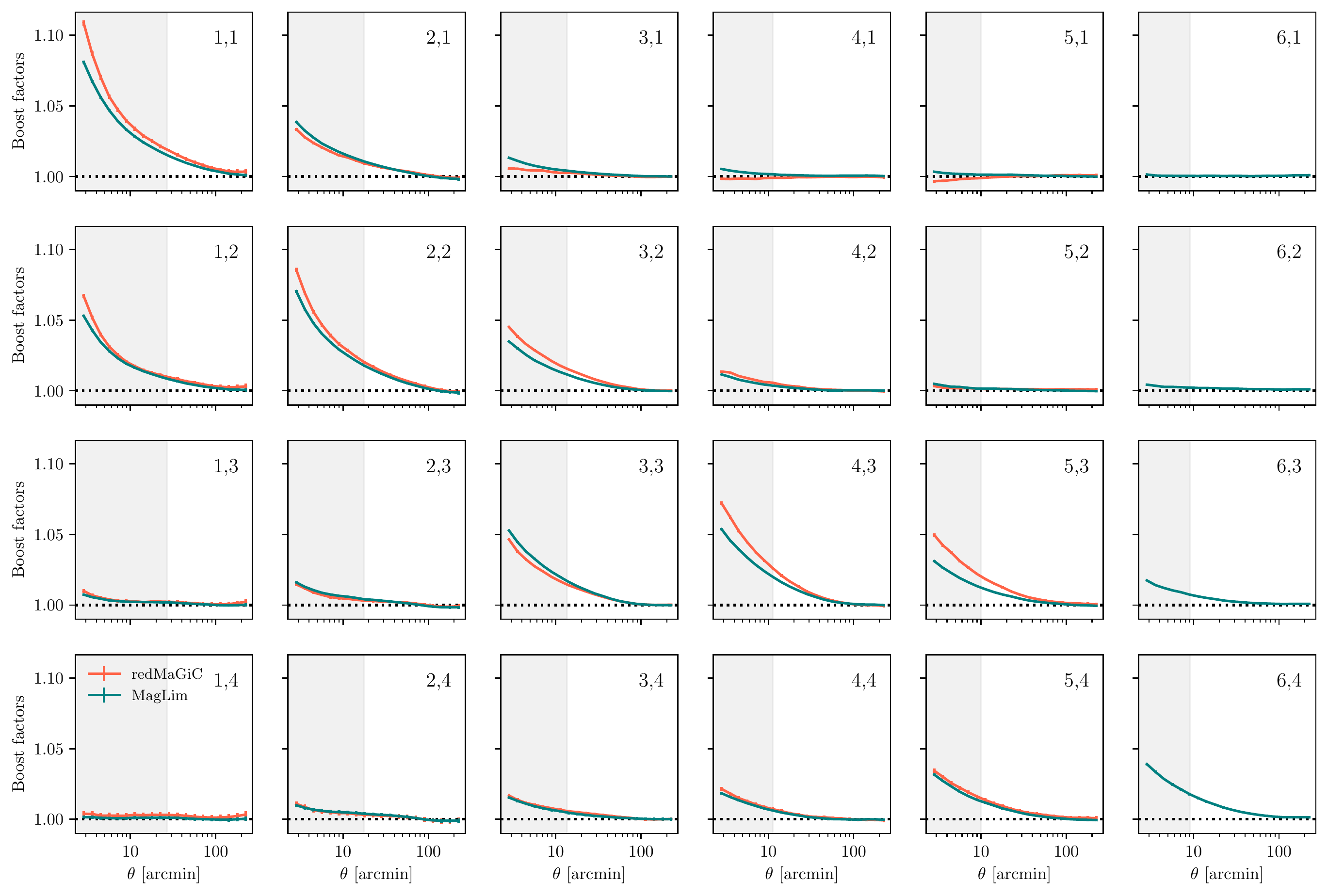}
\caption{Boost factor measurements for both lens samples with uncertainties from the jackknife method. Shaded regions correspond to scales below 6~Mpc/$h$ excluded for the galaxy-galaxy lensing part of the 3$\times$2pt analysis. Note that some of the scales below 6~Mpc/$h$ \textit{are} used for the shear-ratio (SR) analysis, see \citet*{y3-shearratio} for the exact SR scale cuts.}
\label{fig:boostfactors}
\end{center}
\end{figure*}

In this work we choose to correct for this effect using the boost factors since they are both accurate and easy to implement on photometric data.
We can express the boost factors in terms of standard estimators for galaxy clustering and $\gamma_t$. We can rewrite the simplest standard tangential shear estimator from Eq.~(\ref{eq:simplest_gt}) with \textit{no boost factors} (no bf) as
{\small
\begin{align}
&\gamma_{t, \text{no bf}} \, (\theta) = 
   \left( \frac{\sum_l w_{l}}{\sum_r w_{r}} \frac{{\sum_{RS} w_{RS}(\theta)}}{{\sum_{LS} w_{LS}(\theta)}}\right) \left(\frac{\sum_r w_{r}}{\sum_l w_{l}}  \frac{\sum_{LS} w_{LS} \, e_{t, LS}(\theta)}{\sum_{RS} w_{RS}(\theta)} \right),
\end{align}}
where $w_{RS} = w_r\, w_s$ is the weight associated with each random-source pair, with $w_r=1$ for all random points. The  second  factor  on  the  right-hand side of the equation is  what our tangential model predicts  when  using  the  mean $n(z)$ across the survey footprint (including relevant higher-order effects as discussed in \ref{sec:model_validation}). The first factor, which accounts for the excess unlensed sources, defines the inverse of the boost factor $B(\theta)$ and is just a simple version of the projected correlation function between lenses and sources $\omega_{LS}(\theta)$:
\begin{equation}\label{eq:boost_factors}
    B(\theta) = 1+\omega_{LS}(\theta) \equiv \frac{\sum_r w_{r}}{\sum_l w_{l}}
    \frac{{\sum_{LS} w_{LS}(\theta)}}{{\sum_{RS} w_{RS}(\theta)}},
\end{equation}
The ratio between the sum of random points weights and lens galaxies weights normalizes the boosts accounting for the fact that the sample of random points is usually larger than the sample of lenses to decrease shot noise.  We show the measured boost factors in Fig.~\ref{fig:boostfactors} for each lens-source combination. They produce a maximum correction of $\sim10\%$ at the smallest measured angular scale, and of $\sim2\%$ at the smallest scale used in the 3$\times$2pt cosmological analysis (6~Mpc/$h$). We estimate the uncertainty of the boost factors using the jackknife method described in Sec.~\ref{sec:pipeline_technical_details}. 

A major advantage of measuring the boost factors in this way is that it is independent of the estimated redshift distributions, and in particular of the tails of the redshift distributions, which need to be very well characterized to measure the overlap between lenses and sources accurately. Also, the boost factors measured from data naturally include all effects that can impact lens and source pair counts, such as lens and source magnification. In this analysis we model lens magnification but we do not include source magnification, which is a much smaller effect for galaxy-galaxy lensing. Also, we discuss the general impact of both lens and source magnification on galaxy-galaxy lensing in Sections \ref{sec:lens_mag} and \ref{sec:red_shear_etc}, respectively. The estimator for the tangential shear that includes boost factors (bf) to match the theoretical prediction given some mean $n(z)$ is:
\begin{align}\label{eq:gt_boosted}
\gamma_{t, \text{bf}} \, (\theta) &= B(\theta)\ \gamma_{t, \text{no bf}}
    = \frac{\sum_r w_{r}}{\sum_l w_{l}} \frac{\sum_{LS} w_{LS} \, e_{t, LS}(\theta)}{\sum_{RS} w_{RS}(\theta)} 
\end{align}
which in the end is just the usual tangential shear estimator normalized by the sum of random-source weights instead of lens-source weights, taking into account the ratio between the total sum of weights for the whole sample of random points and lenses. 

The $\Delta \chi^2$ between the tangential shear estimator with boost factors and without them is $\sim$9.8 from \textsc{MagLim} ($\sim$6.6 for  \redmagic) for the whole range of scales  and $\sim$0.2 ($\sim$0.1) for the scales used in the 3$\times$2pt combination (above 6~Mpc/$h$), so it is negligible for the large scales. We still apply the boost factor correction at all scales to be consistent with the small scales used in the shear-ratio analysis where the correction becomes more important. We show the impact of the boost factor correction on the datavector in Fig.~\ref{fig:estimator_tests}.

\subsection{Random point subtraction}\label{sec:random_point_sub}
One advantage of galaxy-shear cross-correlations over shear-shear correlations is that additive shear systematics average to zero in the tangential coordinate system. However, this cancellation only occurs when sources are distributed isotropically around the lens and additive shear is spatially constant, two assumptions that are not accurate in practice, especially near the survey edge or in heavily masked regions, where there is a lack of symmetry on the source distribution around the lens. To remove additive systematics robustly, we also measure the tangential shear around random points. Such points have no net lensing signal (see Appendix~\ref{appendix_randompoints}), yet they sample the survey edge and masked regions in the same way as the lenses. Another advantage of removing the tangential shear measurement around random points is that it removes a term in the covariance due to performing the measurement using the over-density field instead of the density field, as was found in \citet{Singh_2017}. Our estimator  including boost factors and random point subtraction becomes:
\begin{equation}
\gamma_t (\theta) = \frac{\sum_r w_{r}}{\sum_l w_{l}} \frac{\sum_{LS} w_{LS} \, e_{t, LS}(\theta)}{\sum_{RS} w_{RS}(\theta)} - \frac{\sum_{RS} w_{RS} \, e_{t, RS}(\theta)}{\sum_{RS} w_{RS}(\theta)}
\end{equation}
Note we only apply the boost factor correction to the lens term, since only the lenses are clustered with the sources. 

In this work we use 40 times as many random points as the number of lens galaxies per tomographic bin for each galaxy sample. We have tested that this number of random points is enough by using two independent sets of random points with $\times$40 randoms each and comparing the results. We have performed this test using the \buzzard \ \citep{y3-simvalidation, DeRose2019, Wechsler2021, DeRose2021, Becker2013} DES Y3 $N$-body simulations using a \redmagic-like sample. The $\Delta \chi^2$ between these two measurements in the simulations is $\sim 16$ for the whole range of scales (400 data points) and $\sim 7.5$ for the scales used in the 3$\times$2pt combination above 6 Mpc/$h$ (248 data points). This level of added noise is not significant for our analysis according to \citet{y3-covariances}. We also show the difference between these two measurements compared with the fiducial uncertainties in Fig.~\ref{fig:estimator_tests}. 
{
\begin{figure*}
\begin{center}
\includegraphics[width=0.99\textwidth]{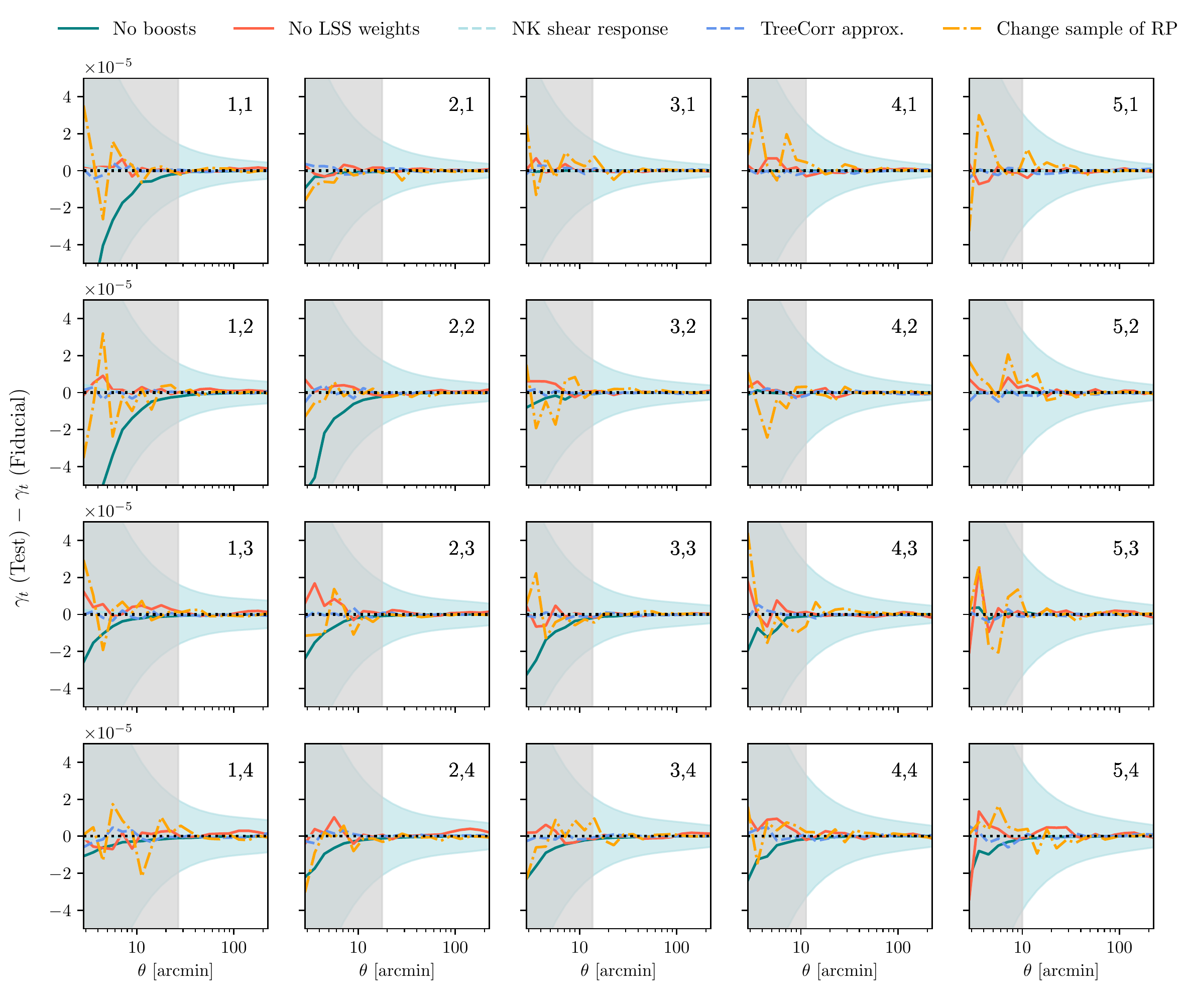}
\caption{Impact of different effects on the tangential shear estimator, for the \redmagic \ sample. The fiducial uncertainties (but without including the point-mass marginalization) are shown shaded in light blue. The scale cuts used in the DES Y3 3$\times$2pt analysis (6~Mpc/$h$) are represented by the gray shading. \textit{No boosts}: Not including the boost factor correction to the estimator, see Sec.~\ref{sec:boost_factors}. \textit{No LSS weights}: Not applying the weights to the lens galaxies to correct for observing conditions, see Sec.~\ref{sec:observing_conditions}. \textit{NK shear response}: Using the scale dependent shear responses instead of the mean response, see more details in Sec.~\ref{sec:responses_approx}. \textit{TreeCorr approx.}: Impact of the approximation \texttt{TreeCorr} uses to increase speed, see Sec.~\ref{sec:pipeline_technical_details}. \textit{Change sample of RP}:  Compares the random-subtracted fiducial measurements using a sample of random points 40$\times$ larger than the number of lenses, with measurements using another sample of random points, see
Sec.~\ref{sec:random_point_sub}. }
\label{fig:estimator_tests}
\end{center}
\end{figure*}

\begin{table*}
\centering
\begin{tabular}{cccccccc}
\hline 

 & \multicolumn{3}{c}{$\Delta \chi^2$ \textsc{redMaGiC}} & & \multicolumn{3}{c}{$\Delta \chi^2$ \textsc{MagLim}}   \\

 & All scales & $R < 6h^{-1}$Mpc & $R > 6h^{-1}$Mpc & & All scales & $R < 6h^{-1}$Mpc & $R > 6h^{-1}$Mpc\\

\hline 

Boost factor (Included) & 6.6 & 6.5 & 0.1 & & 9.8 & 9.8 & 0.2  \\
LSS weights (Included)  & 4.2 & 1.1 & 3.1 & & 5.4 & 0.69 & 4.8  \\
NK shear response & 0.0078 & 0.0076 & 0.0002 & & 0.0071 & 0.0068 & 0.0006 \\
\texttt{TreeCorr} Approximation & $\sim$1.5 & $\sim$0.5 & $\sim$1 & &  $\sim$1.5 & $\sim$0.5 & $\sim$1  \\

\bottomrule

\end{tabular}
\caption{Difference in $\chi^2$ of several measurement effects with respect to the fiducial measurements, using the \textsc{CosmoLike} theoretical covariance (without point-mass marginalization). The impact of these effects is also shown in Fig.~\ref{fig:estimator_tests} for the \redmagic \ sample. The \redmagic \ datavector has 400 data points, 152 at small scales (below $6h^{-1}$Mpc) and 248 at large scales (above $6h^{-1}$Mpc). The \textsc{MagLim} one has 480 data points, 176 at small angular scales and 304 at large scales. }\label{tab:summary_chi2s}
\end{table*}
}    

\subsection{Shape measurement calibration: Response factors}
In this work we use the \textsc{Metacalibration} shape catalog \citep{Huff2017, Sheldon2017}, which has the advantage of being able to self calibrate the mean shear measurement using the data themselves, via the so-called \textit{response} factor. In this section we describe the methodology to correct the mean shear, and in particular the mean tangential shear, for potential biases that arise in the process of using the mean of noisy and model-dependent individual ellipticity measurements as an estimator for the mean shear. The two-component ellipticity can be written as a function of the two-component shear $\boldsymbol{e}(\boldsymbol{\gamma})$ and Taylor-expanded around zero shear as:
\begin{equation}
\boldsymbol{e}(\boldsymbol{\gamma}) = \left. \boldsymbol{e}\right|_{\gamma=0} + \left.  \frac{\partial \boldsymbol{e}}{\partial \boldsymbol{\gamma}}\right|_{\gamma=0} \boldsymbol{\gamma} + ... \equiv  \left. \boldsymbol{e}\right|_{\gamma=0} +  \boldsymbol{R}_\gamma \boldsymbol{\gamma} + ... \, ,
\end{equation}
where we have defined the shear response $R_\gamma$ as the first derivative of the ellipticity with respect to shear. This quantity is useful since it allows us to obtain the unbiased relation between the mean ellipticity and the mean shear at first order, assuming the intrinsic ellipticity of galaxies are randomly oriented. This can be seen by averaging the equation above over an ensemble of galaxies:
\begin{equation}
\left< \boldsymbol{e} \right> \approx \left< \boldsymbol{R}_\gamma \boldsymbol{\gamma} \right>, 
\end{equation}
and inverting the relation:
\begin{equation}
\left< \boldsymbol{\gamma} \right> \approx \left< \boldsymbol{R}_\gamma\right>^{-1} \left< \boldsymbol{e} \right>. 
\end{equation}
For the tangential shear, we can apply the tangential rotation defined in Eq.~(\ref{eq:gammat_projection}) to each of the quantities, yielding: 
\begin{equation}\label{eq:responses_tangential}
\left< \gamma_t \right> \approx \left< \boldsymbol{R}_{\gamma_t}\right>^{-1} \left< e_t \right>. 
\end{equation}
Next we describe how to compute the response factors. The shear response can be measured for each galaxy by artificially shearing the images in a particular direction $j$ and remeasuring the ellipticity: 
\begin{equation}
R_{\gamma, i, j} =  \frac{e_i^+ - e_i^-}{\Delta \gamma_j}
\end{equation}
where $e_i^+$, $e_i^-$ are the ellipticity measurements on the component $i$ made on an image sheared by $+\gamma_j$ and $-\gamma_j$, respectively, and  $\Delta \gamma_j = 2\gamma_j$. In this work we use $\Delta \gamma_j =0.02$. Also, notice that $R_{\gamma, i, j}$ is a $2\times 2$ matrix and if the estimator of the ellipticity is unbiased the mean response matrix will be equal to the identity matrix. 

\subsubsection{Selection response}
Besides the shear response correction described above, in the \metacal\  framework, when making a selection on the original full catalog using a quantity that could modify the distribution of ellipticities,
for instance a cut in redshift, it is possible to correct for
selection effects via the so-called \textit{selection response}, defined as:
\begin{equation}
\left<R_{S,i,j} \right>  = \frac{\left< e_i\right>^{S_+}-\left< e_i\right>^{S_-}}{\Delta \gamma_j},
\end{equation}
where $\left<e_i\right>^{S_+}$ represents the mean of the $i$-component of ellipticities measured on images without applied shearing in component $j$, over the group of galaxies selected using the parameters extracted from positively sheared images. $\left<e_i\right>^{S_-}$ is the analogue quantity for negatively sheared images. In the absence of selection biases, $\left<\boldsymbol{R}_S \right>$ would be zero. Otherwise, the full response is given by the sum of the shear and selection response: 
\begin{equation}
\left<\boldsymbol{R} \right> =  \left<\boldsymbol{R}_\gamma \right> + \left<\boldsymbol{R}_S \right>.
\end{equation}

In this work we compute the selection response due to selection  effects produced when dividing the galaxies into tomographic bins. The results of the mean response for each redshift bin are shown in Table~\ref{tab:responses}.

\begin{table}
\centering
\setlength{\tabcolsep}{5pt}
\begin{tabular}{cccccc}
\hline 
$z_s$-bin & $\left< R\right>$ & $ \left< R_{\gamma} \right>$ & $\left< R_S\right>$ & $\left<R_{11}\right>$ &  $\left<R_{22}\right>$   \\
\hline 
1 & 0.7682 & 0.7636 & 0.0046 & 0.7669 & 0.7695\\
2 & 0.7266 & 0.7182 & 0.0083 & 0.7258 & 0.7273\\
3 & 0.7014 & 0.6887 & 0.0126 & 0.7006 & 0.7022\\
4 & 0.6299 & 0.6154 & 0.0145 & 0.6296 & 0.6302\\
\bottomrule

\end{tabular}
\caption{$\left< R\right>$  is the mean total \textsc{Metacalibration} response for each of the source tomographic bins. $ \left< R_{\gamma} \right>$ is the mean shear response and $\left< R_S\right>$ the mean selection response. $R_{11}$ and $R_{22}$ are the diagonal elements of the mean response matrix, i.e., the mean response for each ellipticity component with the artificial shear applied in the same direction. }\label{tab:responses}
\end{table}

\subsubsection{Response factor approximations for the tangential shear estimator} \label{sec:responses_approx}
In order to simplify the calculation of the response factors and reduce the computing time, in this work we make use of two approximations:
\begin{itemize}
   \item We assume the correction to be independent of the relative orientation of galaxies, i.e., we do not rotate the response matrix as it is done with the shears, which are projected to the tangential component. That means we do not apply Eq.~(\ref{eq:responses_tangential}), which would be the exact correction. We find it is safe to not project the response matrix since the difference between the values for each of the two diagonal elements $R_{11}$ and $R_{22}$ is between 0.01\% and 0.1\%, as shown in Table~\ref{tab:responses}. Then, since the response matrix is diagonal to good approximation, we just take the average of these components for each galaxy and therefore the response correction becomes just a scalar instead of a matrix:
   \begin{equation}\label{eq:scalar_response}
    R \approx \frac{R_{11} + R_{22}}{2}
   \end{equation}
    \item We assume it is sufficient to average the individual scalar responses over the ensemble of galaxies for each redshift bin, instead of over the source galaxies used in each angular bin, specifically that is assuming that:
\begin{equation}
\left< R \right> = \frac{\sum_s w_s R_s}{\sum_s w_s}  
\approx \frac{\sum_{LS} w_{LS} R_{LS}(\theta)}{\sum_{LS} w_{LS} (\theta)} 
\approx \frac{\sum_{RS} w_{RS} R_{RS}(\theta)}{\sum_{RS} w_{RS} (\theta)}
\end{equation}
where $R_s$ is the scalar response for each source galaxy $s$ as computed in Eq.~(\ref{eq:scalar_response}), not to be confused with the selection response $R_S$. $LS$ and $RS$ are the same summation indexes used in Sec.~\ref{sec:boost_factors} and Sec.~\ref{sec:random_point_sub}, running over all the lens-source pairs or random-source pairs respectively, in each angular bin $\theta$. If instead we wanted to perform the exact correction averaging the response of the galaxies that fall into each angular bin, the tangential shear estimator would take this form:
\begin{equation}\label{eq:responses_NK}
\gamma_t (\theta) =  B(\theta) \frac{\sum_{LS} w_{LS}  \, e_{t, LS}(\theta)}{\sum_{LS} w_{LS}R_{t, LS}(\theta)} - \frac{\sum_{RS} w_{RS} \, e_{t, RS}(\theta)}{\sum_{RS} w_{RS} R_{t, RS}(\theta)}
\end{equation}
We find the $\Delta \chi^2$ between the measurement using Eq.~(\ref{eq:responses_NK}) (except applying the tangential rotation to the response) and the fiducial estimator using the mean response written in Eq.~(\ref{eq:gt_fullestimator}) to be $\sim$0.01 for the whole range of scales for the \textsc{MagLim} sample ($\sim$0.0006 for large scales above 6~Mpc/$h$) and therefore negligible for our analysis. See Table~\ref{tab:summary_chi2s} for the rest of $\Delta \chi^2$ results. A visualization of this test is also shown in Fig.~\ref{fig:estimator_tests}.
\end{itemize}

\subsection{Final tangential shear estimator}
Using the response approximations described above, the application of the boost factors and the random point subtraction, the complete tangential shear estimator used in this analysis can be written as:
\begin{equation}\label{eq:gt_fullestimator2}
\gamma_t (\theta) = \frac{1}{\left<R\right>}\left[ B(\theta) \frac{\sum_{LS} w_{LS} \, e_{t, LS}(\theta)}{\sum_{LS} w_{LS}(\theta)} - \frac{\sum_{RS} w_{RS} \, e_{t, RS}(\theta)}{\sum_{RS} w_{RS}(\theta)}\right]
\end{equation}
where $\left<R\right>$ is the weighted average \metacal\  response in the corresponding source redshift bin, i.e. $\left<R\right>= \sum_s w_s R_s/\sum_s w_s$. Expanding the boost term, our final estimator can alternatively be written as
\begin{equation}\label{eq:gt_fullestimator}
\gamma_t (\theta) = \frac{1}{\left<R\right>}\left[ \frac{\sum_r w_{r}}{\sum_l w_{l}} \frac{\sum_{LS} w_{LS} \, e_{t, LS}(\theta)}{\sum_{RS} w_{RS}(\theta)} - \frac{\sum_{RS} w_{RS} \, e_{t, RS}(\theta)}{\sum_{RS} w_{RS}(\theta)}\right],
\end{equation}
The tangential shear measurements using this estimator are shown in Fig.~\ref{fig:gammat_measurements}. 

\setlength{\arrayrulewidth}{0.3mm}
\setlength{\tabcolsep}{10pt}
\renewcommand{\arraystretch}{1.4}

\subsection{Measurement pipeline technical details and code comparison}\label{sec:pipeline_technical_details}

In this section we specify the details of our fiducial measurement pipeline. This includes the description of pertinent optimizations we have used to reduce the memory and increase the speed of our code, given the large number of lens-source (and especially random-source) pairs that can be found in the DES Y3 samples. We also describe the  details of the successful comparison of the results of the fiducial code (internally referred to as \texttt{xcorr}) with an independent pipeline (internally referred to as \texttt{2pt\_pipeline}). 

Our measurement pipeline is based on the software package \texttt{TreeCorr}\footnote{\texttt{https://github.com/rmjarvis/TreeCorr}} \citep{Jarvis_2004} to measure the different two-point correlation functions present in Eq.~(\ref{eq:gt_fullestimator2}). Specifically, we use the \texttt{NGCorrelation} class from \texttt{TreeCorr} to perform the shape-position correlations. We set the \texttt{bslop} parameter from \texttt{TreeCorr} to zero in all our measurements, which ensures there is no variance between different users in how galaxy pairs are assigned into angular bins. Both for performance optimization purposes and to obtain a Jackknife covariance, we split the lens galaxies and random points into 150 regions using the \texttt{kmeans}\footnote{\texttt{https://github.com/esheldon/kmeans\_radec}} algorithm, which given the footprint area of $\sim4150$~deg$^2$ yields regions of approximately $5 \mathrm{deg}^2$ or $\sim 300$ arcmin of length assuming a square geometry (the largest angular scale we measure is 250 arcmin). We then call \texttt{TreeCorr} to perform the \texttt{NGCorrelation} between each of the lens (and random) patches and selected sources \textit{around} each lens patch. Once we have the measurement in each of the lens and random patches, we sum all the correlations appropriately following Eq.~(\ref{eq:gt_fullestimator2}) to obtain our fiducial tangential shear measurements. We also use the measurements in the different patches to obtain a Jackknife covariance  for the boost factor measurement and the corresponding diagonal uncertainties used in Figure~\ref{fig:boostfactors}.

The selection of sources around each lens patch significantly reduces the amount of memory needed to complete this calculation, and is achieved by building a \texttt{healpix}\footnote{\texttt{https://healpix.sourceforge.io/}} grid of \texttt{nside}=4 for the source galaxies and selecting the pixel in this grid corresponding to the center of each lens patch together with all its surrounding \texttt{healpix} pixels. Then, we apply a further mask using a matching function from \texttt{Astropy}  \citep{astropy:2018} to only select source galaxies that are within a distance of 1.5 times the maximum angular separation we are interested in measuring. We do this in a two-step process to minimize the amount of memory used and increase speed, since the \texttt{Astropy} matching is more precise but requires more memory and is slower. We have tested that using this optimization does not result in any loss of lens-source pairs. However, note that if a different catalog is given to \texttt{TreeCorr} to build the tree, even if the eventual number of pairs used for the measurements is exactly the same, this will result in a small difference in the measurements. This can be avoided using the brute force option\footnote{For \texttt{NN} and \texttt{KK} correlations, \texttt{bin\_slop=0} should always be identical to the brute force calculation. However, for \texttt{NG} (or \texttt{GG}) correlations they will not be identical. The results will depend on the tree construction, which divides galaxies into cells. Each shear in a tree cell is projected onto the line joining the centers of the two cells, not the line joining it with each point like in the full brute force calculation. This effect can be alleviated using thinner angular bins.} within \texttt{TreeCorr}, which is nonetheless much slower. This approximation produces a $\Delta \chi^2 \sim 1$ for our setup. We also show the impact of using this approximation in Fig.~\ref{fig:estimator_tests}, where we can visualize the difference between the two tangential shear measurements. Due to the increase in speed and decrease in memory we achieve using this approximation, and the very low significance of the effect, we use it in our fiducial measurements.

We have compared the results of  our  fiducial  measurement pipeline applied and obtained a $\Delta \chi^2 \lesssim  1$ for both galaxy samples, with 400 data points for \texttt{redMaGiC} (or 480 for \texttt{MagLim)}. We consider this result successful and also want to take this opportunity to stress the importance of comparing measurement pipelines in future analyses as well, given that in our analysis it was very effective in identifying bugs and sources of error we were not initially considering. After this code comparison we compared with a third pipeline (with the caveat that is also based on \texttt{TreeCorr}) and also obtained a $\Delta \chi^2 \lesssim  1$ to both of our previous pipelines. The reason for these remaining differences is that the different pipelines were building the ``trees'' within \texttt{TreeCorr} in a different way. 

In this whole section all the quoted $\Delta \chi$'s are computed using the theoretical covariance \textit{without} including the point-mass marginalization, therefore the real impact of these effects on the 3$\times$2pt cosmological analysis could actually be smaller given the effective increase of the covariance due to point-mass marginalization, which is especially important at small and intermediate scales, see Sec~\ref{sec:point-mass} for more details.

\subsection{Blinding}\label{sec:blinding}
In this work and within the 3$\times$2pt analysis we use a two-level blinding scheme that consists of having:
\begin{enumerate}
    \item \textbf{Blinding at the catalog level}: An unknown multiplicative factor has been applied to the ellipticity measurements of all the source galaxies used in this work until the moment of unblinding.
    \item \textbf{Blinding at the two-point level}: Using the method described in \citet{Muir_2020} we modify the tangential-shear two-point correlation function measurements, effectively shifting them by a cosmology-dependent factor. The shifted, and thus blinded, two-point function has the property of preferentially looking like the correlation function of another cosmological model. 
\end{enumerate}

More details on the blinding criteria can be found in \citet{y3-3x2ptkp}.

\section{Modeling the tangential shear}
\label{sec:theory}

The tangential shear is the main measurement used in this paper as detailed in the previous section, and here we describe how we model it in this work and within the DES Y3 3$\times$2pt cosmological analysis. See also the DES Y3 3$\times$2pt methodology paper \citep{y3-generalmethods} for further descriptions and the modeling of the other two-point correlation functions. We start by describing the basic modeling scheme, and then discuss the addition of several effects to our model, such as the removal of small scale information using the point-mass marginalization scheme, lens magnification, intrinsic alignments and a description of the galaxy bias model. At the end we detail the comparison of our fiducial modeling pipeline with an independent code. 

\subsection{Basic tangential shear modeling}\label{sec:basic_gt_model}

The tangential shear two-point correlation function is a transformation of the 2D galaxy-matter angular cross-power spectrum $C_{gm}$, which in this work we perform using the curved sky projection as detailed later in Eq.~(\ref{eq:curved_sky}). First we will describe how we can model $C_{gm}$ and express it as a projection of the 3D galaxy-matter power spectrum $P_{gm}$. For a lens redshift bin $i$ and a source redshift bin $j$, under the Limber approximation \citep{Limber53, Limber_LoVerde2008} and  assuming a flat Universe cosmology we can write:
\begin{equation}\label{eq:C_gm}
    C_{gm}^{ij}(\ell) = \int d\chi \frac{N_l^i(\chi)\, q_s^j(\chi)}{\chi^2}P_{gm}\left(k = \frac{\ell+1/2}{\chi},z(\chi)\right)\,,
\end{equation}
where $k$ is the 3D wavenumber, $\ell$ is the 2D multipole moment, $\chi$ is the comoving distance to redshift $z$, and $N_l^i(\chi)$ and $q_s^j(\chi)$ are the window functions of the given lens and source populations of galaxies used in Limber's approximation, which holds if the 3D galaxy overdensity field of the lenses and the 3D matter overdensity field at the redshift of the source galaxies vary on length scales much smaller than the typical length scale of their respective window functions in the line of sight direction. The lens window function is defined as:
\begin{equation}
N_l^i(\chi) = \frac{n^i_l\,(z)}{\bar{n}^i_l}\frac{dz}{d\chi}, 
\end{equation}
where $n^i_l$ is the lens redshift distribution and $\bar{n}^i_l$ is the mean number density of the lens galaxies. The lensing window function of the source galaxies is:
\begin{equation} \label{eq:lensing_window}
q_s^j(\chi) = \frac{3H^2_0 \Omega_m }{2c^2} \frac{\chi}{a(\chi)} g(\chi)
\end{equation}
where $a$ is the scale factor and $g(\chi)$ is the lensing efficiency kernel: 
\begin{equation}
 g(\chi) = \int_\chi^{\chi_\text{lim}} d \chi'  \frac{n^j_s\,(z) \, dz/d\chi'  }{\bar{n}^j_s}\frac{\chi'- \chi}{\chi'}
\end{equation}
with $n_s^j(z)$ being the redshift distribution of the source galaxies, $\bar{n}^j_s$ the mean number density of the source galaxies and $\chi_\mathrm{lim}$ the limiting comoving distance of the source galaxy sample.

Ultimately we want to relate the galaxy-matter power spectrum to the matter power spectrum. In our fiducial model we assume that lens galaxies trace the mass distribution following a simple linear biasing model ($\delta_g = b \:\delta_m$), so the galaxy-matter power spectrum relates to the matter power spectrum by a multiplicative galaxy bias factor:
\begin{equation} \label{eq:linearbias}
 P^{ij}_{gm} = b^{i}  P^{ij}_{mm}.
\end{equation}
We summarize the tests we have performed to make this modeling choice in Sec.~\ref{sec:galaxy_bias_tests}, and see \citet{y3-2x2ptbiasmodelling} for an extended description. We compute the non-linear matter power spectrum using the \citet{Takahashi_2012} version of \texttt{Halofit} and the linear power spectrum is computed with \texttt{CAMB}\footnote{https://camb.info/}.

\subsubsection{Angular bin averaging and full sky projection}

\begin{figure}
\begin{center}
\includegraphics[width=0.45\textwidth]{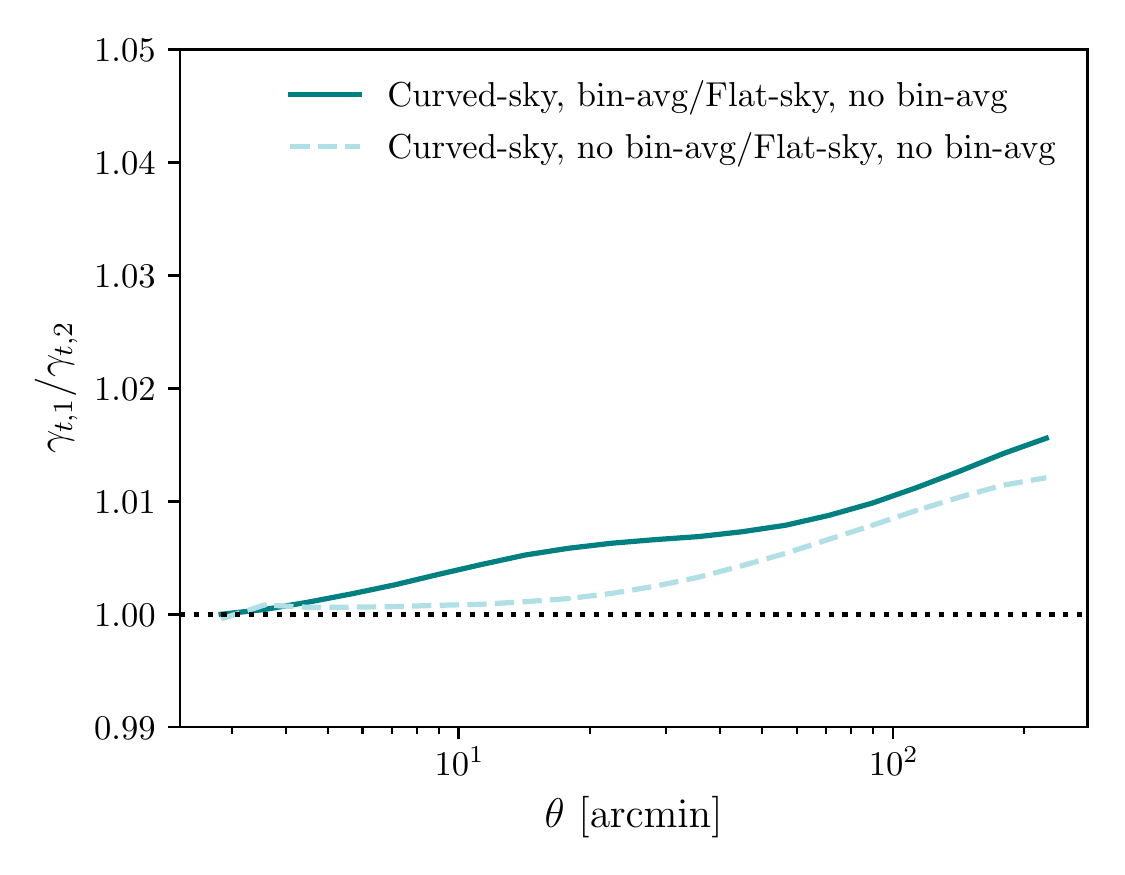}
\caption{Illustration of the curved sky and angular bin averaging effect to the tangential shear modeling. In the DES Y3 3$\times$2pt cosmological analysis we model the tangential shear using the curved sky transform (vs the flat sky approximation, used in the DES Y1 3$\times$2pt analysis for instance) and we properly average within the range of scales falling in each angular bin (vs just picking one value in the middle of the bin, also employed in the Y1 modeling). The effects are shown for the first lens tomographic bin and second source tomographic bin, but they are not very dependent on redshift.}
\label{fig:flat_sky}
\end{center}
\end{figure}

Given the galaxy-matter angular power spectra we can obtain the tangential shear quantity via the following transformation on the curved sky, as a function of angular scale $\theta$ between lens and source galaxies:
\begin{equation}\label{eq:curved_sky}
\gamma_t^{ij}(\theta) = \sum_\ell \frac{2\ell+1}{4\pi\ell(\ell+1)}P^2_\ell(\cos\theta) \, C^{ij}_{gm}(\ell)~,
\end{equation}
where $P_\ell^2$ are the associated Legendre polynomial. We calculate the correlation functions within an angular bin $[\theta_{\rm min},\theta_{\rm max}]$, by carrying out the average over the angular bin, i.e., replacing $P_\ell^2(\cos\theta)$ with their bin-averaged function $\overline{P^2_\ell}$, from \citet{Fang_2020}:
\begin{align}\label{eq:bin-average}
    &\overline{P^2_\ell}\left(\theta_{\rm min},\theta_{\rm max}\right) \equiv \frac{\int_{\cos\theta_{\rm min}}^{\cos\theta_{\rm max}}dx\,P^2_\ell(x)}{\cos\theta_{\rm max}-\cos\theta_{\rm min}}\nonumber=\\
    &= \frac{\left[\left(\ell + \frac{2}{2\ell + 1}\right) P_{\ell -1} (x) + (2-\ell) \, x\,  P_\ell (x) - \frac{2}{2\ell + 1} P_{\ell + 1} (x) \right]^{\cos{\theta_{\rm max}}}_{\cos{\theta_{\rm min}}} }{\cos\theta_{\rm max}-\cos\theta_{\rm min}}
\end{align}
We show the effect of including the full-sky transform and the bin-averaging implementation, given they are both new in the Y3 modeling with respect to the Y1 one, versus using the flat-sky approximation and no averaging in scales within each angular bin in Fig.~\ref{fig:flat_sky}.

Note there is an additional effect from the variation in the pair counts due to the survey geometry not taken into account in Eq.~(\ref{eq:curved_sky}), which we have found negligible for the DES Y3 analysis setup. See App.~\ref{sec:mask_effects} for more details. 

\subsection{Removing small-scale information: Point-mass marginalization} \label{sec:point-mass}

\begin{figure}
\begin{center}
\includegraphics[width=0.47\textwidth]{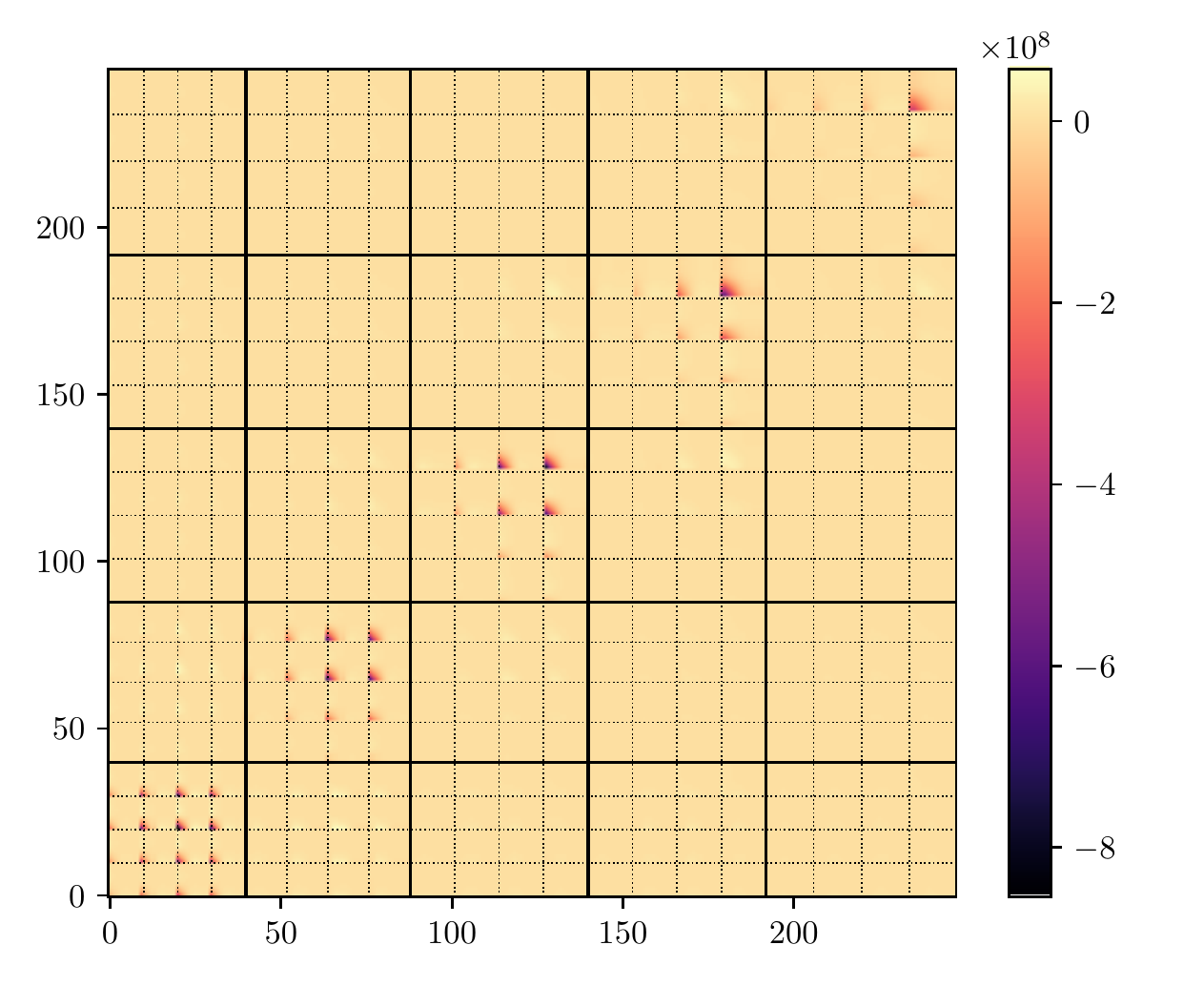}
\caption{Difference in the tangential shear inverse covariance matrix between the cases assuming including the point-mass correlation or not,  $\mathbf{C}^{-1,+PM}-\mathbf{C}^{-1} $. We illustrate this using the \redmagic \  sample and the >6 Mpc/$h$ scale cut. The difference is especially noticeable at small scales. Solid lines separate lens tomographic bins and dotted lines source tomographic bins. }
\label{fig:invcov}
\end{center}
\end{figure}

The tangential shear is a non-local quantity. This can be appreciated expressing the tangential shear of a single lens-source pair as a function of the excess surface mass density $\Delta \Sigma$:
\begin{equation}
\gamma_t \, (\theta) = \frac{\Delta \Sigma \, (\theta)}{\Sigma_\mathrm{crit}},
\end{equation}
where $\Sigma_\mathrm{crit}$ is a geometrical factor that depends on the angular diameter distances to the lens galaxy $D_{l}$, the one between the lens and the source $D_{ls}$ and the one to the source galaxy $D_s$, and is defined as:
\begin{equation}\label{eq:inverse_sigma_crit}
\Sigma_{\mathrm{crit}}^{-1} (z_{\rm l}, z_{\rm s}) = \frac{4\pi G}{c^2} \frac{D_{ls} \, D_l} {D_s} \qquad \mathrm{if} z_s > z_l,
\end{equation}
and zero otherwise. Also, $\Delta \Sigma$ can be expanded as the difference between the mean surface mass density \textit{below} a certain angular scale $\theta$ and the surface mass density \textit{at} this given scale:
\begin{equation}
\Delta \Sigma\,  (\theta) = \overline{\Sigma} \, (<\theta) - \Sigma\,  (\theta),
\end{equation}
where the non-locality of the tangential shear quantity becomes apparent, since the tangential shear defined at some $\theta$ value will always carry information of all the scales below this value. This is the reason the scale cuts in the DES Y1 3$\times$2pt cosmological analysis were higher for the galaxy-galaxy lensing part (12 Mpc/$h$) than for the galaxy clustering part (8 Mpc/$h$). In this analysis we would need to apply an even more stringent cut due to the smaller statistical uncertainties. Alternatively, it is possible to localize the tangential shear measurement. For instance \citet{park2020localizing} suggested applying a linear transformation to the tangential shear observable to remove this non-locality. In this work and in the context of the 3$\times$2pt DES Y3 cosmological analysis we decide to account for this instead following \citet{MacCrann_2019}. Internal tests for the Y3 analysis have shown both methods yielding very similar results in the recovered cosmological constraints. \citet{MacCrann_2019} proposes to analytically marginalize over a point-mass (PM) scaling as $R^{-2}$ with physical separation $R$ between the lens and the source galaxy, including some additional terms in the tangential shear covariance coming from the uncertainty in the model prediction of galaxy-matter correlation function below a given scale. Starting by expressing the point-mass term as an addition to the tangential shear model for a given lens redshift bin $i$ and source redshift bin $j$ as a function of angular separation:
\begin{equation}\label{eq:point-mass}
\gamma_t^{ij} \, (\theta) = \gamma_{t, \rm model}^{ij} \, (\theta) + \frac{ A^{ij}}{\theta^2}\, ,  
\end{equation}
where $A^{ij}$ is the following function:
\begin{equation}
A^{ij} = \int  dz_l \int \ dz_s \ n^i_l (z) \ n^j_s (z) \ B^i(z_l) \ \Sigma^{-1}_{\rm crit}(z_l,z_s) \ D^{-2}(z_l) \, ,
\end{equation}
that depends on the point-mass  $B^i$ we want to marginalize over. In general $B^i$ can evolve within the lens bin but given the tomographic binning scheme of our lens sample, we can assume the lens redshift bins are narrow enough so that we can approximate the previous equation to:
\begin{align}
A^{ij} &\approx  B^i \int  dz_l \int \ dz_s \ n^i_l (z) \ n^j_s (z) \ \Sigma^{-1}_{\rm crit}(z_l,z_s) \ D^{-2}(z_l) \\
&\equiv B^i \beta_{ij}
\end{align}
This is advantageous because in this case the $\beta_{ij}$ parameters can be naturally constrained from the data itself via implicit shear-ratio information. In other words, some of the constraining power of the tangential shear measurements, and in particular the geometrical information, is naturally used within the 3$\times$2pt combination to constrain the $\beta_{ij}$ parameters. Then, given the simple form of this contamination model (e.g. the scale dependence is not dependent on
cosmology or the lens galaxy properties), this term can be analytically marginalized, i.e. we only need to add some terms to the tangential shear covariance matrix to effectively ``remove'' information below the angular scale $\theta$. We perform an analytic marginalization over all $B_i$, which can be done by adding the following terms to the original tangential shear covariance matrix $\mathbf{C}$ to become \citep{Bridle2002, MacCrann_2019}:
\begin{equation}
\mathbf{C}^{+PM}_{i j \theta, i'j'\theta'} = \left\{
                \begin{array}{ll}
                    \mathbf{C}_{i j \theta, i'j'\theta'} +  \sigma^2_{B_i}  \beta_{ij}/\theta^2 \cdot \sigma^2_{B_i'}  \beta_{i'j'}/ \theta'^2 \quad \mathrm{if} \  i=i' \\
                 \mathbf{C}_{i j \theta, i'j'\theta'}  \qquad \qquad \qquad  \qquad \quad  \ \ \  \qquad \mathrm{if} \ i \neq i'
                \end{array}
              \right.
\end{equation}
under the narrow lens bin approximation. $\sigma^2_{B_i}$ is the width of the Gaussian prior on $B_i$ we want to marginalize over. In this work, we choose to adopt an uninformative prior and take the limit $\sigma^2_{B_i} \rightarrow \infty$. This is because for the chosen scale cuts of 6~Mpc/$h$ the point-mass is dominated by the 2-halo regime (see Appendix A from \citealt{y3-2x2ptbiasmodelling}). In the 3$\times$2pt likelihood  we will eventually need the inverse of the covariance matrix, instead of the covariance matrix itself. For the infinite prior case on  $\sigma^2_{B_i}$, the inverse covariance matrix can be written as \citep{MacCrann_2019}:
\begin{equation}\label{eq:invcov_PM}
\mathbf{C}^{-1,+PM} = \mathbf{C}^{-1} 
- \mathbf{C}^{-1} \, \mathbf{V} \, (\mathbf{V}^\mathbf{T}\mathbf{C}^{-1}\,  \mathbf{V})^{-1} \, \mathbf{V}^\mathbf{T}\,  \mathbf{C}^{-1} 
\end{equation}
where $\mathbf{V}$ is a $N_d \times N_\mathrm{lens}$ matrix with the $i$th column being $\beta_{ij}/\theta^2$ and $N_d = N_\theta N_\mathrm{lens} N_\mathrm{source}$ being the number of elements in the datavector, $N_\mathrm{lens}$ the number of lens redshift bins and $N_\mathrm{source}$ the number of source tomographic bins. In Fig.~\ref{fig:invcov} we show the change in the tangential shear inverse covariance matrix this produces. The changes in the inverse covariance are larger for the lower lens redshift  bins due to the fact that at lower redshift a given angular scale corresponds to a larger physical scale than at higher redshift. The changes are also larger where the signal is bigger, i.e. for the lens-source combinations which are more separated in redshift. The S/N of the tangential shear measurements changes from $\sim$55 to $\sim$28 for the \redmagic \ sample when the point-mass marginalization is applied to the inverse covariance\footnote{The S/N is computed here and elsewhere in the paper as $\sqrt{\chi^2 - N_{\text{dp}}}$, where $\chi^2= \gamma_{t,{\text{data}}} \mathbf{C}^{-1} \gamma_{t,{\text{data}}}$, $N_{\text{dp}}$ is the number of data points in the data vector and $\mathbf{C}^{-1}$ is the inverse theoretical covariance.}. For \textsc{MagLim} the change in S/N is from $\sim$67 to $\sim$32. See also \citet{y3-2x2ptbiasmodelling} for further details on the implementation of the point-mass marginalization in the DES Y3 3$\times$2pt analysis.

\begin{figure*}
\begin{center}
\includegraphics[width=0.99\textwidth]{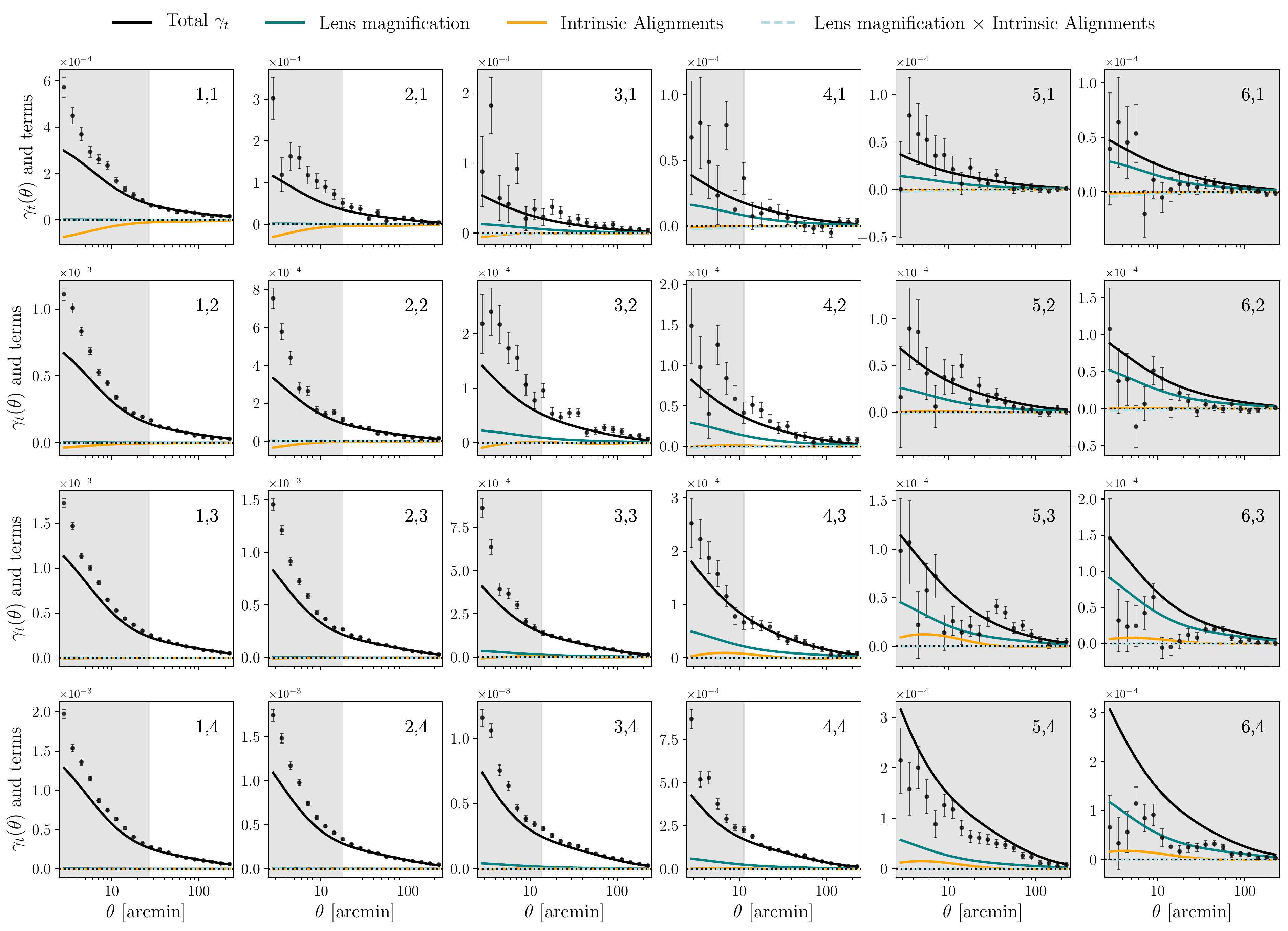}
\caption{This plots shows the contribution from each of the components of our model at the best-fit values from the 3$\times$2pt result for the \textsc{MagLim} sample and the total model (black line), as obtained from Eq.~(\ref{eq:all_terms}). The contribution from the coupling of IA and lens magnification is very close to zero. }
\label{fig:terms_model}
\end{center}
\end{figure*}

\subsection{Lens magnification} \label{sec:lens_mag}

Lens magnification is the effect of magnification produced on the lens galaxy sample by the structure that is between the lens galaxies and the observer. In this section we describe how lens magnification affects the galaxy-galaxy measurements, how significant the effect is for the tangential shear probe, and how we model it. See  \citet*{y3-2x2ptmagnification} for further details regarding lens magnification within the DES Y3 analysis. This effect has also been studied for galaxy-galaxy lensing recently in \citet{Unruh_2020}.

In the weak gravitational lensing picture, besides having shape distortions described by the shear, the solid angle spanned by the image is changed compared to the solid angle covered by the source by the so-called \textit{magnification factor} $\mu$. This change in solid angle can alter the number density of a given sample via two different mechanisms: (1) The number density decreases by a factor $\mu$ due the sky being locally stretched by the same factor and (2) Since the area increases but the surface brightness is conserved, the flux of individual galaxies rises, and some galaxies that would otherwise not have been detected pass the relevant flux threshold for a particular sample. These are two competing effects and the dominant one depends on the specifics of the galaxy sample. Then, to understand how lens magnification affects the tangential shear measurements, it is useful to express the observed density contrast for the lens sample as the sum of the intrinsic galaxy density contrast and the ``artificial'' one produced by lens magnification:
\begin{equation}
    \delta_g^\text{obs} =  \delta_g^\text{int} + \delta_g^\text{mag} .
\end{equation}
Then we can make the assumption that the change in number density produced by magnification is  proportional to the convergence \citep*{y3-2x2ptmagnification}. In that case, we can write
\begin{equation}\label{eq:definition_C}
\delta_g^\text{mag} (\theta) = C \kappa_l (\theta) \, , 
\end{equation}
where $\kappa_l$ is the convergence field at the lens redshift and $C$ is just a proportionality factor. At this point we can separate the area effect and the flux effect on the number density change: $C_\mathrm{total} = C_\mathrm{area} + C_\mathrm{flux}$, since it can be shown that $C_\mathrm{area}= - 2$ \citep*{y3-2x2ptmagnification} while $C_\mathrm{flux}$ will depend on the sample. That is why this proportionality factor is usually written in the literature as  $C_\mathrm{total} = 2 (\alpha -1)$, where $\alpha$ is a property of the sample and is equivalent to $C_\mathrm{flux}/2$. From now on we will adopt the ``$\alpha$ notation'' since it is more commonly used.
 
Lens magnification becomes relevant because the change in number density produced to the lens sample is correlated with the large scale structure that is between the lens galaxies and the observer. That means that for a given sample of lens galaxies, some lines-of-sight with, for instance, more matter between the lens galaxies and us could be over-sampled if $\alpha >1$, or down-sampled if $\alpha<1$, and the tangential shear measurement would be biased, as seen in the following equation:
\begin{equation}
\left< \delta_g^\text{obs} \gamma \right> = \left< \delta_g^\text{int} \gamma \right> + 2 (\alpha -1)\left< \kappa_l\gamma \right> = \left< \delta_g^\text{int} \gamma \right> + 2 (\alpha -1)\left< \kappa_l \gamma \right>.
\end{equation}
The first term is just the usual galaxy-galaxy lensing signal, modeled for the tangential shear as given by Eqs.~(\ref{eq:C_gm}) and (\ref{eq:curved_sky}), and the additional lens magnification term is modeled in the following way before performing the projection to real space:
\begin{equation}
    C_{mm}^{ij}(\ell) = \int d\chi \frac{q_l^i(\chi)\, q_s^j(\chi)}{\chi^2}P_{mm}\left(k = \frac{\ell+1/2}{\chi},z(\chi)\right)\,,
\end{equation}
where the lensing window functions $q_s$ is defined in Eq.~(\ref{eq:lensing_window}), and the analogous window function for the lens sample is given by $q_l$. The $i$ index represents the lens tomographic bin and $j$ the source one. The tangential shear model including the lens magnification term can be written as:
\begin{equation}
\gamma_t^{ij}(\theta) = \sum_\ell \frac{2\ell+1}{4\pi\ell(\ell+1)}P^2_\ell(\cos\theta)  \left[ C^{ij}_{gm}(\ell) + 2 (\alpha^i -1) C_{mm}^{ij}(\ell) \right] ~,
\end{equation}
following the curved sky projection. 

The $\alpha^i$ parameters have been carefully measured and extensively checked for both of the lens samples used in this work in \citet*{y3-2x2ptmagnification}, using realistic $N$-body simulations and \textsc{Balrog} image simulations \citep{y3-balrog}. In this work and within 3$\times$2pt, we use the \textsc{Balrog} $\alpha^i$ estimates for the fiducial model, shown in Table~\ref{tab:samples}.

\subsection{Intrinsic alignment model} \label{sec:IA}

In our tangential shear estimator from Eq.~(\ref{eq:gt_fullestimator}) we are averaging the ellipticity components to extract the shear. However the  observed ellipticity of a galaxy $e^{\text{obs}}$, is related to the shear by \citep{Seitz_1997}:
\begin{align}
    e^{\text{obs}} = \frac{e^{\text{int}}+g}{1+g^{*} e^{\text{int}}}
\end{align}
where in this equation all variables are complex numbers, $g = \gamma/(1 - \kappa)$ is the reduced shear and $g^*$ is the complex  conjugate of $g$. In the weak lensing regime $\kappa \ll 1 $, $\gamma\ll 1$ and we can then approximate the above equation to (we test this approximation in  Sec.~\ref{sec:red_shear}):
\begin{align}
e^{\text{obs}} \approx \gamma + e^{\text{int}}.
\end{align}
Thus, when averaging observed ellipticities we will only recover the shear if the intrinsic component of the ellipticity vanishes after averaging over many lens-source pairs. However this is not the case since the intrinsic component of the ellipticity, that is, the orientation of the source galaxies themselves, is correlated with the underlying large scale structure, and therefore with the lenses tracing this structure. We call this effect \textit{intrinsic alignments}. This effect is only present in galaxy-galaxy lensing measurements if the lens and source galaxies overlap in redshift. To model galaxy intrinsic alignments, we employ the TATT (Tidal Alignment and Tidal Torquing, \citealt{Blazek_2019})  and NLA (Non-linear linear alignment, \citealt{Hirata2004}) models.

It is typically assumed that the correlated component of intrinsic galaxy shapes is determined by the large scale cosmological tidal field $s$. The simplest relationship, which should dominate on large scales and for central galaxies, is when galaxy shapes align linearly with the background tidal field. This is what the NLA model is based on. More complex alignment processes, including tidal torquing, are relevant for determining the angular momentum of spiral galaxies and therefore their intrinsic orientation. The TATT model includes this additional component and is therefore better suited to describe the IA effects in a source sample that includes both red and blue galaxies. In nonlinear cosmological perturbation theory, we can write the intrinsic galaxy shape field, measured at the location of source galaxies, as an expansion of the density and tidal fields:
\begin{align}\label{eq:gamma_IA}
e^{\rm int}_{ij} = A_1 s_{ij} +A_{1\delta} \delta s_{ij} + A_2 s_{ik}s_{kj} + \cdots, 
\end{align}
where only here we use the $i,j,k$ letters to label the indices for a spin-2 tensor (elsewhere they denote redshift bins). In this expansion, using only the first ``linear'' $A_1$ term corresponds to the NLA model (when the nonlinear power spectrum is used for density correlations), while using all three parameters corresponds to the TATT model. $A_2$ captures the quadratic contribution from tidal torquing and $A_{1\delta}$ can be seen as a contribution from ``density weighting'' the tidal alignment contribution: we only observe IAs where there are galaxies, which contributes this additional term at next to leading order. The relevant two-point correlation for galaxy-galaxy lensing is expressed through the galaxy-intrinsic power spectrum:
\begin{align}
P_{gI} = b P_{\text{GI}},
\end{align}
where $b$ is the linear bias of the lens galaxies. While there are terms involving the correlation of IA and nonlinear galaxy bias, they are not included in our analysis here. These terms should be subdominant in the context of the TATT model and can be largely captured through the free $b_{TA}$ parameter defined in Eq.~(\ref{eq:b_ta}) (see, e.g., \citealt{Blazek_2015}).
 $P_{\text{GI}}(k)$ is the lensing-intrinsic power spectrum which we will write for both the NLA and TATT models. For the NLA model, cross correlating the tangential component $e^{\rm int}$ from the first term of Eq.~(\ref{eq:gamma_IA}) with the lens galaxy density field we can write the lensing-intrinsic power spectrum:
\begin{align}
    P_{\text{GI}}(k, z) = \left< \delta_g e^{\rm int}_t \right> = 
    A_1 \left< \delta_g s_E \right> = A_1  P_{mm}(k,z)\, ,  \quad [\textbf{NLA}]
\end{align}
where $s_E$ is the $E$-mode of the tidal field, and the last step is only possible because $P_{mm}$ is actually the same as $P_{m s_E}$ (but not in real space). Then, in the NLA model the IA power spectra are of the same shape as the matter power spectrum, but subject to a redshift-dependent rescaling, since we parametrize $A_1$ as $A_1(z)$, as  defined below. For the TATT model, we  perform the same expansion but now using all the terms from Eq.~(\ref{eq:gamma_IA}) to reach
\begin{align}
\begin{split}
P_{\text{GI}}(k, z) &= A_1 \left< \delta_g s_E \right> + A_{1\delta} \left<\delta_g \, \delta s_E\right>
+ A_2\left<\delta_g \,  s_E s_E\right> = \\
&= A_1 P_{mm}(k,z) + A_{1\delta} P_{0|0E}(k,z)
+ A_2 P_{0|E2}(k,z) . \\ 
&\qquad \qquad \qquad \qquad \qquad \qquad  \qquad \qquad \ \ [\textbf{TATT}]
\end{split}
\end{align}
In this work, these terms are evaluated using \texttt{FAST-PT} (\citealt{mcewen16, fang17}), as implemented in \textsc{CosmoSIS}. The full expressions for these power spectra can be found in \citet{Blazek_2019} (see equations 37-39 and their appendix A). In our TATT model implementation $A_1$, $A_2$, and $A_{1\delta}$ are all redshift-dependent quantities, defined as:
\begin{equation}\label{eq:tatt_c1}
    A_1(z) = -a_1 \bar{C}_{1} \frac{\rho_{\rm crit}\Omega_{\rm m}}{D(z)} \left(\frac{1+z}{1+z_{0}}\right)^{\eta_1}
\end{equation}
\begin{equation}\label{eq:tatt_c2}
    A_2(z) = 5 a_2 \bar{C}_{1} \frac{\rho_{\rm crit}\Omega_{\rm m}}{D^2(z)} \left(\frac{1+z}{1+z_{0}}\right)^{\eta_2}
\end{equation}
\begin{equation}\label{eq:b_ta}
A_{1\delta} (z) = b_{\mathrm{TA}} A_1 (z),
\end{equation}
where $\bar{C}_1$ is a normalisation constant, by convention fixed at a value $\bar{C}_1=5\times10^{-14}M_\odot h^{-2} \mathrm{Mpc}^2$, obtained from SuperCOSMOS (see \citealt{brown02}). The denominator $z_{0}$ is a pivot redshift, which we fix to the value 0.62, the same as the value used in DES Y1 3$\times$2pt analysis. The dimensionless amplitudes $(a_1, a_2)$ and power law indices $(\eta_1,\eta_2)$ are free parameters in the TATT model, as well as the $b_{\text{TA}}$ parameter which accounts for the fact that the shape field is preferentially sampled in overdense regions. 

Finally, the angular power spectrum of this IA contribution to galaxy-galaxy lensing is the relevant line-of-sight integral:
\begin{equation}
    C_{gI}^{ij}(\ell) = \int d\chi \frac{N_l^i(\chi)\, N_{s}^j(\chi)}{\chi^2}P_{gI}\left(k = \frac{\ell+1/2}{\chi},z(\chi)\right)\,.
\end{equation}

\subsubsection{Lens magnification $\times$ intrinsic alignments term}
Similarly, there is the contribution from the correlation between lens magnification and source intrinsic alignments, which is also included in our fiducial model:
\begin{equation}
    C_{mI}^{ij}(\ell) = \int d\chi \frac{q_l^i(\chi)\, N_{s}^j(\chi)}{\chi^2}P_{mI}\left(k = \frac{\ell+1/2}{\chi},z(\chi)\right)\,,
\end{equation}
where $P_{mI}(k) = P_{\text{GI}}(k)$. 

\subsection{Full tangential shear model}

Our tangential shear fiducial model includes the lens magnification term, intrinsic alignments and cross-terms between lens magnification and IA and  can be written as:
\begin{align}
\begin{split}
\gamma_t^{ij}(\theta) &= \sum_\ell \frac{2\ell+1}{4\pi\ell(\ell+1)}P^2_\ell(\cos\theta) \notag \\  
&\times \left[ C^{ij}_{gm}(\ell) + 2 (\alpha^i -1) \, \left(C_{mm}^{ij}(\ell)+ C_{mI}^{ij}(\ell)\right) + C_{gI}^{ij}(\ell) \right] ~,
\end{split}
\end{align}
following the curved sky projection. We show the different contributions to our model in Fig.~\ref{fig:terms_model} with the free parameters evaluated at the 3$\times$2pt best-fit. For the IA parameters these correspond to $A_1 = 0.60, \, A_2 = -0.16, \,  \alpha_1 = 4.2,\,  \alpha_2 = 3.8,\,  b_{TA}= 0.074$.

\subsection{Modeling pipeline technical details and code comparison}
We use the \textsc{CosmoSIS} framework \citep{Zuntz_2015} to compute the theoretical modelling. The output from \textsc{CosmoSIS} has been compared with that of \textsc{CosmoLike} \citep{Krause_2017} and reached an agreement of $\Delta \chi^2 < 0.1$ for the tangential shear part after scale cuts (>6 Mpc/$h$), which includes 248 points. The main differences between the two codes are that (1) \textsc{CosmoSIS} uses \texttt{CAMB} while \textsc{CosmoLike} uses \texttt{CLASS}, even though they are interchangeable for the DES Y3 3$\times$2pt analysis and (2) they use completely independent interpolation and integration schemes. Equivalently as for the measurement code, we stress the importance of performing such comparisons due to its effectiveness in identifying unexpected sources of error.

\section{Model validation}\label{sec:model_validation}

\begin{figure*}
\begin{center}
\includegraphics[width=0.9\textwidth]{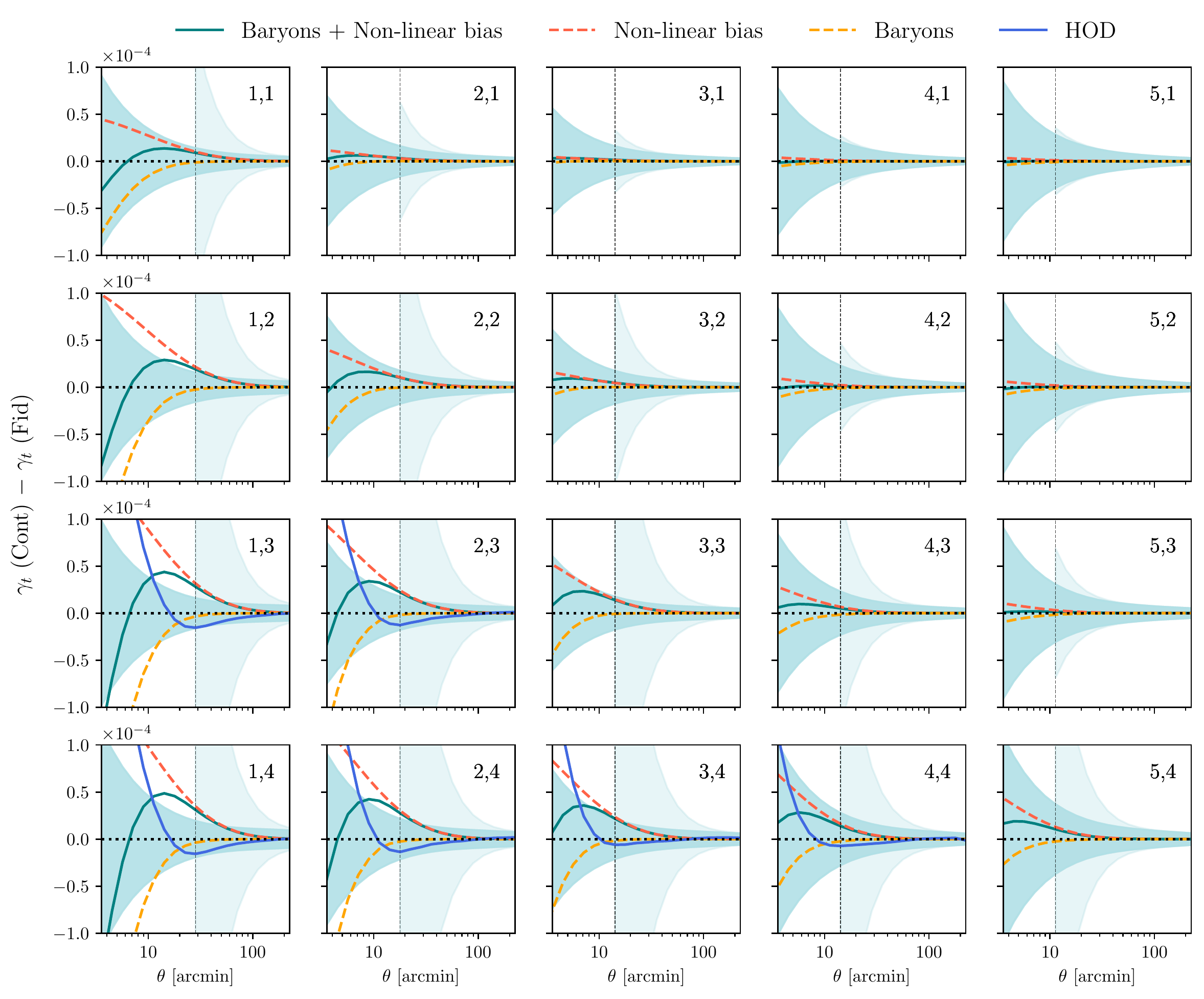}
\caption{In this figure we show the difference between a simulated datavector contaminated with baryonic effects and non-linear galaxy bias with respect to the fiducial model (linear bias and \texttt{Halofit} non-linear matter power spectrum), and the equivalent difference for an HOD contaminated datavector using the model and results from \citet{y3-HOD}. The darker blue shaded regions indicate the  uncertainties coming from the theory covariance without point-mass marginalization and the lighter ones including the point-mass marginalization. The dashed vertical black lines indicate the 6~Mpc/$h$ scale cuts. For more details see Sec~\ref{sec:galaxy_bias_tests} and Sec.~\ref{sec:hod}}
\label{fig:nlb}
\end{center}
\end{figure*}

We now summarize the validation of the model for the  galaxy-galaxy lensing signal described in \citet{y3-generalmethods} for all the probes, by exploring and illustrating the impact of several modeling effects and choices that are relevant to galaxy-galaxy lensing. The fiducial model, which includes several effects such as intrinsic alignments or lens magnification, is described in Section \ref{sec:theory}. We explore the impact of several effects that are not included in the fiducial modeling, in particular those concerning non-linear galaxy bias modeling, baryonic effects on the power spectrum, the effect of reduced shear, source magnification and source clustering, and their interplay. Within the DES Y3 3$\times$2pt analysis, we have adopted the threshold of 0.3$\sigma$ changes in the $\Omega_m - S_8$ plane to decide whether some effect is significant enough to be included in the fiducial model before unblinding. 

\subsection{Galaxy bias model and baryonic effects} \label{sec:galaxy_bias_tests}

In our fiducial model we assume a linear galaxy bias relation between the matter power spectrum and the galaxy-matter cross-power spectrum, as written in Eq.~(\ref{eq:linearbias}). Also we do not include baryonic contributions to the non-linear matter spectrum we assume, given by the \citet{Takahashi_2012} version from \texttt{Halofit}. In order to validate the applicability of both of these choices on scales greater than 6~Mpc/$h$,  we analyze a simulated galaxy-galaxy lensing datavector that receives contributions from non-linearities due to non-linear galaxy biasing and baryonic feedback. The non-linear bias contribution to this contaminated simulated datavector was generated using 1-Loop Perturbation Theory (see \citealt{Desjacques_2018} for a review) at parameter values motivated from analyzing 3D statistics in MICE simulations (see \citealt{pandey2020perturbation} and \citealt{y3-2x2ptbiasmodelling} for more details). In Fig.~\ref{fig:nlb} we illustrate the difference between the simulated datavector contaminated by baryonic effects and non-linear galaxy biasing as detailed above, and the fiducial vector, in comparison with the theoretical uncertainties, for illustrative purposes. We also show each of the effects separately in the same figure. When compared with the uncertainties without point-mass marginalization (with darker shade in that figure), we find they are not large enough to account for the differences between the two vectors, but once the point-mass marginalization is in place, the difference is always smaller than the uncertainties. Here the uncertainties from point-mass are obtained using a finite point-mass of $5\times10^{13} M_\odot$ --- otherwise the inverse covariance from Eq.~(\ref{eq:invcov_PM}) is not invertible. Then, this contaminated galaxy-galaxy lensing datavector is  analyzed with the fiducial linear bias model in conjunction with galaxy clustering and cosmic shear. The bias in recovered cosmological parameters is less than 0.3$\sigma$ from the input truth values (see \citealt{y3-2x2ptbiasmodelling} and \citealt{y3-generalmethods} for more details).

\subsection{Halo occupation distribution model}\label{sec:hod}
In Figure~\ref{fig:nlb} we also show a simulated datavector produced with the halo occupation distribution model (HOD) developed in \citet{y3-HOD}. In their paper they fit the HOD model to tangential shear measurements of the \redmagic \ and the \textsc{MagLim} sample from 0.25 to 250 arcmin, divided into 30 angular bins. In \citet{y3-HOD} only the highest S/N lens-source redshift bins combinations are used to fit the HOD model, which are the ones where the lens and the source galaxies are more separated in redshift. In Fig.~\ref{fig:nlb} the HOD line corresponds to their best-fit HOD model of the \redmagic \ sample, which we compare with the fiducial model used in the 3$\times$2pt cosmological analysis. As expected, the HOD and the 3$\times$2pt model agree on large scales but they show strong deviations at smaller scales. Also note the data-informed HOD model shows smaller differences with respect to the fiducial model than the baryons + Non-linear bias contaminated data vector which has been used to define the scale cuts, validating it as a conservative choice.

\subsection{Reduced shear, source magnification and source clustering}\label{sec:red_shear_etc}

We now consider the impact of the reduced shear approximation, and the source magnification and source clustering effects, which are all connected to each other as well as to the lens magnification and IA terms which we described in Sec.~\ref{sec:lens_mag} and Sec.~\ref{sec:IA} respectively. In this section we will write the contribution to position-shape correlations of all these effects. In Sec.~\ref{sec:red_shear} we will describe in more detail the reduced shear approximation and the tests we have performed to validate it, and in Sec.~\ref{sec:source_mag_source_clu} we focus on source magnification and source clustering. This work has been performed following  \citet{y3-generalmethods}, which studies second-order effects not only to galaxy-galaxy lensing but to the other correlation functions and where the full expressions for each of the effects can be found. Here we summarize their conclusions affecting the galaxy-galaxy lensing observable and illustrate some of the effects at the two-point function level. We also expand on the relation of the source magnification and source clustering effects with the tangential shear estimator presented in this work.

We can start by writing the observed lens galaxy density as we derived in Sec.~\ref{sec:lens_mag}, including the lens magnification term:
\begin{equation}\label{eq:lens_mag_2}
      \delta_g^\text{obs} =  \delta_g^\text{int} + \delta_g^\text{mag} =  \delta_g^\text{int} +  C_l \kappa_l,
\end{equation}
and then we can also write the observed ellipticity $e^{\text{obs}}$ as the following expression, which includes the higher-order effects of reduced shear, with a (1+$\kappa_s$) factor after using a Taylor expansion, where $\kappa_s$ is the convergence field at the redshift of the sources, intrinsic alignments produced by the intrinsic ellipticity $e^{\rm int}$, source clustering represented by $\delta_s$, source magnification $C_s \kappa_s$ (following the analogous notation as for lens magnification):
\begin{equation}\label{eq:gamma_obs_full}
    e^{\text{obs}} =  (\gamma\, (1 + \kappa_s)+ e^{\rm int}) (1  + \delta_s + C_s \kappa_s).
\end{equation}
Correlating these two fields gives:
\begin{equation}
\label{eq:all_terms}
\begin{split}
 &\left< \delta_g^\text{obs} e^{\text{obs}} \right> = 
 \left< (\delta_g^\text{int} +  C_l \kappa_l)   (\gamma \, (1 + \kappa_s)+ e^{\rm int}) (1  + \delta_s + C_s \kappa_s)\right> =  \\
 &\qquad= \underbrace{\left< \delta_g^\text{int} \gamma\right>}_{\text{signal}} 
 + \, \underbrace{C_l\left< \kappa_l  \gamma \right>}_{\text{lens mag}} + 
 \, \underbrace{\left< \delta_g^\text{int} e^{\rm int}\right>}_{\text{IA}} + 
 \, \underbrace{C_l\left< \kappa_l  e^{\rm int} \right>}_{\text{lens mag + IA}}  + \\
 &\qquad+ \, \underbrace{\left< \delta_g^\text{int}  e^{\rm int} \delta_s \right>}_\text{IA + source clu } +  
 \,\underbrace{ (1+ C_s) \left< \delta_g^\text{int}  \gamma \kappa_s \right>}_\text{red. shear or source mag } \, + 
 \, \underbrace{\left< \delta_g^\text{int} \gamma   \delta_s \right>}_{\text{source clu}} \, + \\ 
 &\qquad+ \,\underbrace{ C_l \, (1+ C_s) \left< \kappa_l   \gamma  \kappa_s \right>}_\text{lens mag + (red. shear or source mag)}  + 
 \, \underbrace{  C_s \left< \delta_g^\text{int} e^{\rm int} \kappa_s \right>}_\text{IA + source mag } \, 
 + \, \cdots
\end{split}
\end{equation}
where the first terms in the expansion are included in our model, that is, lens magnification, IA, and lens magnification coupled with IA, and IA coupled to source clustering. Then, we have computed the next term that appears, which includes contributions from reduced shear and source magnification independently (they are only grouped together since the terms have the same form). We have also estimated the source clustering term and found it negligible \citep{y3-generalmethods}. Importantly, as discussed in \cite{y3-generalmethods}, these correlations must be calculated in three dimensions and then projected along the line of sight. We have not computed the rest of the terms, but given that we find the reduced shear and source magnification terms negligible with the current uncertainties, we expect them to also be negligible, being even smaller than the terms we have computed. Also, we have omitted terms which are not written in the equation above involving correlations of four fields, as well as two terms involving lens magnification coupled with source clustering and IA, which we expect to be very small.

\subsubsection{Reduced shear}\label{sec:red_shear}

\begin{figure*}
\begin{center}
\includegraphics[width=0.9\textwidth]{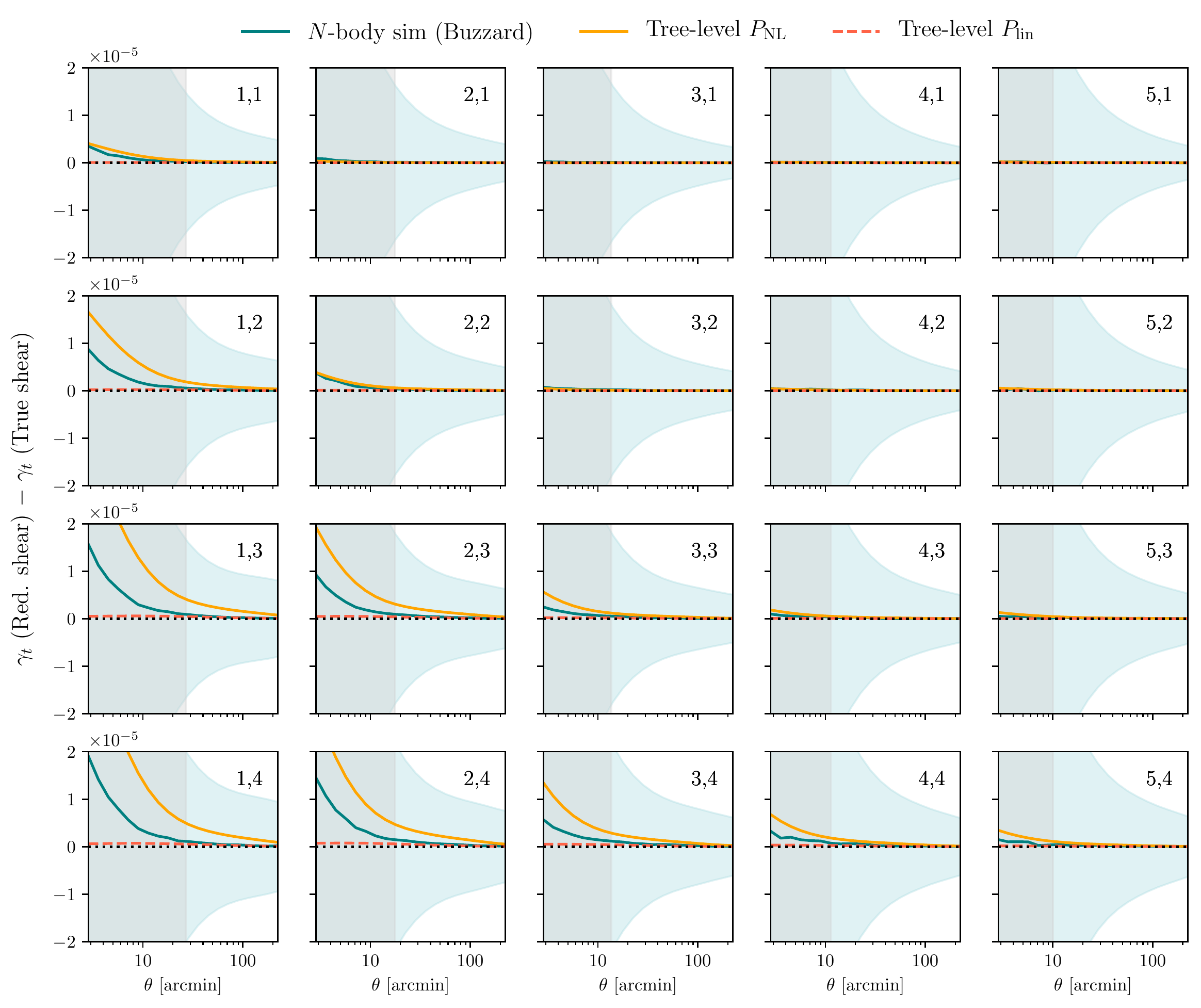}
\caption{Reduced shear impact on the tangential shear quantity for the \redmagic \ sample. We compare the results from $N$-body \textsc{Buzzard} simulations to the ones using theoretical predictions from \textsc{CosmoLike} both with Tree-level non-linear power spectrum, labeled as $P_{\rm{NL}}$, and with linear theory, labelled as $P_{\rm{lin}}$. See \citet{y3-generalmethods} for the theoretical expressions used for these predictions. The vertical gray shading corresponds to the 6 Mpc/$h$ scale cuts and the blue shading to the tangential shear uncertainties. The numbers in each panel correspond to the lens and source redshift bins, in that order.}
\label{fig:reduced_shear}
\end{center}
\end{figure*}

When a galaxy is weakly lensed, the change in its observed ellipticity is proportional to the reduced shear, $g$, which is related to both the shear and the convergence as: 
\begin{equation}
    g = \frac{\gamma}{1-\kappa} \simeq  \gamma \, (1 + \kappa),
\end{equation}
using a Taylor expansion in the last step. Since $|\gamma|,  |\kappa| \ll 1$ for individual galaxies in the weak lensing regime, the reduced shear is typically approximated by the shear, in what is known as the reduced shear approximation. In this work we make use of this approximation, and here we test whether that is sufficient for the current analysis, given DES Y3 uncertainties. 

After performing the expansion correlating with the observed density field in Eq.~(\ref{eq:all_terms}), we have computed the reduced shear term $\left< \delta_g^\text{int} \, \gamma \,  \kappa_s \right>$ using \textsc{CosmoLike} -- see Eq. 42 from \citet{y3-generalmethods} for the expression with the expanded integrals using Tree-level perturbation theory. We have also estimated the impact of the reduced shear approximation using the \textsc{Buzzard} $N$-body simulations, directly comparing the tangential shear measurements obtained with true shear compared with the shear contaminated with the $(1-\kappa)^{-1}$ factor. In Fig.~\ref{fig:reduced_shear} we compare the different estimates of the reduced shear effect to the tangential shear estimator, including two theoretical estimates using a tree-level bispectrum based on the non-linear power spectrum $P_\text{nl}$ or on the linear power spectrum $P_\text{lin}$. The tree-level bispectrum with $P_\text{nl}$ is known to not be an accurate model and the numbers obtained from that are useful as an upper limit only. The \textsc{Buzzard} estimate is expected to be the most accurate at large scales and intermediate scales, with the only the limiting factors being limited resolution at the smallest scales, especially for the low lens redshift bins, and some level of noise from the measurements. Still, to perform the robustness tests we use the 
the largest estimate of the three to be conservative. Comparing the addition of reduced shear using the theoretical estimate with the Tree-level $P_\text{nl}$ with the fiducial modeling, we find a $\Delta \chi^2 = 0.45$ after the scale cuts of 6~Mpc/$h$ without point mass marginalization and $\Delta \chi^2 = 0.15$ with PM marginalization. This translates to a shift of 0.07$\sigma$ in the 2D $\Omega_m - S_8$ plane \citep{y3-generalmethods}. Within the DES Y3 3$\times$2pt analysis, we have adopted the threshold of 0.3$\sigma$ shifts in the $\Omega_m - S_8$ plane to decide whether some effect is significant enough to be included in the fiducial model. Therefore we found the reduced shear approximation to be good enough for the 3$\times$2pt DES Y3 analysis. In this work we have not computed the term that comes out of the coupling between lens magnification and reduced shear since it would be smaller than the reduced shear only term, and therefore negligible for our analysis. However, this term might become important in  future analyses.

\subsubsection{Source magnification and source clustering}\label{sec:source_mag_source_clu}

Here we consider the effects of source magnification and source clustering, which both impact the observed source number density.
Given our choice of estimator for the tangential shear signal, Eq.~(\ref{eq:gt_fullestimator}),
we are sensitive to the density of source galaxies in three ways: (1) the boost factors; (2) intrinsic alignments; and (3) the relative weighting of lenses and sources in the sample given that we are averaging the tangential shear in lens-source galaxy pairs. The boost factors, which come from the excess number of lens-source pairs due to clustering, are discussed in Sec.~\ref{sec:boost_factors}. We correct for this effect on the measurement side to match the theoretical predictions for the tangential shear signal that use the mean survey $n(z)$. The impact of intrinsic alignments is modulated by the number of observed source galaxies. Thus, any correlation between intrinsic galaxy ellipticity and observed source density can appear in the signal. 

The third effect above arises because the tangential shear signal is weighted by the source positions, both their angular positions and redshifts. Lenses with more observed background sources will receive more weight in the signal. This effect can potentially bias the tangential shear signal when the source observed density is correlated with the lens density, for instance via magnification or when lenses and sources overlap in redshift. It could be partially removed if we averaged the tangential shear for each lens galaxy and \textit{then} we averaged again over all lenses to ensure that they are weighted equally, modulo the lens weights themselves (e.g. \citealt{Taylor_2020}). However, we choose to average the tangential shear in lens-source pairs because of the significant increase in S/N this method yields, due to optimal handling of shape noise. 

We note that if we had access to the true scale-dependent $n(z, \theta)$, giving the relevant source number density as a function of separation from lens positions, we could accurately model the tangential shear signal, including the impact of source magnification and source clustering, and without needing any boost factor correction in the measurement since the impact of lens-source clustering on the redshift distributions would naturally be accounted for in the model. However, this information is not readily available in photometric surveys, and thus we instead test how significant these contributions are. 

\textbf{Source clustering} refers to the clustering of source galaxies due to large scale structure. This implies we are more likely to find a galaxy for shear estimation in regions that are overdense in the underlying density field. As long as the source and the lens redshift are well-separated, the large scale structure at the source redshift is not correlated with that at lens redshift, and therefore, even if we will still be weighting the lens galaxies in front of these overdensities more, this will not bias our signal. Alternatively, if there is some correlation between the large scale structure at the redshift of the source galaxy $\delta_s$ and the one at lens redshift $\delta_l$ this can potentially bias our tangential shear estimator. To test the impact of source clustering, we use the following transformation when computing the integrals developed in \citet{y3-generalmethods}:
 \begin{align} \label{eq:source_clustering_nz}
    n_{\mathrm{s}}(\chi)\rightarrow n_{\mathrm{s}}(\chi)
\left[1+{\delta^{(3\mathrm D)}_{\mathrm{s}}(\hat{\mathbf n} \chi, \chi)}
\textcolor{black}{} \right] \, , 
 \end{align}
 which applies the transformation at the source redshift distribution level, with $\hat{n}$ being a line of sight unity vector.
 
 The resulting contribution to the lensing correlations is very small, and indeed it vanishes in the Limber approximation, because sources at the lens redshift are not lensed. However, an analogous contribution exists for the source clustering-IA term, which is more important since IAs arise when lenses and sources are physically nearby, i.e.\ the same regime where they are clustered. We account for this in our fiducial TATT model perturbatively using the $b_{TA}$ parameter defined in Eq.~({\ref{eq:b_ta}}) (also see \citealt{Blazek_2015}). 
 
 Note that the contributions discussed here are different from the boost factor correction, which must be applied for Eq.~(\ref{eq:all_terms}) to hold -- i.e. it is written assuming $\left< \delta_g^\text{obs} e^{\text{obs}} \right>$ is normalized by the ``random-random'' number of pairs in the tangential shear estimator since the terms are computed using the mean survey $n(z)$'s.

\textbf{Source magnification} refers to the magnification produced to source galaxies by the large scale structure in front of them. Because of magnification, the number density of source galaxies will be influenced by the overdensities or underdensities present at the lens redshift bin in particular. Thus, given our tangential shear estimator, lines of sight with higher matter densities will be weighted differently than those with less matter, potentially biasing the tangential shear signal. The impact depends on the characteristics of the source sample, specifically on whether the magnification factor $C_s$ (analog to the one defined in Eq.~\ref{eq:lens_mag_2} for the lens sample) is positive or negative. For the same reason as for the source clustering case, we also model the correction at the three-dimensional $n(z)$ level. When combining both effects this leads to \citep{y3-generalmethods}:
 \begin{align}
 \label{eq:source_clustering_mag_nz}
    n_{\mathrm{s}}(\chi)\rightarrow n_{\mathrm{s}}(\chi)
\left[1+{\delta^{(3\mathrm D)}_{\mathrm{s}}(\hat{\mathbf n} \chi, \chi)} \right] \left[1+ C_\mathrm{s} \kappa \left(\hat{\mathbf n}, z \right)\right]\,.
 \end{align}
Using \cosmolike \ we have computed the term that includes both reduced shear and source magnification, which has the same form as the term with only reduced shear but also including the factor $C_s$ that determines the strength of source magnification and is sample dependent. Analogously as for the lens sample (see Sec.~\ref{sec:lens_mag}) we can change the notation to: $C_s = 2 (\alpha_s -1)$. \citet*{y3-2x2ptmagnification} has measured $\alpha_s$ for the DES Y3 shape catalog, using \texttt{Balrog} \citep{y3-balrog} and obtained the values shown in Table~\ref{tab:samples}, for each of the source bins. Using these estimates for the magnification coefficients, we obtained a $\Delta \chi^2=1.8$ for the tangential shear part after scales cuts of >6~Mpc/$h$ without point-mass marginalization (1.3 with PM marginalization) and a corresponding shift of 0.128$\sigma$ in the 2D plane of $\Omega_m - S_8$. These estimates are based on the Tree-level bispectrum models using the non-linear power spectrum. We therefore do not find the combination of source magnification and reduced shear significant for this analysis, but it is possible this already becomes relevant for DES Y6 data. Regarding the coupling between lens magnification, source magnification and reduced shear, we have not computed this term since it will be smaller than the one we have found negligible in the current analysis.

\subsection{Deflection effects}\label{sec:deflection_effects}
Galaxies at $z\sim 1$ are typically deflected $\sim 1$ arcmin by the large scale structure in front of them \citep{Chang_2014}. This could in principle significantly affect our estimation of the galaxy positions, both lenses and sources, and therefore the estimated angular separation between a given lens-source pair. However, it is important to note that for source galaxies, which are the ones that will generally be experiencing more deflection, the original position does not actually matter. The only relevant position is where the light from the source galaxy is lensed by the foreground galaxy; thus we only need to consider the deflection experienced between the redshift of the lens and the observer. Lens galaxies will also be deflected. Hence, the error that will propagate into the estimation of the angular separation between a given lens-source pair comes from the difference between the deflection angles of the source and lens galaxies between the lens redshift and us. This difference, $\Delta \vec{\alpha}$, will generally be larger for larger angular separations, since the gravitational potential along more separated lines of sight will differ more. The relevant component of $\Delta \vec{\alpha}$ (the one propagating to angular separations) can be as large as $\sim 0.6$ arcmin for $z_l=1$ at the maximum angular separation used in this analysis, of 250 arcmin, and lower for smaller separations and lens redshift. At lens redshift $z_l = 1$ and $\theta =10$ arcmin, the error on the spatial separation between the source and the lens is about 1\%, and  0.2\% for $\theta =250$ arcmin \citep{Chang_2014}. We do not expect such errors to  significantly impact the results presented in this paper nor in the DES Y3 cosmological analysis, but might need to be considered in future generation surveys.

There is a second effect to be considered. Since the relative position of a given lens-source plane will be affected by the difference in the deflection angles, this induces an error on the projection of the Cartesian ellipticity components to the tangential, as illustrated in Figure 3 of \citet{Chang_2014}. This is also a second-order effect and is still below a percent \citep{Chang_2014}. For the cross-component of the shear this becomes a first-order effect. Since we do not measure any signal in $\gamma_\times$ (see Sec.~\ref{sec:gammax}) we conclude this is not significantly impacting our tangential shear measurements but might become relevant in future data sets.

\section{Measurement robustness tests}
\label{sec:sys_tests}

In the previous section we have explored the impact of several effects on our modeling of the galaxy-galaxy lensing signal. Similarly, we can assess the robustness of the galaxy-galaxy lensing measurement using DES Y3 data. In this section, we perform a series of tests that should produce a null signal when applied to true gravitational shear, but whose non-zero measurement, if significant, would be an indication of systematic errors leaking into the main galaxy-galaxy lensing observable. Other tests of the shear measurement, but not specific to galaxy-galaxy lensing, are presented in \citet*{y3-shapecatalog}. 

\subsection{Cross-component}\label{sec:gammax}
The mean cross-component of the shear $\gamma_\times$, which is rotated 45 degrees with respect to the tangential shear $\gamma_t$, is expected to be compatible with zero if the shear is produced by gravitational lensing alone, because all the galaxy-galaxy lensing signal is captured by the tangential shear. The cross-component should also vanish in the presence of systematic effects that are invariant under parity.

\begin{figure*}
\begin{center}
\includegraphics[width=0.99\textwidth]{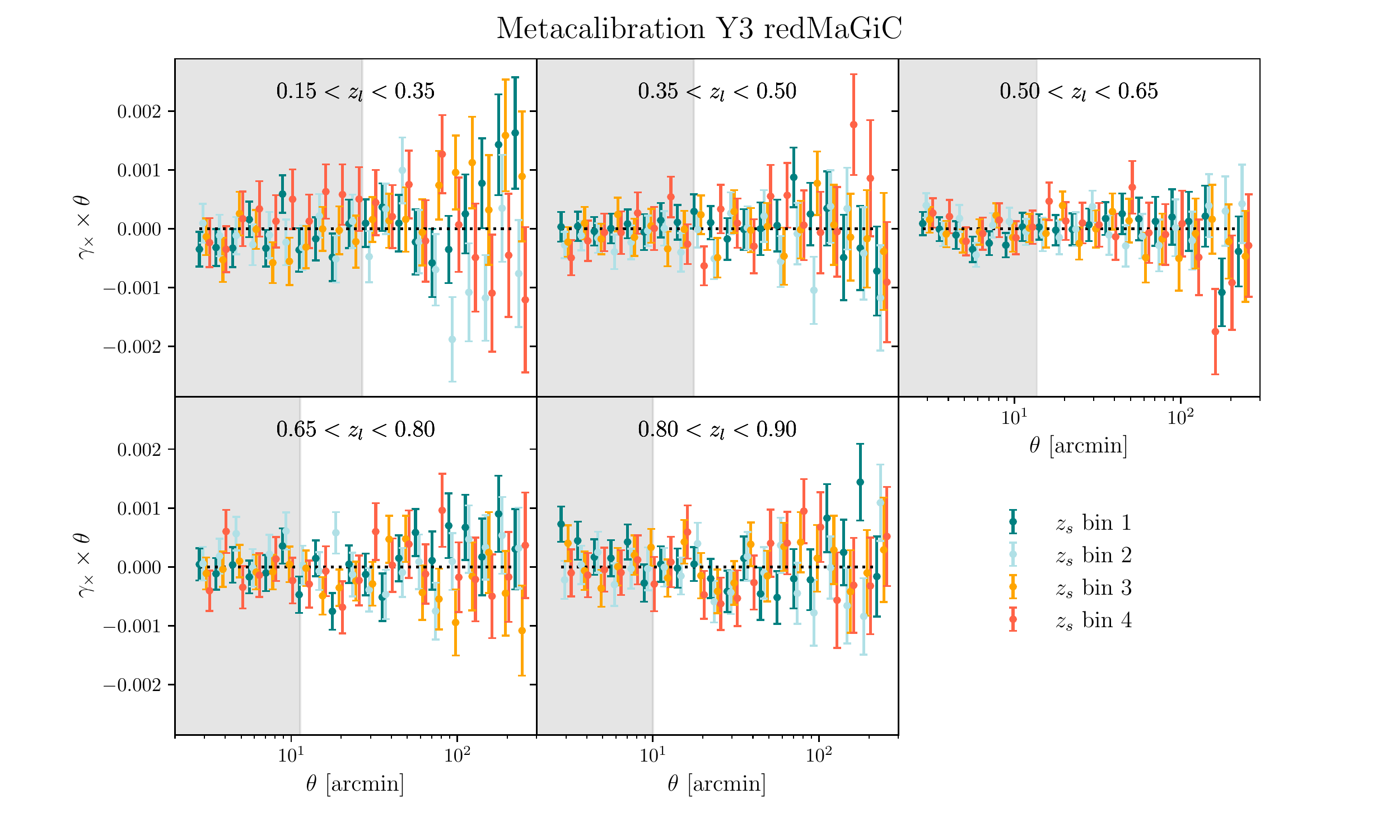}
\includegraphics[width=0.99\textwidth]{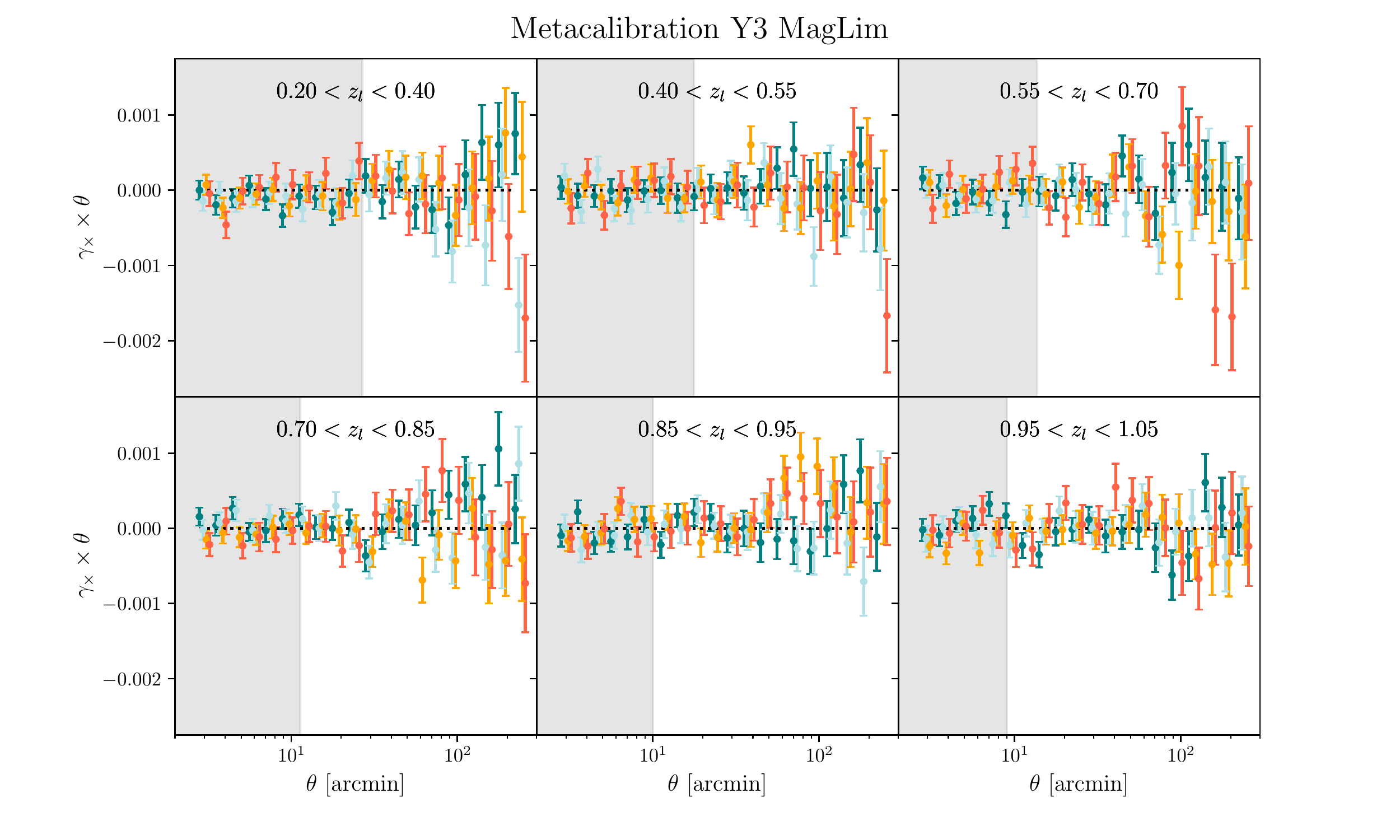}

\caption{Cross-component of the galaxy-galaxy lensing signal around \redmagic\ (top) and  \textsc{MagLim} (bottom) lenses. The same lens-source bin combinations used for the tangential galaxy shear measurements have been considered.}
\label{fig:gammax_allbins}
\end{center}
\end{figure*}

\begin{figure}
\begin{center}
\includegraphics[width=0.4\textwidth]{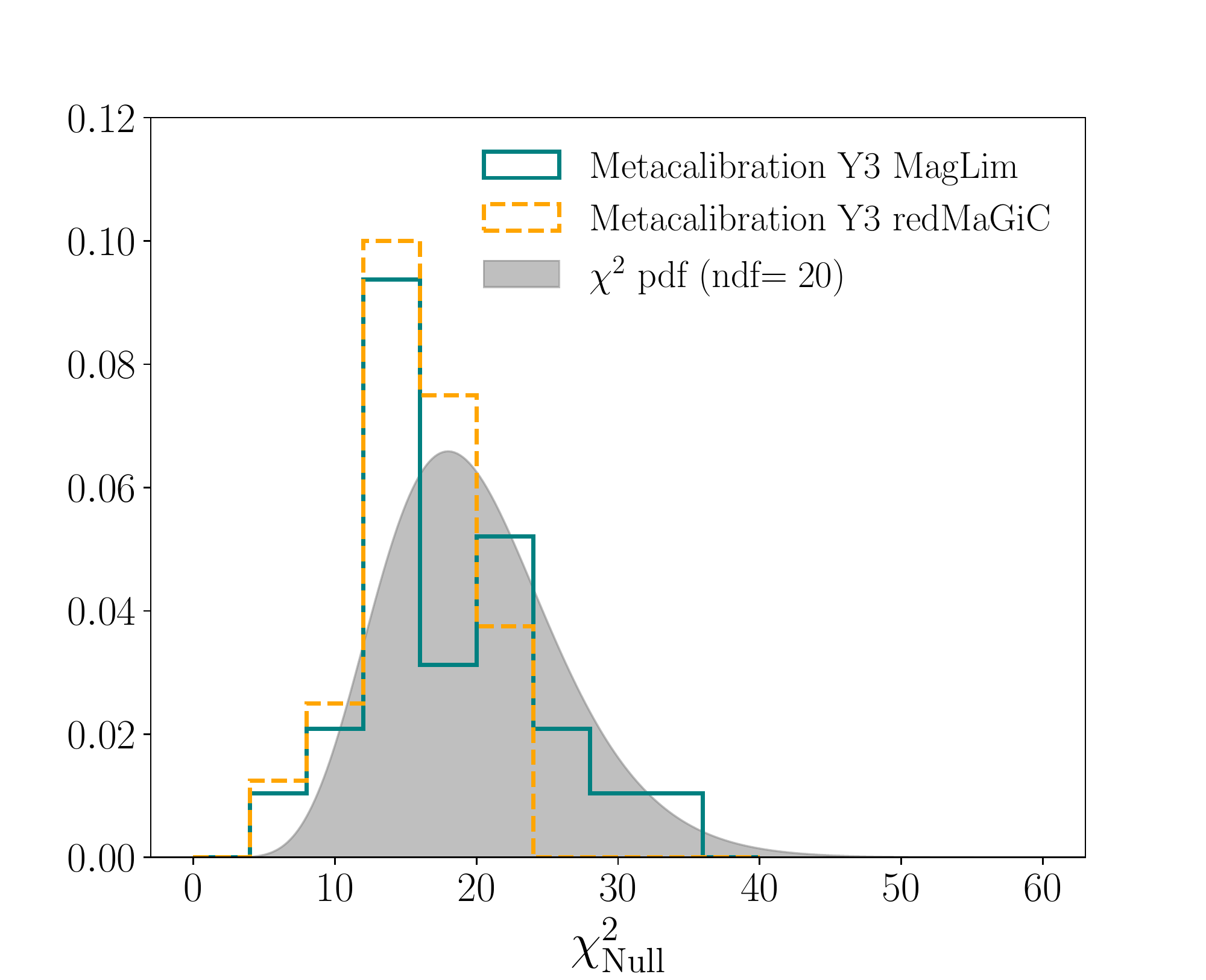}
\caption{Null $\chi^2$ distribution for all the lens and source redshift bin combinations of the cross-component of the galaxy-galaxy lensing signal compared to the expected $\chi^2$ for 20 degrees of freedom. The blue histogram represents the \textsc{MagLim} lenses, while the orange histogram stands for \redmagic\ lenses. In both cases the histogram is comparable to the expected $\chi^2$ distribution. We conclude that the $\gamma_\times$ measurements are compatible with a null signal.} 
\label{fig:gammax}
\end{center}
\end{figure}

In Fig.~\ref{fig:gammax_allbins} we show the resulting cross-shear measured around \textsc{MagLim} and \redmagic\ lenses (including random point subtraction, but not boost factors) for the exact same bin specifications used for the tangential shear. In that figure we use jackknife uncertainties (see Appendix~\ref{appendix_randompoints} for a comparison of jackknife uncertainties to theoretical ones). In order to quantitatively assess the compatibility of the cross-shear with a null signal, we compute the $\chi^2$ of our measurements against a null signal. In Fig.~\ref{fig:gammax} we present the null $\chi^2$ histogram coming from all the lens-source bin combinations of $\gamma_\times$, computed using the jackknife covariance for each lens-source bin. Note that we neglect any cross-covariance that might exist between different lens-source pairs (see Appendix\,\ref{appendix_gammax} for a justification based on lognormal simulations). We consider both the \textsc{MagLim} and \redmagic\ samples. In order to compute the $\chi^2=\gamma_\times^TC^{-1}\gamma_\times$ we need an estimate of the inverse of the covariance matrix. Given the fact that jackknife covariances contain a significant level of noise, we correct for the biased estimation of the covariance matrix with the Hartlap correcting factor, that, while not mathematically exact in the case of nonindependent realisations, it was shown by \citet{Hartlap_2007} to
yield accurate results also in this case. Therefore, we multiply the inverse covariance matrix by $(N_{\rm JK}-p-2)/(N_{\rm JK}-1)$, where $N_{\rm JK}$ is the number of jackknife regions (150 in our case) and $p$ the number of angular bins (20 in our case). As can be seen from the figure, our results are consistent with a null signal. Note that for this test we have used all angular scales. Therefore, our cross-shear measurements are compatible with zero at all scales, not only at the ones used for the galaxy-galaxy lensing probe in the DES Y3 3$\times$2pt cosmological analysis. However, we have also checked the results considering only the largest scales (above one degree) and obtained a very good agreement with a null signal.

\subsection{PSF residuals}
\label{sec:psf}

The estimation of source galaxy shapes involves their modeling after being convolved with the point spread function (PSF) pattern, which depends on the atmosphere and the telescope optics and which is characterized using stars in our data sample \citep*{y3-shapecatalog}. Here we test the impact of PSF modeling residuals on the galaxy-galaxy lensing estimator, and their compatibility with a null signal.

\begin{figure}
\begin{center}
\includegraphics[width=0.5\textwidth]{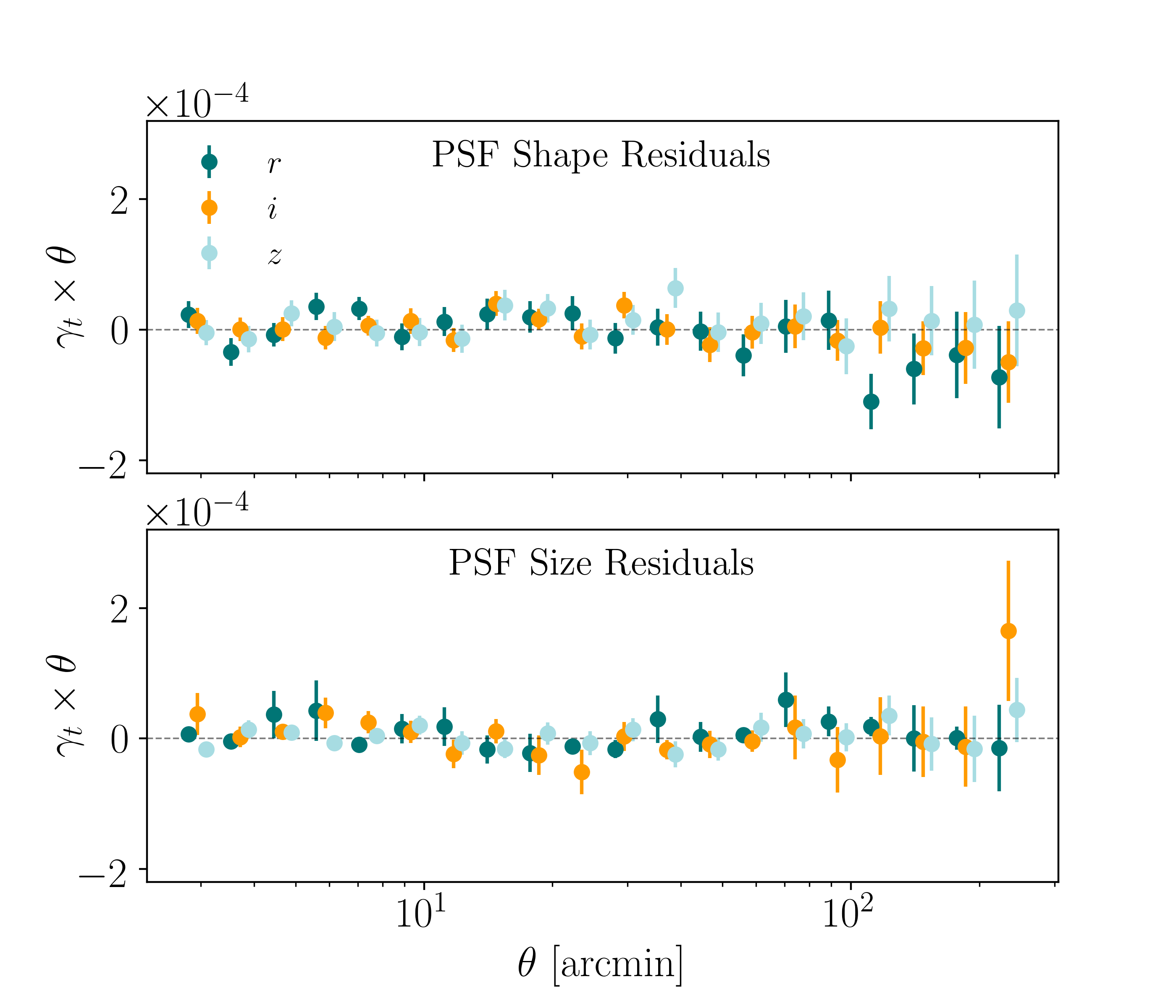}
\caption{Measurements of PSF shape and size residuals in $r,i,z$-bands, as described in \S \ref{sec:psf}. The measurements are found to be consistent with the null hypothesis. }
\label{fig:psfresiduals}
\end{center}
\end{figure}

In particular, we consider two kinds of PSF residuals. On the one hand, we look at PSF shape residuals which are the differences between the measured shape of the (reserved) stars and the \texttt{PIFF} model shape \citep{y3-piff} at those same locations. On the other hand, the PSF size residuals are computed by rescaling the size of the measured PSF to match the difference in PSF size between the measurement and the model of the PSF, but keeping the PSF shape to its measured value. In Figure \ref{fig:psfresiduals} we show the measurement mean of the tangential component of the two PSF residuals we just described around \textsc{redMaGiC} galaxies, including the subtraction of the same quantity around random points, in the same manner as for the tangential shear signal. For the PSF shape residuals, we obtain the following null hypothesis $\chi^2$ values, using a jackknife covariance, for the PSF measurements in the three bands considered ($r$, $i$, $z$): 27.1, 21.5, 14.3 (for 20 data points). Correspondingly, we find the following $\chi^2$ values for the PSF size residuals: 15.0, 13.5, 13.6 (for 20 data points). In summary, we find no sign of contamination from PSF residuals.

\subsection{Impact of observing conditions} \label{sec:observing_conditions}

\begin{table*}
\centering
\pmb{Included in the DES Y3 GGL fiducial setup} \\
\vspace{2mm}
\setlength\extrarowheight{-3pt}
\setlength{\tabcolsep}{5pt}
\begin{tabular}{clccc}
  &    &  Contribution ($\Delta \chi^2$ with vs. w/o) & Uncertainty ($\Delta \chi^2$) & Uncertainty Propagated? \\ [0.02in]
\hline 
\\ [-0.08in]
\multirow{4}{*}{Measurement} & LSS weights & 3.1 (3.0 with PM) & 1.3 (1.2 with PM) & No\\
                             & Shear response & 405 (125 with PM) & Captured by the Mult. Shear bias & Yes\\
                             & TreeCorr approx. & $\sim 1 $  & 0 & No \\
                             & Boost factors & 0.1 & Negligible & No\\[0.02in]
                             
\hline 
\\ [-0.08in]
\multirow{5}{*}{Model}    & Lens magnification & 20 \ (9.4 with PM) & 7.7 \ (3.1 with PM) & No\\
                             & Intrinsic alignments & 46 \ (38 with PM) &  22 \  (20 with PM) & Yes (within TATT) \\
                             & Source redshifts & - & 20 \ (11 with PM) & Yes (on mean)\\
                             & Lens redshifts & - & 4.8 \ (2.4 with PM) & Yes (on mean and width) \\
                             & Mult. Shear bias & 3.7 \ (1.8 with PM) & 0.65 \ (0.54 with PM) & Yes \\ 
\bottomrule
\end{tabular}
\\
\vspace{5mm}
\pmb{Not included in DES Y3 GGL fiducial setup} \\
\vspace{2mm}
\begin{tabular}{clcc}
  &    &  Contribution ($\Delta \chi^2$ with vs. w/o) & Uncertainty ($\Delta \chi^2$)  \\ [0.02in]
\hline 
\\ [-0.08in]
Measurement & NK shear response & 0.0002 & Negligible  \\
\hline 
\\ [-0.08in]
\multirow{5}{*}{Model} & Non-linear galaxy bias & 22 \ (0.42 with PM) & $\sim$17 \ ($\sim$0.16 with PM)  \\
                          & Baryonic effects & 10 \ (1.7 with PM) & $\sim$10 ($\sim$1.7 with PM) \\
                          & Reduced shear & 0.45 \ (0.15 with PM)  & $\sim 0.25 $ \\
                          & Source mag. + red. shear & $\sim$1.8 \ ($\sim$1.3 with PM) & $\sim 1 $ \\
\bottomrule
\end{tabular}
\caption{\color{Black} Summary table of the effects included in the DES Y3 3$\times$2pt fiducial measurements and model that are relevant for the galaxy-galaxy lensing probe (\textit{top}) and the ones that are not included (\textit{bottom}) but that we test in both this paper and the DES Y3 3$\times$2pt methodology paper in \citet{y3-generalmethods}. In the first column we show the contribution of each of these effects in either the measurement or the best-fit fiducial DES Y3 3$\times$2pt model. The contribution is estimated by computing the $\Delta \chi^2$ between the best-fit model and the same model removing the corresponding contribution. In the second column we show the uncertainty in each of the effects,  estimated computing the $\Delta \chi^2$ between the best-fit model and the model with a $2\sigma$ deviation in the corresponding effect. In the third column of the upper part we indicate whether the uncertainty is propagated to the cosmological contours. In this table we consider the \redmagic \ lens sample. We use the inverse of the theoretical covariance with and without the point-mass marginalization (PM) to estimate the $\Delta \chi^2$'s, considering only the large scales used in the cosmological analysis for the galaxy-galaxy lensing part (above 6 Mpc/$h$), which include 248 points. See Section~\ref{sec:discussion_future} for more details and discussion about future prospects.}
\label{tab:all_effects}
\end{table*}

Time-dependent observing conditions are intrinsic to photometric surveys, and they may impact the derived galaxy catalogs, for instance, introducing galaxy density variations across the survey footprint. The dependence of galaxy density on observing conditions introduces a spurious clustering signal that can have a strong impact on some of the observables used in the DES Y3 cosmological analysis, particularly on galaxy clustering. 

In order to correct for such dependence, a weighting scheme has been developed in \citet{y3-galaxyclustering} developed to remove the dependence of lens galaxy density on observing conditions. The scheme utilizes the maps of observing conditions such as exposure time, airmass, seeing and others, to then produce a set of weights which when applied to the galaxy population no correlations are observed between galaxy density and such observational properties. The impact of such a weighting scheme is significant for galaxy clustering measurements, as expected \citep{y3-galaxyclustering}. However, since galaxy-galaxy lensing is a cross-correlation between lens and source galaxies, we expect the impact of varying observing conditions, and hence of the weighting scheme, to be less important \citep*{Y1GGL}. In Figure~\ref{fig:estimator_tests} we show the impact of the weighting on the galaxy-galaxy lensing measurements, and we report a $\Delta \chi^2 = 4.2$ for all the scales (corresponding to 400 data points) and a $\Delta \chi^2 = 3.1$ for scales above 6~Mpc/$h$ used directly in the cosmological analysis for the \redmagic \ sample (see Table~\ref{tab:summary_chi2s} for similar results of the \textsc{MagLim} sample). 

Besides this, in Fig.~\ref{fig:errors} from Appendix~\ref{appendix_randompoints} we compare the Jackknife uncertainties using 150 patches to the theoretical uncertainties using the fiducial covariance used in the 3$\times$2pt analysis from \citet{y3-covariances}. The fact they agree provides some evidence that the tangential shear measurements presented in this work do not present stronger variations across the footprint than expected.

\section{Summary of measurement and modeling  uncertainties} \label{sec:discussion_future}

In this section we discuss the contribution of each of the components of the model and the measurement as well providing an estimate of their uncertainty. We present this in Table~\ref{tab:all_effects}. In the top part of the table we summarize the effects which are included in our fiducial model and in the bottom the ones which are not included in the fiducial model but whose impact we have estimated. We also classify the effects depending on whether they are measurement or model components. Here we detail how we compute the uncertainty column shown in the table for each of the effects and point to the  part of the paper where each effect is explained: 
\begin{itemize}
    \item \textbf{LSS weights}: See Sec.~\ref{sec:observing_conditions}. We determine how the uncertainty on the LSS weights propagates to the tangential shear measurements by comparing the fiducial set of weights with an alternative version that uses a different methodology. The fiducial set of weights, validated in \citet{y3-galaxyclustering}, is obtained using a principal component analysis of the 107 observing conditions maps, using the first 50 identified modes as the basis (labelled as \texttt{ISD-PC<50} in \citealt{y3-galaxyclustering}). We compare the impact of using the fiducial weights to applying the ones labeled as \texttt{ISD-STD34} in \citet{y3-galaxyclustering}, which were obtained using 34 observing conditions maps as a basis instead. 
    \item \textbf{Shear response}: See Sec.~\ref{sec:responses_approx}. The uncertainty on the shear response correction is determined using image simulations in \citet{y3-imagesims} and propagated to the analysis using the multiplicative bias parameters, which are marginalized over in the cosmological analysis. 
    \item \textbf{TreeCorr approximation}: See Sec.~\ref{sec:pipeline_technical_details}.  
    \item \textbf{Boost factors}: See Sec.~\ref{sec:boost_factors}.
    \item \textbf{Lens magnification}:  See Sec.~\ref{sec:lens_mag}. We determine how the uncertainty on the lens magnification model component propagates to the tangential shear total model by comparing two different set of magnification coefficients. We compare the fiducial values which are fixed in the 3$\times$2pt analysis (displayed in Table~\ref{tab:samples} and obtained using \texttt{Balrog} in \citet*{y3-2x2ptmagnification}) to the values obtained from the data themselves, displayed in Table 2 from \citet*{y3-2x2ptmagnification}. 
    \item \textbf{Intrinsic alignments (IA)}: See Sec.~\ref{sec:IA}. We determine the uncertainty in the tangential shear model coming from the uncertainty in the IA model by comparing the best-fit theory curve to the one generated using different IA values, chosen from a point in the 3$\times$2pt chain that is at around $2\sigma$ from the best-fit values: $A_1 = 1.02, A_2 = -1.22, \alpha_1 = -0.016, \alpha_2 = 0.41, b_{TA}= 0.14$ (vs $A_1 = 0.60, A_2 = -0.16, \alpha_1 = 4.2, \alpha_2 = 3.8, b_{TA}= 0.074$ for the best-fit). We do not include uncertainty coming beyond the TATT model.
    \item \textbf{Source redshifts}: We do not show the contribution of the source redshifts to the model since they are essential, i.e., the model cannot be computed without an estimated redshift distribution. We compute the uncertainty comparing the best-fit model to values in the source redshift parameters that are $2\sigma$ away from the best-fit values in the 3$\times$2pt posterior. 
    \item \textbf{Lens redshifts}: Analogous to the source redshifts. 
    \item \textbf{Multiplicative shear bias}: The uncertainty is computed analogously to one for redshifts. 
    \item \textbf{NK shear response}: See Sec.~\ref{sec:responses_approx}. This test corresponds to using the scale dependent response factors using the NK correlations within TreeCorr. 
    \item \textbf{Non-linear galaxy bias}: See Sec.~\ref{sec:galaxy_bias_tests}. The contribution from higher-order terms to the fiducial linear galaxy bias model together with the baryonic effects described below was the main limitation to define scale cuts. We estimate its uncertainty by comparing the fiducial non-linear bias model used for scale cuts (and shown in Fig.~\ref{fig:nlb}) with the same model generated with different values for the higher-order $b_2$ term. The values from the $b_2$ term have been obtained from a point in the 3$\times$2pt chain assuming non-linear bias that is separated $\sim$2$\sigma$ from the best-fit values (specifically $0.83, 1.04, -0.38, 0.17, 3.72$, for each of the lens redshift bins, in comparison with the original $0.38, 0.37, 0.44, 0.72, 0.90$ values). 
    \item \textbf{Baryonic effects}: See Sec.~\ref{sec:galaxy_bias_tests}. To estimate the uncertainty in the baryonic effects on the galaxy-galaxy lensing probe we compare the fiducial contamination obtained from the OWLS hydrodynamic simulation \citep{OWLS, vanDaalen11} to contamination from the EAGLE simulation \citep{eagle}. The contamination coming from  EAGLE is much smaller than the OWLS one and actually almost negligible over the scales that we use for the cosmology analysis. That is the reason why in the table the contribution and uncertainty have a similar value.
    \item \textbf{Reduced shear}: See Sec.~\ref{sec:red_shear}. We estimate the uncertainty in this higher-order effect using the differences between the theoretical model for the reduced shear labeled as Tree-level $P_\text{NL}$ in Fig.~\ref{fig:reduced_shear} and the one estimated from the \textsc{Buzzard} $N$-body simulation, also shown in that figure. 
    \item \textbf{Source magnification + Reduced shear}: See Sec.~\ref{sec:source_mag_source_clu}. We estimate its uncertainty scaling the uncertainty we obtain from the reduced shear effect (since the source magnification term is computed using the same base integral). 
\end{itemize}

Analyzing in detail the contribution and uncertainties of the current analysis is also useful to help us make predictions for future analyses, including understanding better what the limitations will be. A critical question for larger lensing datasets, such as DES Y6, and the Euclid, LSST and WFIRST lensing surveys, is how the control of uncertainties will be improved. This improvement is required to keep them subdominant to statistical errors. While we have not  studied this challenging problem here, the results summarized in Table~\ref{tab:all_effects} provide a basis for figuring out the prospects for galaxy-galaxy lensing. A number of sources of uncertainty are small enough that we can be confident they will remain subdominant for a survey with S/N that is 2-4 times larger (e.g. Boost factors with DES Y6 and LSST Year 1 data). Other sources of uncertainty, such as source redshifts may require improved calibration, while astrophysical effects such as intrinsic alignments may require improved theoretical modeling coupled with empirical constraints. We leave this exercise for future work.

\section{Conclusions}
\label{sec:conclusions}

We obtain and validate the galaxy-galaxy lensing measurements that are used in the DES Y3 3$\times$2pt analysis \citep{y3-3x2ptkp}. They are also used in the 2$\times$2pt analyses \citep{y3-2x2ptbiasmodelling, y3-2x2ptaltlensresults} and to obtain the small scale lensing ratios described in \citet*{y3-shearratio} that are then used in the cosmic shear analyses (\citealt{y3-cosmicshear1}; \citealt*{y3-cosmicshear2}). We measure the mean tangential shear between 2.5 and 250 arcmin for two different lens galaxy samples: a sample of photometrically selected luminous red galaxies with excellent photometric redshifts (the so-called \redmagic \ sample; \citealt{Rozo2015, y3-galaxyclustering}) and a four times denser flux limited sample (\textsc{MagLim}; \citealt{y3-2x2maglimforecast}), which is used as fiducial in the 3$\times$2pt analysis. For source galaxy shears we use the DES Y3 \metacal \ catalog described in \citet*{y3-shapecatalog}. We validate the measurements both in the large-scale regime used in the cosmological analysis (above $6h^{-1}$ Mpc) and in the small scale regime (below $6h^{-1}$ Mpc) which is used for the shear-ratio analysis \citep*{y3-shearratio}. The same measurement methodology and testing we develop in this paper is also used in \citet{y3-HOD} to extend the measurements to smaller scales (down to 0.25 arcmin) in order to fit them with a halo occupation distribution (HOD) model. We also present and illustrate the different components of our fiducial model, which was defined in \citet{y3-generalmethods}, and discuss the impact of higher-order lensing effects. 

Our fiducial mean tangential shear measurements are the highest signal-to-noise galaxy-galaxy lensing measurements to date. For the magnitude-limited sample we obtain a S/N of $\sim$148 ($\sim$120 for \textsc{redMaGic}). The S/N becomes $\sim$67 ($\sim$55) after applying the scale cut of 6 Mpc/$h$ and removing the two highest redshift bins for the \textsc{MagLim} sample, which are excluded from the DES Y3 3x2pt cosmological analysis. After applying the point-mass marginalization scheme developed in \citet{MacCrann_2019} to localize the tangential shear measurements the S/N becomes $\sim$32 for \textsc{MagLim} and $\sim$28 for \redmagic. Our fiducial measurements include boost factors, random point subtraction and a correction for the mean shear \metacal \ response. We find that the approximation of using the mean shear response for each source redshift bin --- instead of averaging the response for lens-source pairs falling in each angular bin --- is highly accurate given the current uncertainties. Therefore scale-dependent shear responses are not needed in this analysis and will likely not be necessary for future data sets either. In this analysis we use a sample of random points which is 40 times more numerous than the lens sample. We find that this adds a minor level of noise but recommend using more random points in future analysis to further minimize the impact of this effect. We find that the boost factors, which correct for lens-source clustering effects on the redshift distributions, are negligible for large scales but become relevant at small scales. We also conclude the tangential shear measurements are robust to observing conditions and PSF model residuals, as well as obtaining that the cross-component of the shear is compatible with the null. 

The fiducial model used in the DES Y3 3$\times$2pt analysis is based on the non-linear matter power spectrum from \texttt{Halofit} \citep{Takahashi_2012} with a linear galaxy bias model validated with higher-order effects \citep{y3-2x2ptbiasmodelling}, a Fourier-to-real space curved-sky projection and angular bin averaging. To account for the fact that the mean tangential shear quantity is non-local, we analytically marginalize over a point-mass following the procedure described in \citet{MacCrann_2019}. We also include effects from lens magnification, with the constants of proportionality determined from \texttt{Balrog} image simulations in \citet*{y3-2x2ptmagnification}, a five-parameter intrinsic alignment (IA) model that includes tidal-alignment and tidal-torquing terms (TATT) and source galaxy bias effects, and terms including the interplay between lens magnification and IA effects. We have performed an extensive code comparison of our fiducial model pipeline, \textsc{CosmoSIS}, with the \textsc{CosmoLike} code. We find this model to be a decent fit to the data with a $\chi^2$ of 236.3 for 192 data points for \textsc{MagLim} and a $\chi^2$ of 285.7 for 248 data points for \redmagic, for the tangential shear part. 

In this work we also explore and illustrate the impact of source magnification, source clustering and reduced shear, and how they interplay with each other and with the other effects already included in our fiducial model. We discuss how these effects depend on the chosen estimator, in this case the mean tangential shear averaged over lens-source pairs. In this work together with \citet{y3-generalmethods} we find none of the higher-order effects or their combinations will bias our cosmological constraints by more than 0.3$\sigma$ in the $\Omega_m-\sigma_8$ plane. 

Overall, we show that the high S/N tangential shear measurements presented in this work are free of systematic effects and ready to be used in the companion papers showing the combination of clustering and galaxy-galaxy lensing in \citet{y3-2x2ptbiasmodelling, y3-2x2ptaltlensresults} and the combination with cosmic shear in \citet{y3-3x2ptkp}. The low statistical uncertainties of the measurements presented in this work have motivated us to perform a thorough study of several approximations that are commonly used to measure and model the mean tangential shear quantity. The impact of such effects will only become more important in the future with larger and deeper data sets. Thus, the methodology developed in this work lays the foundation for upcoming analyses, e.g. for the final DES Y6 data and future galaxy surveys such as LSST or Euclid.


\section*{Acknowledgements}

JP is supported by DOE grant DE-SC0021429. Funding for the DES Projects has been provided by the U.S. Department of Energy, the U.S. National Science Foundation, the Ministry of Science and Education of Spain, 
the Science and Technology Facilities Council of the United Kingdom, the Higher Education Funding Council for England, the National Center for Supercomputing 
Applications at the University of Illinois at Urbana-Champaign, the Kavli Institute of Cosmological Physics at the University of Chicago, 
the Center for Cosmology and Astro-Particle Physics at the Ohio State University,
the Mitchell Institute for Fundamental Physics and Astronomy at Texas A\&M University, Financiadora de Estudos e Projetos, 
Funda{\c c}{\~a}o Carlos Chagas Filho de Amparo {\`a} Pesquisa do Estado do Rio de Janeiro, Conselho Nacional de Desenvolvimento Cient{\'i}fico e Tecnol{\'o}gico and 
the Minist{\'e}rio da Ci{\^e}ncia, Tecnologia e Inova{\c c}{\~a}o, the Deutsche Forschungsgemeinschaft and the Collaborating Institutions in the Dark Energy Survey. 

The Collaborating Institutions are Argonne National Laboratory, the University of California at Santa Cruz, the University of Cambridge, Centro de Investigaciones Energ{\'e}ticas, 
Medioambientales y Tecnol{\'o}gicas-Madrid, the University of Chicago, University College London, the DES-Brazil Consortium, the University of Edinburgh, 
the Eidgen{\"o}ssische Technische Hochschule (ETH) Z{\"u}rich, 
Fermi National Accelerator Laboratory, the University of Illinois at Urbana-Champaign, the Institut de Ci{\`e}ncies de l'Espai (IEEC/CSIC), 
the Institut de F{\'i}sica d'Altes Energies, Lawrence Berkeley National Laboratory, the Ludwig-Maximilians Universit{\"a}t M{\"u}nchen and the associated Excellence Cluster Universe, 
the University of Michigan, the National Optical Astronomy Observatory, the University of Nottingham, The Ohio State University, the University of Pennsylvania, the University of Portsmouth, 
SLAC National Accelerator Laboratory, Stanford University, the University of Sussex, Texas A\&M University, and the OzDES Membership Consortium.

Based in part on observations at Cerro Tololo Inter-American Observatory, National Optical Astronomy Observatory, which is operated by the Association of 
Universities for Research in Astronomy (AURA) under a cooperative agreement with the National Science Foundation.

The DES data management system is supported by the National Science Foundation under Grant Numbers AST-1138766 and AST-1536171.
The DES participants from Spanish institutions are partially supported by MINECO under grants AYA2015-71825, ESP2015-88861, FPA2015-68048, SEV-2012-0234, SEV-2016-0597, and MDM-2015-0509, 
some of which include ERDF funds from the European Union. IFAE is partially funded by the CERCA program of the Generalitat de Catalunya.
Research leading to these results has received funding from the European Research
Council under the European Union's Seventh Framework Program (FP7/2007-2013) including ERC grant agreements 240672, 291329, and 306478.
We  acknowledge support from the Australian Research Council Centre of Excellence for All-sky Astrophysics (CAASTRO), through project number CE110001020.

This manuscript has been authored by Fermi Research Alliance, LLC under Contract No. DE-AC02-07CH11359 with the U.S. Department of Energy, Office of Science, Office of High Energy Physics. The United States Government retains and the publisher, by accepting the article for publication, acknowledges that the United States Government retains a non-exclusive, paid-up, irrevocable, world-wide license to publish or reproduce the published form of this manuscript, or allow others to do so, for United States Government purposes.


\bibliography{library}
\bibliographystyle{mnras_2author}

\appendix

\section{Cross-covariances in the cross-component analysis}
\label{appendix_gammax}
In Sect.\,\ref{sec:gammax} we have shown that our cross-component measurements are compatible with a null signal neglecting any cross-covariance that might exist between different lens-source bin pairs. In order to support this assumption we have used lognormal simulations. This kind of simulation has shown good agreement with $N$-body simulations and real data up to nonlinear scales\,\citep{Kayo_2001,Hilbert2011,2004LRR.....7....8L} and have previously been used in galaxy-galaxy lensing analyses\,\citep*{Y1GGL}. We use the publicly available FLASK\,\citep{10.1093/mnras/stw874} code to generate 1799 realizations of shear and density mock catalogs consistent with our lens and source samples. We limit ourselves to \redmagic\ lenses, for simplicity, and we refer the reader to \citet{y3-covariances} for all the details regarding the generation of the simulations. We then measure the cross-component in each one of these realizations and derive the covariance matrix. In order to be less sensitive to the exact setup used when generating the simulations and capture potential effects in the data, we combine the covariance obtained from simulations with the uncertainties obtained with the jackknife resampling. Following \citet{2017MNRAS.465..746S}, we normalize the simulations-derived covariance with the diagonal elements of the jackknife covariance:
\begin{equation}
    \text{Cov}_{\theta_i,\theta_j}^{\rm comb} = \text{Corr}_{\theta_i,\theta_j}^{\rm FLASK}\sigma_{\theta_i}^{\rm JK}\sigma_{\theta_j}^{\rm JK}\,,
\end{equation}
where Corr stands for the correlation matrix.

\begin{figure}
\begin{center}
\includegraphics[width=0.45\textwidth]{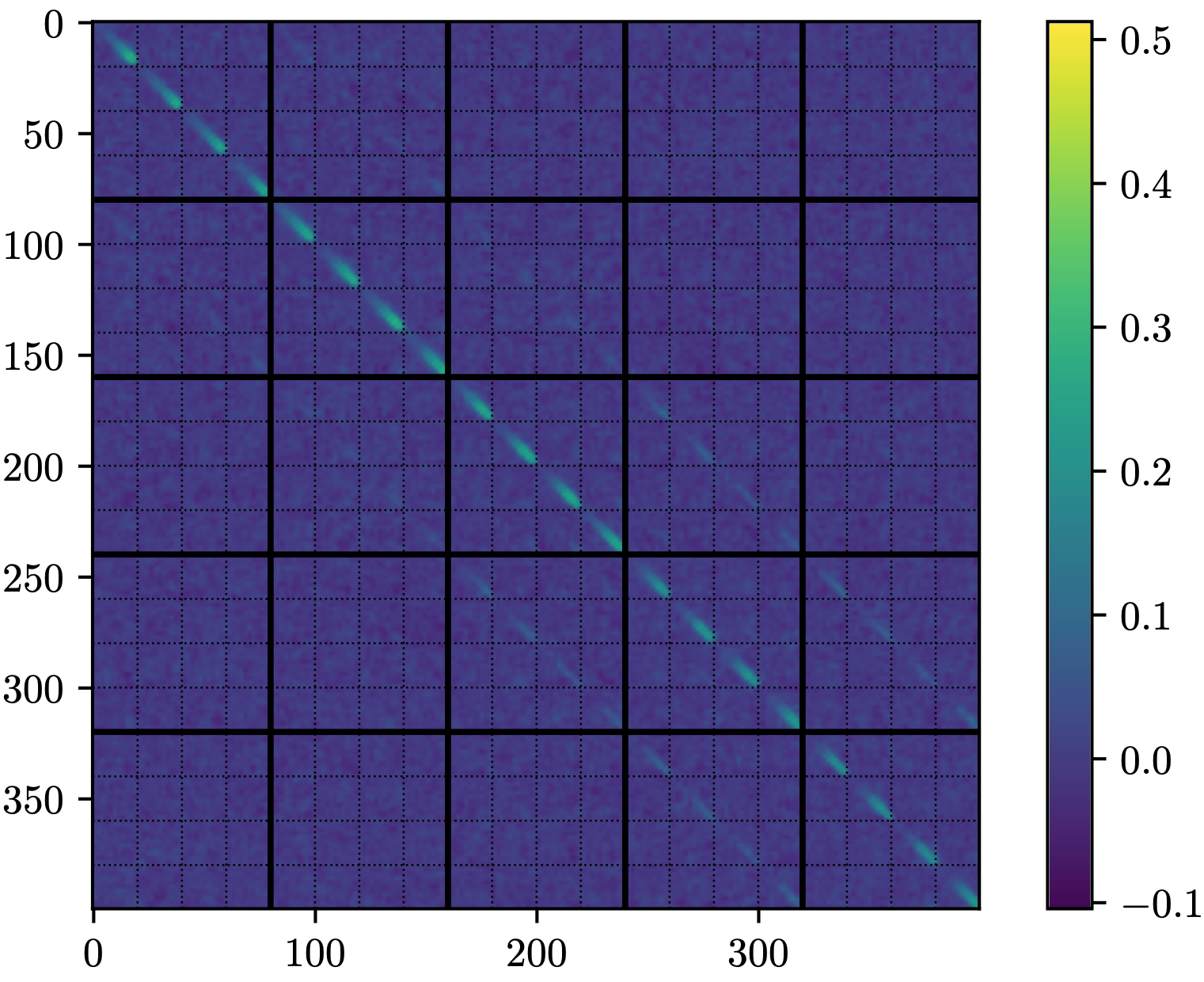}
\caption{Correlation matrix of the cross-component for the \redmagic\ sample accounting for all lens-source pairs cross-covariance using 1799 lognormal simulations combined with jackknife resampling (see the text for details). The figure shows the correlation matrix minus its diagonal for illustrative purposes. Each one of the large squares corresponds to one of the five lens bins, while the small squares correspond to one of the four source bins. Beyond the autocorrelations in each lens-source pair, we can see that bin pairs sharing the same sources and with adjacent lens bins are also correlated.}
\label{fig:gammax_fullcorr}
\end{center}
\end{figure}

The full combined correlation matrix is shown in Fig.\,\ref{fig:gammax_fullcorr}. We can appreciate some cross-covariance between different lens-source pairs. In more detail, we observe that pairs of lens-source bins sharing the same sources and with adjacent lenses are correlated. This is due to the same shape noise realization (of the same sources) and the overlap between adjacent lens bins that can be observed in Fig.\,\ref{fig:nzs}. The lens-lens clustering increases the probability (above random) to have a lens in each bin near the same angular location and therefore getting the same cross-component contribution from noise.

As can be seen in Fig.\,\ref{fig:gammax_fullcorr}, the absolute value of these cross-covariances is much smaller than the main correlations in the diagonal $(\sim 30\%)$. However, in order to confirm whether or not the cross-covariances between different lens-source pairs can be neglected, we compute the total $\chi^2$ of the measured $\gamma_{\times}$ data vector with respect to a null value using this combined covariance matrix. The final value is $\chi_{\rm Null}^2=339$ for a data vector of 400 values. Therefore, the joint analysis also shows that our measurements of the cross-component are compatible with a vanishing signal. Note that in this case we have applied the Hartlap factor with 1799 simulations and 400 angular bins. Just for completeness, we have redone the analysis considering only the largest scales (above one degree). In this case we have obtained a final value of $\chi_{\rm Null}^2=117$ for a data vector of 120 values, showing that also the large-scale measurements are compatible with a vanishing signal. 

\section{Bin-averaging and mask effects}\label{sec:mask_effects}

\begin{figure}
\begin{center}
\includegraphics[width=0.49\textwidth]{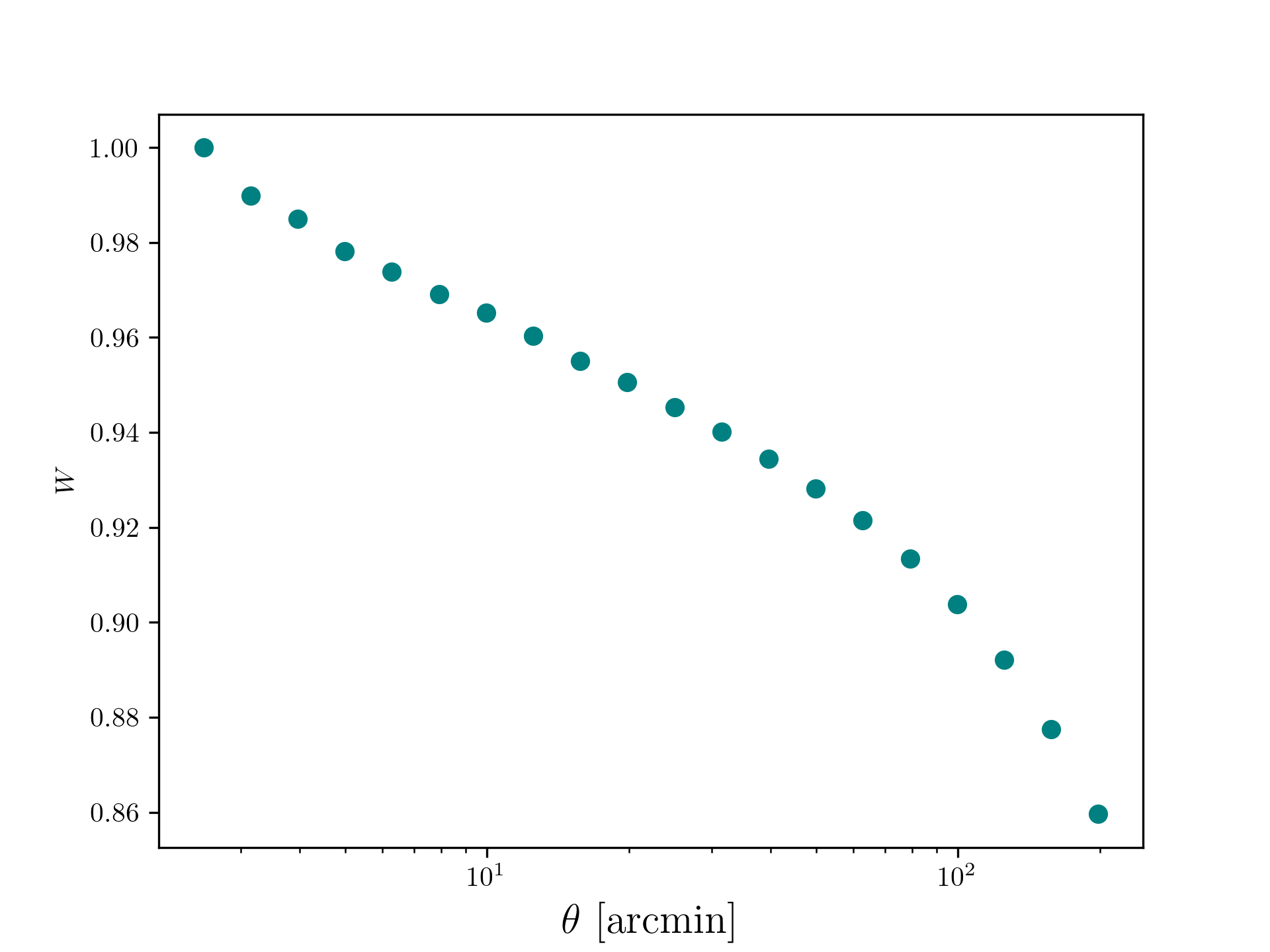}
\caption{The survey window function $W(\theta)$ as computed from the mask auto-correlation function from Eq.~(\ref{eq:mask_autocorrelation}), with the value at the smallest
bin normalized to unity.}
\label{fig:window_function}
\end{center}
\end{figure}

When performing the angular bin-averaging in Eqs.~(\ref{eq:curved_sky}) and (\ref{eq:bin-average}) we have not taken into account the variation in the pairs counts due to the survey geometry, which is expected to affect mostly large scales. This effect has been considered in previous analyses such as \citet{Asgari2019} and \citet{Singh2020}. Here we have estimated its impact computing the window auto-correlation function as
\begin{equation}\label{eq:mask_autocorrelation}
W(\theta) = \sum_\ell \frac{2\ell + 1}{4\pi} P_\ell(\cos \theta) \, C(\ell)
\end{equation}
with $C(\ell)$ being the Fourier space correlation function of the mask. The window function is shown in Fig.~\ref{fig:window_function}, with the value at the smallest scale normalized to unity. The variation of the window function within an angular bin provides an upper limit on the impact of this effect to the modeling of the tangential shear. We have computed the $\Delta \chi^2$ between the fiducial tangential shear model and the fiducial model scaled by the quantity $W(\theta_i,\textrm{min})/W(\theta_i,\textrm{max})$, where $i$ labels a given angular bin. We have found it has negligible impact, with a result of 0.18 using the covariance without point mass marginalization and 0.13 including the point mass, for the 248 data points considered in the cosmological analysis for the \redmagic \ sample. 

\section{Tangential shear around random points and Jackknife covariance tests}
\label{appendix_randompoints}

\begin{figure*}
\begin{center}
\includegraphics[width=0.82\textwidth]{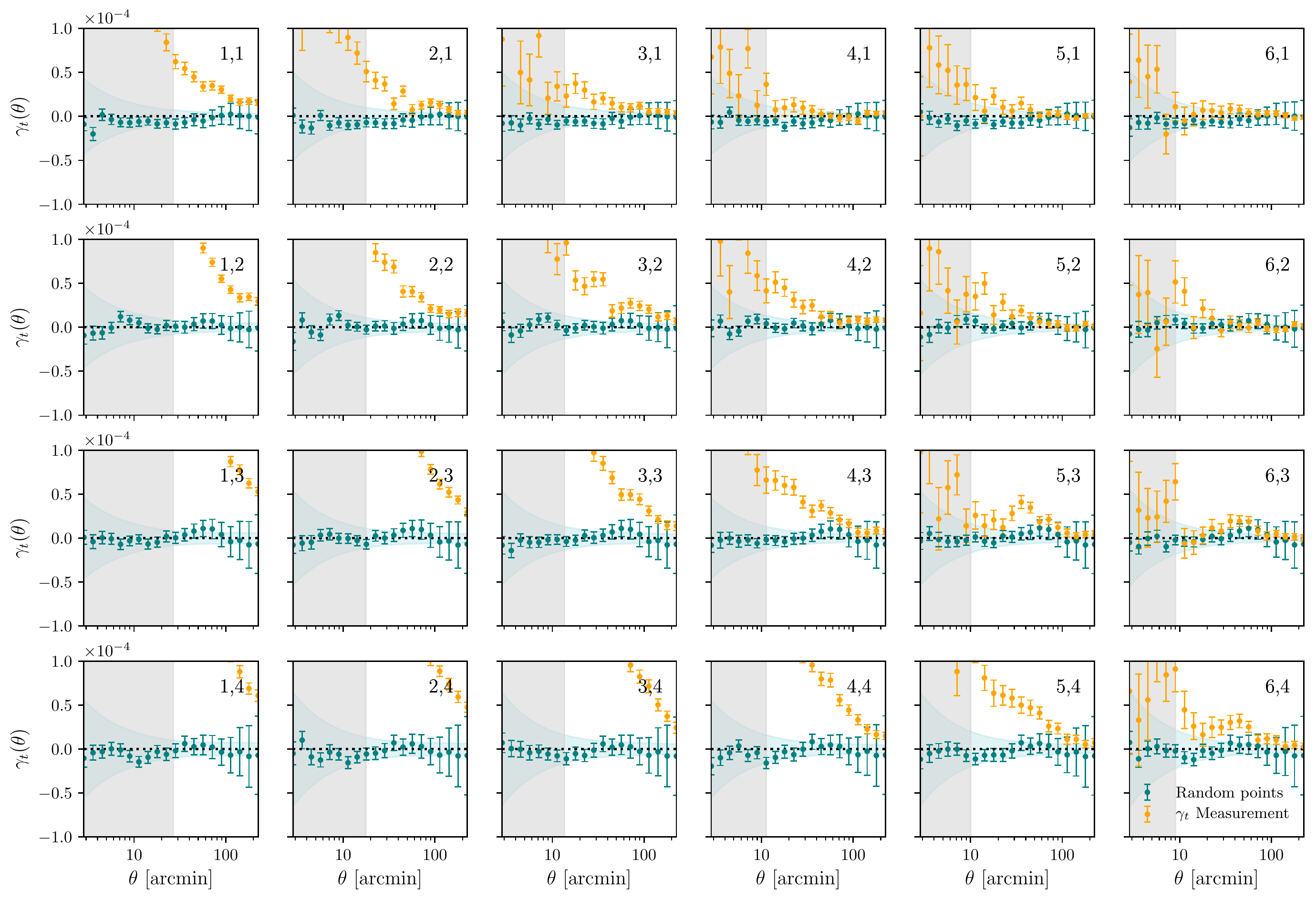}
\caption{Tangential shear around random points using the \textsc{MagLim} sample as lenses in comparison with the signal, with jackknife errorbars in both cases, and comparing with the theoretical uncertainties shown in the blue bands.}
\label{fig:randoms}
\end{center}
\end{figure*}

\begin{figure*}
\begin{center}
\includegraphics[width=0.82\textwidth]{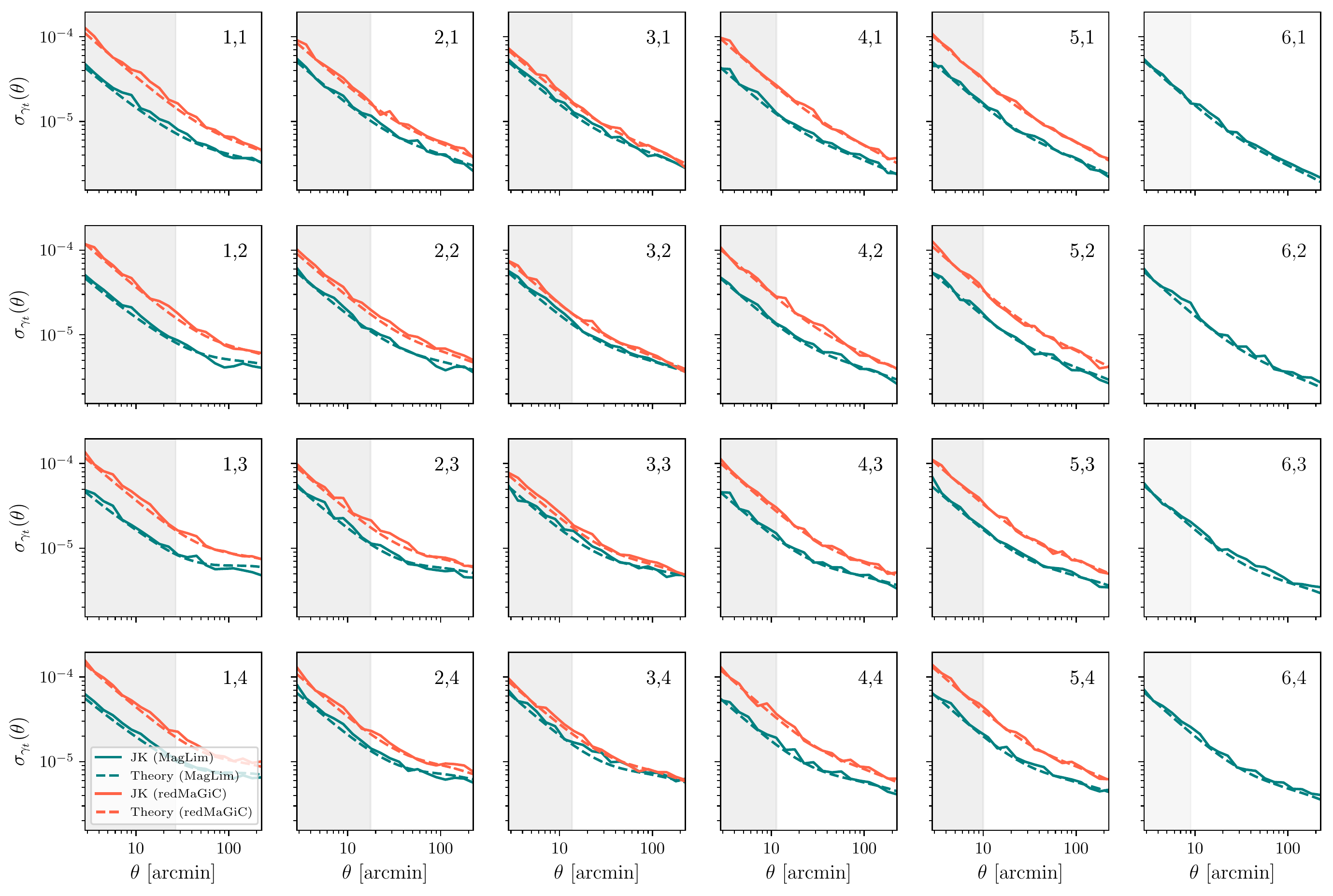}
\caption{Comparison of the Jackknife errorbars (JK) computed in this work as described in Sec.~\ref{sec:pipeline_technical_details} with the theory errorbars from \citet{y3-covariances}.}
\label{fig:errors}
\end{center}
\end{figure*}

The mean tangential shear around random points tests the importance of geometrical and mask effects in the signal. Although our estimator of galaxy-galaxy lensing includes the subtraction of tangential shear measurement around random points, it is useful to check that this correction is small, which is shown in Fig.~\ref{fig:randoms}, especially for the bins with the highest signal. The uncertainties in that plot are obtained from the jackknife method, implemented as described in Sec.~\ref{sec:pipeline_technical_details}. We compare the jackknife uncertainties (JK) to the theoretical uncertainties obtained in \citet{y3-covariances} using a halo model covariance \citep{y3-covariances} in Figs.~\ref{fig:errors} and \ref{fig:covs}. We find the diagonal elements are in agreement to the 10--20\% level. We also compare the uncertainties between the \textsc{MagLim} and the \redmagic samples, finding the \textsc{MagLim} uncertainties are significantly smaller, due to the larger number density of this sample.

\section{Galaxy-galaxy lensing components in Fourier space}
\begin{figure*}
\begin{center}
\includegraphics[width=0.93\textwidth]{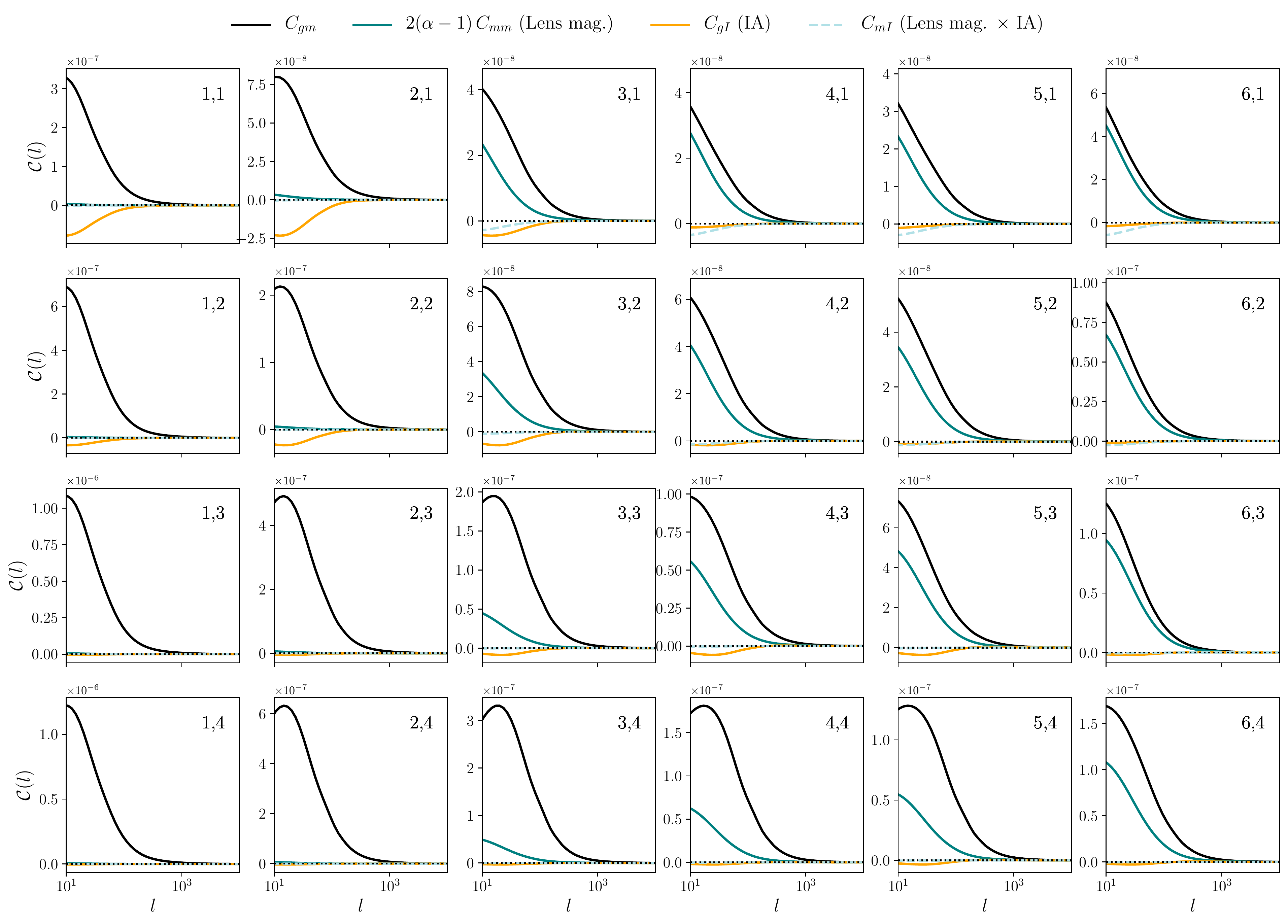}
\caption{This plots shows the contribution from each of the components of our model in Fourier space at the best-fit values from the 3$\times$2pt results for the \textsc{MagLim} sample.}
\label{fig:terms_model_fourier}
\end{center}
\end{figure*}

\begin{figure}
\begin{center}
\includegraphics[width=0.49\textwidth]{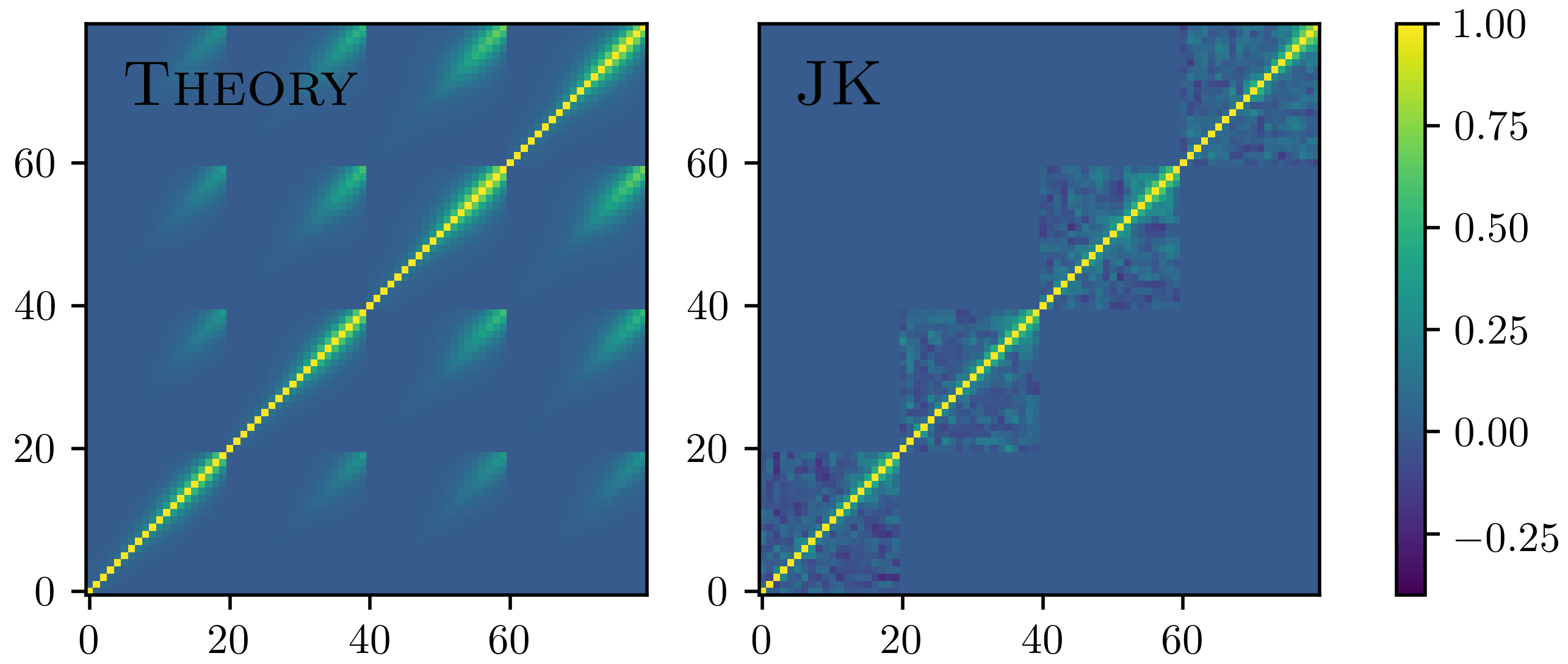}
\caption{Correlation matrix for the \redmagic \ sample using the halo-model theory covariance from \citet{y3-covariances} (left) and the jackknife method (right). This is a subset of the covariance for the second lens bin and each of the four source redshift bins.}
\label{fig:covs}
\end{center}
\end{figure}

In Fig.~\ref{fig:terms_model_fourier} we show the importance of each component of the model at the best-fit values of the 3$\times$2pt cosmology, analogously to Fig.~\ref{fig:terms_model} but now in Fourier space. 

\label{lastpage}
\end{document}